\pdfoutput=1
\documentclass[12pt,a4paper]{article}
\usepackage{amsmath}
\usepackage{epsfig}
\def\tbl#1#2{\begin{center}#2\end{center}\caption{#1}}
\def\abstracts#1{\begin{abstract}\noindent #1\par\end{abstract}}
\let\mycite=\cite
\def\cite#1{ \mycite{#1}}
\def\epspdffile#1{\leavevmode\epsffile{#1.pdf}}
\font\tenbb=msbm10
\font\sevenbb=msbm7
\font\fivebb=msbm5
  \newfam\bbfam
  \def\bb{\fam\bbfam\tenbb}
  \textfont\bbfam=\tenbb
  \scriptfont\bbfam=\sevenbb
  \scriptscriptfont\bbfam=\fivebb
\def\Re{\mathop{\rm Re}}	    
\def\tr{\mathop{\rm tr}}	    
\def\identity{{\bb I}}		    
\def\dd#1#2{{\mathchoice{\frac{d#1}{d#2}}%
  {\frac{d#1}{d#2}}%
  {d#1\!/\!d#2}%
  {d#1\!/\!d#2}}}		    
\def\pdd#1#2{{\mathchoice{\frac{\partial#1}{\partial#2}}%
  {\frac{\partial#1}{\partial#2}}%
  {\partial#1\!/\!\partial#2}%
  {\partial#1\!/\!\partial#2}}}	    
\def\implies{\Rightarrow}	    
\def\defn{\equiv}		    
\def\from{\leftarrow}		    
\def\asymp{\sim}		    
\def\rational#1#2{{\mathchoice{\textstyle{\frac{#1}{#2}}}%
  {\scriptstyle{\frac{#1}{#2}}}{\scriptscriptstyle{\frac{#1}{#2}}}{#1/#2}}}
\def\half{\rational12}		    
\def\quarter{\rational14}	    
\def\rmsub#1#2{#1_{\mbox{\tiny #2}}} 
\def\rmsup#1#2{#1^{\mbox{\tiny #2}}} 
\def\Q{{\bb Q}}			    
\def\R{{\bb R}}			    
\def\Z{{\bb Z}}			    
\def\M{\mathcal{M}}		    
\def\smin{\rmsub s{min}}	    
\def\smax{\rmsub s{max}}	    
\def\lambmax{\rmsub\lambda{max}}    
\def\nexp{\rmsub N{exp}}	    
\def\trjlen{\tau}		    
\def\dt{\delta\tau}		    
\def\dH{\delta H}		    
\def\ad{\mathop{\rm ad}\nolimits}   
\def\USW{\rmsub U{SW}}		    
\def\SF{\rmsub SF}		    
\def\SB{\rmsub SB}		    
\def\nth{\ifmmode\rmsup n{th}\else\(\nth\)\fi} 
\def\kth{\ifmmode\rmsup k{th}\else\(\nth\)\fi} 
\def\nopt{\rmsub n{opt}}	    
\def\ml{\rmsub m{ud}}		    
\def\ms{\rmsub m{s}}		    
\def\Pacc{\rmsub P{acc}}	    
\def\Dslash{\mathchoice
    {D\hskip-0.62em\raise0.2ex\hbox{$\displaystyle/$}\hskip0.2em}%
    {D\hskip-0.62em\raise0.2ex\hbox{$\textstyle/$}\hskip0.2em}%
    {D\hskip-0.5em\raise0.15ex\hbox{$\scriptstyle/$}\hskip0.2em}%
    {D\hskip-0.5em\raise0.15ex\hbox{$\scriptscriptstyle/$}\hskip0.2em}}
\def\dslash{\mathchoice
    {\partial\hskip-0.5em\raise0.2ex\hbox{$\displaystyle/$}\hskip0.2em}%
    {\partial\hskip-0.5em\raise0.2ex\hbox{$\textstyle/$}\hskip0.2em}%
    {\partial\hskip-0.4em\raise0.15ex\hbox{$\scriptstyle/$}\hskip0.2em}%
    {\partial\hskip-0.4em\raise0.15ex\hbox{$\scriptscriptstyle/$}\hskip0.2em}}
\def\sgn{\mathop{\rm sgn}\nolimits} 
\def\secref#1{\S\ref{#1}}	    
\def\sn{\mathop{\rm sn}}	    
\def\cshift{{\mathcal P}}	    
\def\Q{{\mathcal Q}}		    
\def\DDW{\rmsub\Dslash{DW}}	    
\def\mres{\rmsub m{res}}	    
\begin{document}
\title{Algorithms for Dynamical Fermions}
\author{A.~D.~Kennedy\thanks{{\tt adk@ph.ed.ac.uk}} \\ [1ex]
  School of Physics and Astronomy,\\
  University of Edinburgh,\\
  King's Buildings, Mayfield Road,\\
  Edinburgh, EH9 3JZ, Scotland \\ [2ex]
  }
\date{30 July 2006 (revised 21 February 2012)}
\maketitle
\abstracts{This is the write-up of three lectures on algorithms for dynamical
fermions that were given at the ILFTN workshop `Perspectives in Lattice QCD' in
Nara during November 2005. The first lecture is on the fundamentals of Markov
Chain Monte Carlo methods and introduces the Hybrid Monte Carlo (HMC) algorithm
and symplectic integrators; the second lecture covers topics in approximation
theory and thereby introduces the Rational Hybrid Monte Carlo (RHMC) algorithm
and ways of evading integrator instabilities by means of multiple pseudofermion
fields; the third lecture introduces on-shell chiral (Ginsparg--Wilson) lattice
fermions and discusses five-dimensional formulations for computing fermion
propagators for such fermions.\parfillskip=0pt\pretolerance=10000}

\section{Introduction}

This is a written version of a set of lectures on algorithms for dynamical
fermions. The organization of these lecture notes is as follows.

In the first lecture (\secref{sec:BuildingBlocks}) we introduce some `building
blocks' from which we can construct Monte Carlo algorithms for the evaluation
of the functional integrals that describe quantum field theories on the
lattice, and in particular for such computations including the dynamical
effects of fermions. After introducing the basic idea of Monte Carlo
integration and proving the Central Limit theorem we introduce Markov chains
(\secref{sec:MCMC}), and prove the basic theorem on their convergence
(\secref{sec:MCconv}). We explain how detailed balance and the Metropolis
algorithm provide a simple way of constructing Markov steps with some
specified fixed point, and then how we can construct composite Markov steps
that are likely to be ergodic.

Next we briefly introduce perfect sampling by the method of `Coupling from the
Past' that in some cases lets us generate completely uncorrelated samples from
the exact fixed point distribution of a Markov chain. Following this we
consider the effects of autocorrelations, and introduce the Hybrid Monte Carlo
(HMC) algorithm as a way of reducing them without having to resort to
approximating the equilibrium distribution. The requirement HMC has for
reversible and area preserving integrators for Hamiltonian systems leads us to
analyze symplectic integrators (\secref{sec:sympint}) using the
Baker--Campbell--Hausdorff (BCH) formula (\secref{sec:BCH}). The amplification
of floating point arithmetic rounding errors by such integrators is considered
in \secref{sec:reverse}, where we also see the effects of the instabilities
(\secref{sec:instability}) that must occur in symplectic integrators when the
step size becomes too large. The use of multiple timescale integration schemes
to avoid these instabilities is discussed in \secref{sec:multitimescale}.
Finally we discuss the use of pseudofermions within the HMC algorithm to handle
dynamical fermions (\secref{sec:dynFer}).

We begin the second lecture by introducing the theory of optimal polynomial
approximation and establishing Chebyshev's criterion. After elucidating the
r\^ole of Chebyshev polynomials (\secref{sec:chebyPoly}) we consider Chebyshev
optimal rational function approximation in \secref{sec:chebyRat}. A discussion
of the significance of rational approximations for functions of matrices is
given in \secref{sec:ratMatApprox}. The use of multiple pseudofermion fields
(\secref{sec:multipseudofermions}) to reduce the fluctuations in the force
exerted by the pseudofermions on the gauge fields was introduced by Hasenbusch
(\secref{sec:hasenbuschery}), and the way in which this may be implemented
using RHMC follows.

The results of numerical comparison of RHMC with the R~algorithm are given
both for finite temperature QCD (\secref{sec:finiteTemperature}) and for
domain wall fermions (\secref{sec:DWF}). Data showing the efficacy of using
multiple timescale integrators for multiple pseudofermions follows, as does
data showing how these methods succeed in `bringing down the Berlin Wall' for
Wilson fermions.

The third lecture (\secref{sec:GW}) is concerned with five-dimensional
algorithms for Ginsparg--Wilson (GW) fermions. We introduce the concept of
chiral lattice fermions in what we believe to be a logical (rather than
historical) approach, starting with L\"uscher's on-shell chiral
transformation, deriving the GW identity from it, and then showing that
Neuberger's operator is essentially the unique solution (up to the choice of
kernel, and assuming \(\gamma_5\) hermiticity).

We then turn to a class of five-dimensional algorithms to invert Neuberger's
operator: in these the Schur complement (\secref{sec:schur}) of a matrix plays
a central r\^ole. The algorithms are characterized by four independent
choices: the kernel of the Neuberger operator (\secref{sec:kernel}); whether
we introduce five-dimensional pseudofermions as dynamical fields in a Markov
process or just view them as constraints in the computation of the inverse of
the Schur complement (\secref{sec:constraint}); the choice of rational
approximation of the \(\sgn\) (signum) function (\secref{sec:approximation});
and the choice of the five dimensional matrix used to linearize the
approximation to the Neuberger operator (\secref{sec:representation}). The
different choices of five dimensional matrices correspond to different ways of
representing a rational functions; as a continued fraction, a partial
fraction, or as a Euclidean Cayley transform. The latter lead directly to the
domain wall fermions formulation and its generalizations (\secref{sec:ECT}).
Finally we consider the characterization of chiral symmetry breaking in these
approaches, and look at the results of preliminary numerical studies.

\section{Building Blocks for Monte Carlo Methods}
\label{sec:BuildingBlocks}

We start by reviewing the basic ideas of Monte Carlo integration and Markov
processes.\cite{Kennedy:1999}

\subsection{Monte Carlo Methods}

The basic idea of Monte Carlo integration is the mathematical identification
of probabilities with integration measures: we evaluate an integral by
sampling the integrand at points selected at random from a probability
distribution proportional to the integration measure.

Of course, there are much better methods for carrying out numerical integration
(quadrature) in low dimensional spaces; however, all these methods become
hopelessly expensive for high dimensional integrals. Since in lattice quantum
field theory there is one integration per degree of freedom, the only practical
way to carry out such integrations is to use Monte Carlo methods.

The fundamental objective of (Euclidean) quantum field theory is to compute the
\emph{expectation value} of some operator \(\Omega(\phi)\) that depends on the
field~\(\phi\)
\begin{displaymath}
  \langle\Omega\rangle = \frac1Z\int d\phi\, e^{-S(\phi)}\, \Omega(\phi),
\end{displaymath}
where the action is \(S(\phi)\), the measure is \(d\phi\), and the partition
function \(Z\) is introduced to impose the normalisation condition \(\langle1
\rangle=1\).

In order to calculate this expectation value, we generate a sequence of field
configurations \((\phi_1,\phi_2,\dots,\phi_t,\ldots,\phi_N)\) chosen from the
probability distribution
\begin{displaymath}
 P(\phi_t)\,d\phi_t = \frac1Z\,e^{-S(\phi_t)};
\end{displaymath}
how this may be done will be explained later (\secref{sec:MCMC}). We then
measure the value of the operator \(\Omega\) on each configuration, and
compute its average over all the configurations
\begin{displaymath}
 \bar{\Omega} \defn \frac1N\,\sum_{t=1}^N\Omega(\phi_t).
\end{displaymath}
This \emph{sample average}, which we denote by writing a bar over the quantity
averaged, is to be contrasted with the expectation value which is denoted by
enclosing the quantity in angle brackets. The \emph{law of large numbers} then
tells us that the configuration average \(\bar\Omega\) tends to the expectation
value \(\langle\Omega\rangle\) as~\(N\), the number of configurations sampled,
tends to infinity,
\begin{displaymath}
 \langle\Omega\rangle = \lim_{N\to\infty}\bar\Omega.
\end{displaymath}

\subsection{Central Limit Theorem}

The Laplace--DeMoivre \emph{central limit theorem} establishes the stronger
result that under very general conditions the sample average tends to become
Gaussian distributed with the expectation value \(\langle\Omega\rangle\) as
its mean and with a standard deviation which falls as \(1/\sqrt N\),
\begin{displaymath}
 P(\bar\Omega) \sim \mbox{constant} \times
  \exp\!\left[-\frac{\left(\bar\Omega-\langle\Omega\rangle\right)^2}{2C_2/N}
   \right],
\end{displaymath}
where \(C_2\defn\left\langle\bigl(\Omega-\langle\Omega\rangle\bigr)^2
\right\rangle\) is the \emph{variance} of the distribution of \(\Omega\). Note
that the central limit theorem is an asymptotic expansion in \(1/\sqrt N\) for
the probability distribution of~\(\bar\Omega\).

In order to prove the central limit theorem let us consider the distribution
of the configuration average: the distribution of the values of a single
sample, \(\omega=\Omega(\phi)\), is
\begin{displaymath}
 P_{\Omega}(\omega)
  \defn \int d\phi\, P(\phi)\, \delta\bigl(\omega-\Omega(\phi)\bigr)
  = \left\langle\vphantom{\Bigl(}\delta\bigl(\omega-\Omega(\phi)\bigr)
    \right\rangle.
\end{displaymath}
If we take the logarithm of the Fourier transform of this we obtain the
generating function for the connected moments (cumulants) of \(\Omega\),
namely
\begin{eqnarray*}
 W_{\Omega}(k)
 &\defn& \ln \int d\omega\, P_\Omega(\omega)\, e^{ik\omega}
 = \ln \int d\phi\, P(\phi) e^{ik\Omega(\phi)} \\
 &=& \ln \langle e^{ik\Omega}\rangle
 \defn \sum_{n=0}^{\infty} \frac{(ik)^n}{n!}\,C_{n},
\end{eqnarray*}
where the first few cumulants\footnote{\(C_1\) and \(C_2\) are the mean and
variance, \(\sigma \defn \sqrt C_2\) is the standard deviation, and
\(C_3/\sigma^3\) and \(C_4/\sigma^4\) are called the skewness and kurtosis.}
are
\begin{displaymath}
 \begin{array}{rcl@{\qquad}rcl}
     C_0 &=& 0,
   & C_3 &=& \bigl\langle\bigl(\Omega-\langle\Omega\rangle\bigr)^3\bigr\rangle,
    \\ 
     C_1 &=& \langle\Omega\rangle,
   & C_4 &=& \bigl\langle\bigl(\Omega-\langle\Omega\rangle\bigr)^4\bigr\rangle
      -3C_2^2. \\
     C_2 &=& \bigl\langle\bigl(\Omega-\langle\Omega\rangle\bigr)^2\bigr\rangle,
   &
 \end{array}
\end{displaymath}

Next we consider the distribution of the value of the average of \(N\) samples,
\begin{displaymath}
 P_{\bar\Omega}(\bar\omega)
  \defn \int d\phi_1 \cdots d\phi_N\, P(\phi_1) \cdots P(\phi_N)\,
    \delta\Bigl(\bar\omega-\frac1N\sum_{t=1}^N\Omega(\phi_t)\Bigr),
\end{displaymath}
and we construct its connected generating function
\begin{eqnarray*}
 W_{\bar{\Omega}}(k)
  &\defn& \ln\int d\bar{\omega}\,P_{\bar\Omega}(\bar\omega)e^{ik\bar{\omega}}\\
  &=& \ln \int d\phi_1\cdots d\phi_N\,P(\phi_1)\cdots P(\phi_N)\,
   \exp\Bigl(\frac{ik}N\sum_{t=1}^N\Omega(\phi_t)\Bigr) \\
  &=& \ln\Bigl[\int d\phi\,P(\phi) e^{ik\Omega(\phi)/N}\Bigr]^N
  = N\ln\left\langle e^{ik\Omega/N}\right\rangle \\
  &=& NW_{\Omega}(k/N)
 = \sum_{n=1}^\infty\frac{(ik)^n}{n!} \frac{C_n}{N^{n-1}}.
\end{eqnarray*}
We may take the inverse Fourier transform of \(W_{\bar\Omega}\) to obtain an
explicit expression for the distribution \(P_{\bar\Omega}\),
\begin{eqnarray*}
 P_{\bar\Omega}(\bar\omega)
 &=& \frac1{2\pi} \int dk\, e^{W_{\bar\Omega}(k)}e^{-ik\bar\omega} \\
 &\sim& \exp\left[\sum_{n=3}^\infty \frac{(-1)^nC_n}{n!N^{n-1}}
   \left(\dd{}{\bar\omega}\right)^n \right] \int\frac{dk}{2\pi}
   \exp\left[ik\langle\Omega\rangle+(ik)^2\frac{C_2}{2N}-ik\bar\omega\right] \\
 &=& \exp\left[-\frac{C_3}{3!N^2} \left(\dd{}{\bar\omega}\right)^3
   + \frac{C_4}{4!N^3} \left(\dd{}{\bar\omega}\right)^4 - \cdots\right]
   \frac{\exp\Bigl[-\frac{(\bar\omega - \langle\Omega\rangle)^2}
     {2C_2/N}\Bigr]}{\sqrt{2\pi C_2/N}}.
\end{eqnarray*}
This is an asymptotic expansion because the cumulant expansion in general only
converges for sufficiently small values of \(|k|\), whereas the integral is
over all values of \(k\). It can be shown that this leads to corrections of
\(O(e^{-\alpha N})\) for some constant \(\alpha>0\) that, for any given value
of \(N\), will exceed the \(1/N^\ell\) term in the series for some
\(\ell\).\cite{erdelyi:1956}

The distribution \(P_{\bar\Omega}\) tends to a \(\delta\) function as
\(N\to\infty\), and in order to see its Gaussian nature it is useful to rescale
its argument to \(\xi\defn\sqrt N\bigl(\bar{\omega}-\langle\Omega\rangle
\bigr)\), in terms of which
\begin{displaymath}
 P_{\bar\Omega}(\bar\omega) = F(\xi) \frac{d\xi}{d\bar\omega}
\end{displaymath}
with
\begin{displaymath}
 F(\xi) = \left[1 + \frac{C_3\xi(\xi^2-3C_2)}{6C_2^3\sqrt N} + \cdots\right]
  \frac{e^{-\xi^2/2C_2}}{\sqrt{2\pi C_2}}.
\end{displaymath}

\begin{figure}[htb]
  \centerline{\epsfxsize=0.9\textwidth\epspdffile{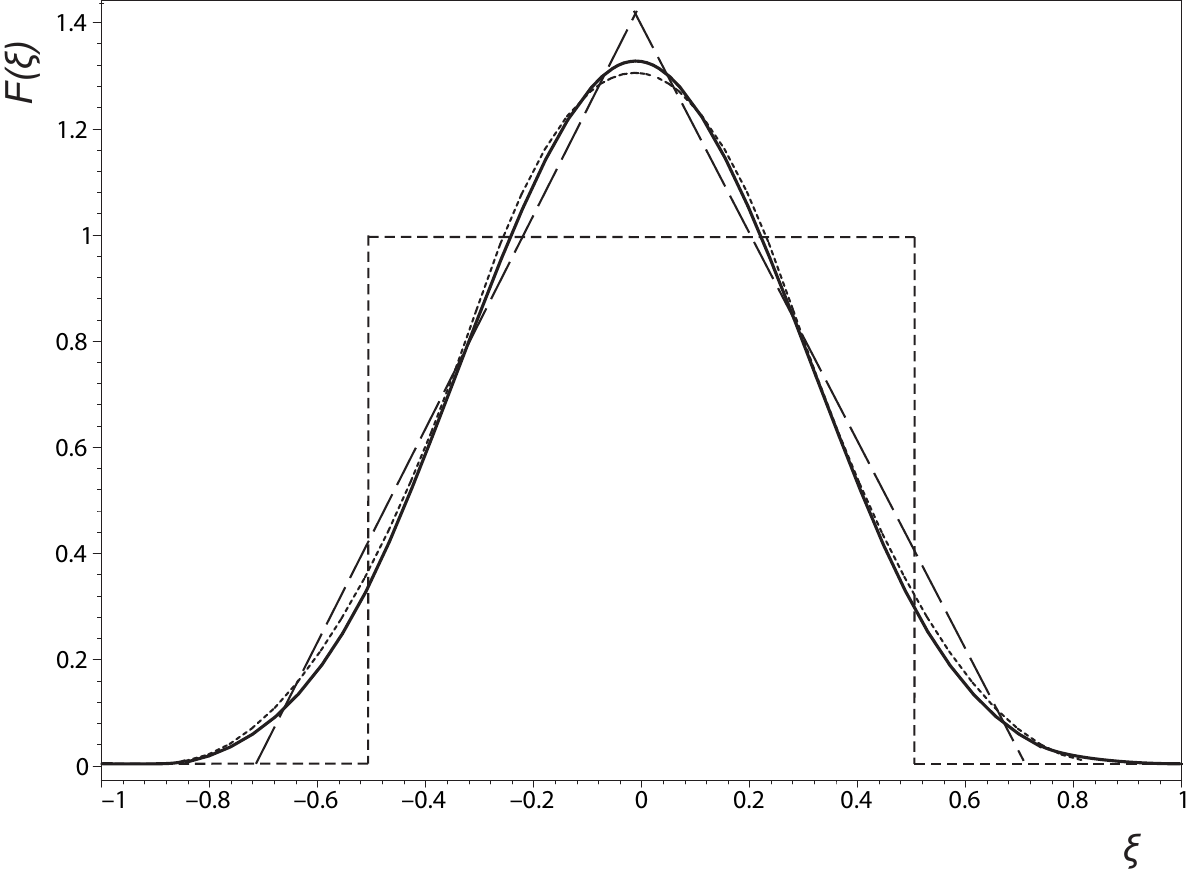}}   
  \caption[Approach of scaled probabiliy distributions to a Gaussian]{A simple
    illustration of how the probability distribution of the rescaled sample
    mean approaches a Gaussian. The four curves show the functions \(F_N(\xi)\)
    of the scaled variable \(\xi=x\sqrt N\) for \(N=1,\ldots,4\). These
    functions are defined by \(F_N(\xi) d\xi = P_N(x) dx\), where the unscaled
    probability distributions are \(P_N(x) \defn \int dx_1\cdots dx_N\,
    P_1(x_1) \cdots P_1(x_N)\, \delta\Bigl(x - \frac1N\sum_{i=1}^N x_i\Bigr)\),
    starting with the very simple (and very non-Gaussian) distribution \(P_1(x)
    = \theta(\half+x) \theta(\half-x)\).}  \label{fig:central_limit_theorem}
\end{figure}

Figure~\ref{fig:central_limit_theorem} illustrates the central limit theorem
by showing how the scaled probability distribution \(F(\xi)\) approaches a
Gaussian distribution for the case where a single sample \(x\) is chosen
uniformly in \(-\half\leq x\leq\half\).

\subsection{Markov Chains}
\label{sec:MCMC}

In order to implement the Monte Carlo integration procedure outlined in the
previous section we need to generate a sequence of configurations distributed
with probability proportional to \(e^{-S}\). The only practicable way that we
can do this is by utilising Markov chains, a procedure that mathematicians
call \emph{Markov Chain Monte Carlo} (MCMC).

Let \(\Omega\) be the state space: for example in the present case each point
in this space is gauge field configuration. We consider an \emph{ergodic}
stochastic mapping \(P:\Omega\to\Omega\); by `stochastic' we mean that
\(P(j\from i)\) gives the \emph{probability} that state \(i\) will be mapped
to state \(j\), but that the actual state that \(i\) is taken to is selected
at random with this probability. By `ergodic' we mean that the probability
getting from any state to any other state is greater than zero. The key to
analyzing the properties of a Markov chain is to think of it as a
deterministic mapping of probability distributions on the state space rather
than as a stochastic mapping on the state space \emph{per se}. A probability
distribution is a mapping \(Q:\Omega\to\R\) which is positive and normalized,
\(Q(x)>0\;\forall x\in\Omega\) and \(\int_\Omega dx\,Q(x)=1\). If we call the
space of all such mappings\footnote{Strictly speaking we define \(Q_\Omega\)
to be the space of equivalence classes of such mappings, as two probability
distributions are to be considered equal if their difference vanishes almost
everywhere: that is, if their difference vanishes except on a set of measure
zero.} \(Q_\Omega\) then the Markov process \(P\) induces a map \(P:Q_\Omega
\to Q_\Omega\).

\subsection{Convergence of Markov Chains}
\label{sec:MCconv}

The principal result in the theory of Markov chains is that whatever
distribution of states we start with (for example we could start with a very
specfic state, such as a totally ordered gauge configuration or `cold
start', or we could choose our starting state entirely at random, a `hot
start'), the sequence of distributions generated by repeatedly applying the
Markov mapping \(P\) converges to unique fixed point distribution
\(\bar{Q}\). The purpose of this section is to establish this result.

To this end we introduce a metric on the space \(Q_\Omega\) of probability
distributions by defining the distance\footnote{It is readily verified that
this satisfies the axioms \(d(x,x)=0\), \(d(x,y)=d(y,x)\), and the
\emph{triangle inequality} \(d(x,y)\leq d(x,z)+d(z,y)\;\forall x,y,z\in
Q_\Omega\); indeed, it just corresponds to the \(L_1\) norm.} between two
probability distributions \(Q_1\) and \(Q_2\)
\begin{equation}
  d(Q_1,Q_2) \defn \int dx\bigl|Q_1(x)-Q_2(x)\bigr|.
  \label{eq:L_1_norm}
\end{equation}
Our approach to the proof is that the Markov process is a \emph{contraction
mapping} with respect to this distance: when we apply \(P\) to two probability
distributions the distance between the resulting distributions is smaller than
the distance between the two original probability distributions
\begin{displaymath}
  d(PQ_1,PQ_2) \leq (1-\alpha) d(Q_1,Q_2),
\end{displaymath}
where \(0<\alpha\leq1\).

Once we have established this, we may argue that \(P\) is a contraction
mapping with respect to the metric of equation~(\ref{eq:L_1_norm}), so the
sequence (\(Q,PQ,P^2Q,P^3Q,\ldots)\) is Cauchy, namely that for any
\(\varepsilon>0\) there is an integer \(N\) such that for all \(n\geq m\geq
N\) we have \(d(P^mQ,P^nQ) < \varepsilon\). This is just the \emph{Banach
fixed-point} theorem, which is easily proved as follows:
\begin{eqnarray*}
  \quad && \hskip-2em
    d(P^mQ,P^nQ) \leq \sum_{j=0}^{n-m-1} d(P^{m+j}Q,P^{m+j+1}Q) \\
  \noalign{\noindent by repeated application of the triangle inequality,}
  &\leq& \sum_{j=0}^{n-m-1} (1-\alpha)^j d(P^mQ,P^{m+1}Q)
  \leq \sum_{j=0}^\infty (1-\alpha)^j d(P^mQ,P^{m+1}Q) \\
  &=& \frac1\alpha d(P^mQ,P^{m+1}Q) \leq \frac{(1-\alpha)^m}\alpha d(Q,PQ)
  \leq \frac{(1-\alpha)^N}\alpha d(Q,PQ) < \varepsilon,
\end{eqnarray*}
where the final inequality holds provided that
\begin{displaymath}
  N > \frac{\ln\left(\displaystyle\frac{\varepsilon\alpha}{d(Q,PQ)}\right)}
   {\ln(1-\alpha)} 
\end{displaymath}
or \(d(Q,PQ)=0\). Note that the condition \(d(PQ_1,PQ_2) < d(Q_1,Q_2)\) is not
strong enough to prove this; indeed, this weaker inequality holds without
requiring the assumption of ergodicity which is in general necessary for the
result to hold. As the space of probability distributions is complete we can
conclude that the sequence converges to a unique fixed point \(\bar Q =
\lim_{n\to\infty}P^nQ\).

In order to show that the Markov process is a contraction mapping, we
proceed as follows:
\begin{eqnarray*}
  d(PQ_1,PQ_2) &=& \int dx\, \bigl|PQ_1(x)-PQ_2(x)\bigr| \\
  &=& \int dx \left|\int dy P(x\from y)Q_1(y)
    - \int dy P(x\from y)Q_2(y)\right| \\
  &=& \int dx \left|\int dy P(x\from y) \Delta Q(y)\right| \\
  \noalign{\noindent where we have defined the quantity \(\Delta Q(y) \defn
  Q_1(y) - Q_2(y)\),}
  &=& \int dx \left|\int dy P(x\from y) \Delta Q(y)
    \Bigl[\theta\bigl(\Delta Q(y)\bigr)
      + \theta\bigl(-\Delta Q(y)\bigr)\Bigr] \right| \\
  \noalign{\noindent upon inserting the trivial step function identity
  \(\theta(y) + \theta(-y)=1\). Next we make use of the identity
  \(\bigl||a|-|b|\bigr| = |a| + |b| - 2\min\bigl(|a|,|b|\bigr)\), which is
  readily established by considering the cases \(|a|\geq|b|\) and \(|a|<|b|\)
  separately,}
  &=& \int dx \int dy\, P(x\from y) \bigl|\Delta Q(y)\bigr| \\
  && \qquad - 2\int dx \min_{\pm} \left|\int dy\, P(x\from y) \Delta Q(y)
    \theta\bigl(\pm \Delta Q(y)\bigr)\right|.
\end{eqnarray*}
The first term simplifies if we note that \(\int dx\, P(x\from y) = 1\) (the
probability of going from state \(y\) to somewhere must be unity), so
\begin{eqnarray*}
  \noalign{\(\displaystyle d(PQ_1,PQ_2)\)}
  \qquad &=& \int dy\, |\Delta Q(y)| 
    - 2 \int dx \min_\pm \left|\int dy\, P(x\leftarrow y) \Delta Q(y)
      \theta\bigl(\pm \Delta Q(y)\bigr)\right| \\
  &\leq& \int dy |\Delta Q(y)|
    - 2 \int dx \inf_y P(x\leftarrow y) \min_\pm
      \left|\int dy\, \Delta Q(y) \theta\bigl(\pm\Delta Q(y)\bigr)\right|.
\end{eqnarray*}
We now observe that
\begin{eqnarray*}
  && \int dy\, \Delta Q(y) \theta\bigl(\Delta Q(y)\bigr) +
  \int dy\, \Delta Q(y) \theta\bigl(-\Delta Q(y)\bigr) \\
  && \qquad = \int dy\, \Delta Q(y) = \int dy\,Q_1(y) - \int dy\,Q_2(y)
  = 1-1 = 0,
\end{eqnarray*}
hence
\begin{displaymath}
  \left|\int dy\, \Delta Q(y) \theta\bigl(\pm\Delta Q(y)\bigr)\right|
    = \half \int dy \bigl|\Delta Q(y)\bigr|.
\end{displaymath}
We thus finally reach the desired result
\begin{eqnarray*}
  d(PQ_1,PQ_2) &\leq& \int dy |\Delta Q(y)|
    - \int dx \inf_y P(x\leftarrow y) \int dy \bigl|\Delta Q(y)\bigr| \\
  &\leq& (1-\alpha) d(Q_1,Q_2),
\end{eqnarray*}
with \(0< \alpha\defn\int dx\,\inf_y P(x\from y) \leq1\).

Suppose that we can construct an ergodic Markov process \(P\) that has some
desired distribution \(\bar Q\) as its fixed point. We then start with an
arbitrary state (`field configuration') and iterate the Markov process until
it has converged (`thermalized'). Thereafter, successive configurations will
be distributed according to \(\bar Q\), but in general they will correlated.

An important point is that we only need the \emph{relative probabilities} of
states \(Q(i)/Q(j)\) to construct \(P\): we do not need to know the absolute
normalization of \(Q\). Conversely, suppose we want to evaluate \(\int_\Omega
dx\, w(x)\) with \(w(x)\geq0\quad\forall x\in\Omega\) by Monte Carlo using a
Markov chain to generate suitably distributed samples of \(x\in\Omega\). The
Markov chain samples \(x\in\Omega\) with probability proportional to \(w(x)\),
but gives us no hint as to what the absolute probability is, so we are unable
to find the value of the integral. In other words, we cannot use Markov chains
to compute integrals directly, but only ratios of integrals of the form
\begin{displaymath}
  \frac{\int_\Omega dx\,w(x)f(x)}{\int_\Omega dx\,w(x)}.
\end{displaymath}
Fortunately this is usually what we want in quantum field theory where we are
not interested in the value of the partition function \emph{per se}.

\subsection{Detailed Balance and the Metropolis Algorithm}

We now consider how to construct a Markov process with a specified fixed point
\begin{displaymath}
 \bar Q(x) = \int dy P(x\from y) \bar Q(y).
\end{displaymath}
A sufficient (but not necessary) condition is to make it satisfy \emph{detailed
balance}
\begin{displaymath}
   P(y\from x) \bar Q(x) = P(x\from y) \bar Q(y):
\end{displaymath}
we can easily show that this implies the fixed point condition by integrating
both sides with respect to \(y\). One simple way of implementing detailed
balance is the \emph{Metropolis algorithm} where we select a candidate state
\(x\in\Omega\) at random\footnote{It is not necessary to choose the candidate
entirely at random; it suffices that the probability of choosing \(x\) when
starting from \(y\) is the same as the probability of choosing \(y\) when
starting from \(x\).} and then accept it with probability
\begin{equation}
   P(x\from y) = \min\left(1,\frac{\bar Q(x)}{\bar Q(y)}\right)
  \label{eq:Metropolis}
\end{equation}
or otherwise keep the initial state \(y\) as the next state in the Markov
chain. We can show that this implies detailed balance (and hence has \(Q\) as
its fixed point) by considering the cases \(\bar Q(x)>\bar Q(y)\) and \(\bar
Q(x)\leq\bar Q(y)\) separately.\footnote{The case \(x=y\) is special as it must
`mop up' all the rejections.}

The particular form of the acceptance probability of
equation~(\ref{eq:Metropolis}) is not unique: other choices are possible,
e.g.,
\begin{displaymath}
  P(x \from y) = \frac{\bar Q(x)}{\bar Q(x)+\bar Q(y)},
\end{displaymath}
but they have lower acceptance.

\subsection{Composition of Markov Steps}

The reason why we can construct bespoke Markov processes from a toolkit of
methods is that we can combine different Markov steps together.  Let \(P_1\)
and \(P_2\) be two Markov steps that both have the desired fixed point
distribution, but are not necessarily ergodic. Then the composition of the two
steps \(P_2\circ P_1\) is a Markov step that also has the desired fixed point,
and it may be ergodic. This trivially generalizes to any (fixed) number of
steps. For the case where \(P_1\) is not ergodic but \(P_1^n\) is the
terminology \emph{weakly} and \emph{strongly} ergodic is sometimes used.

This result justifies `sweeping' through a lattice performing single site
Metropolis updates. Each individual single site update has the desired fixed
point because it satisfies detailed balance; the entire sweep therefore has the
desired fixed point, and furthermore is ergodic. On the other hand, the entire
sweep does not in general satisfy detailed balance; `undoing' the single site
updates would correspond to sweeping through the lattice in the reverse
order. Of course it would satisfy detailed balance if the sites were updated in
a random order, but this is neither necessary nor desirable (because it puts
too much randomness into the system).

\subsection{Coupling from the Past}

It would appear that two of the fundamental limitations of MCMC are that the
distribution of states generated only converges to the desired fixed point
distribution and never exactly reaches it, and that successive states are
necessarily correlated to some extent. Before we turn to methods to alleviate
these problems it is interesting to note that there is a way of sampling a
sequence of completely independent states from the exact fixed point
distribution.

The method of \emph{coupling from the past} or \emph{perfect sampling} was
introduced by Propp and Wilson.\cite{Propp:1996} Imagine that we have some
ergodic Markov chain, where the stochasticity of each step is implemented by
using a book of random numbers. For example, if the system is in state \(i\) at
step \(k\) then we select the \kth~random number \(r_k\in\R\) from our book,
where \(0\leq r_k\leq1\), and set the new state at step \(k+1\) to be \(j\)
where
\begin{displaymath}
  \sum_{\ell=0}^{j-1} P(\ell\from i) < r_k \leq \sum_{\ell=0}^j P(\ell\from i).
\end{displaymath}
We now ask what state \(f\) will the system be in at step \(0\) if it was in
state \(i\) at step \(-N\)? (The use of negative step numbers is just a
notational convenience). This has a well-defined answer that depends on the
random numbers \(r_{-N}, r_{-N+1},\ldots,r_{-1}\) in our book. If \(N\) is
large enough then with probability one \(f\) will be independent of \(i\),
because there is a positive probability at each step that any two states will
map to the same state, and thereafter they will continue through the same
sequence of states since we are using the same sequence of random numbers from
our book (I like to call this the `flypaper principle,' once two states have
coalesced they stay together forever, and ergodicity guarantees that all the
states must coalesce eventually).

Coupling from the Past therefore just consists of finding this state \(f\); it
will be sampled from the exact fixed point distribution because it is the
state that the Markov chain would reach at step \(0\) if it were started at
step \(-\infty\). We can then repeat the entire procedure using a different
book of random numbers to get a completely independent sample.

So far the algorithm is entirely impractical, as it requires following the
sequence of states visited starting with each state \(i\) at step \(-N\), and
in cases of interest the number of states is extremely large. What makes it
practicable in some situations is the existence of a partial ordering of the
states with a largest and a smallest state (this is what mathematicians call a
\emph{lattice}) that is preserved by each Markov step. In other words we have
an ordering\footnote{An \emph{order relation} satisfies the axioms \(x\succeq y
\land y\succeq x \Leftrightarrow x=y\), and \(x\succeq y \land y\succeq z
\implies x\succeq z\: \forall x,y,z\).} such that \(i\succeq j \implies
i'\succeq j'\) where the Markov step takes the unprimed states into the
corresponding primed ones, using the same random number(s) in both cases. The
ordering need only be a partial ordering, so for any pair of states \(i\) and
\(j\) we can have \(i\succeq j\), \(j\succeq i\), both, or
neither. Nevertheless the ordering is a lattice, so there is an \(\smin\) and
an \(\smax\) satisifying \(\smax\succeq i\succeq \smin\:\forall i\). With such
a partial ordering we just need to see if the sequence of states starting with
\(\smin\) and \(\smax\) coalesce: if they do then all states must coalesce and
end at the same state \(f\) at step \(0\); if they do not then we need to
increase \(N\) and repeat the calculation (using the same book of random
numbers, of course).

An interesting non-trivial example where this is practicable is provided by
the Ising model, where a state consists of an array \(s\) of spins each taking
values \(\pm1\), and the desired fixed point distribution is \(Q(s) \propto
\exp\bigl[\beta\sum_{\langle ij\rangle} s_is_j\bigr]\) where the sum is over
all pairs of nearest-neighbour spins and \(\beta>0\). The Markov update step
consists of sweeping through the lattice updating each spin in turn from a
heatbath,\footnote{In general a \emph{heatbath} algorithm is one that directly
samples the desired distrution. This is straightforward for a single-site
update in the Ising model, and with more effort can also be done for
single-link updates in SU(2) gauge theories.}  specifically the new value
\(s'_i\) of the spin at site \(i\) is chosen, independently of the old value
\(s_i\), by taking a random number \(0\leq r\leq1\) and setting
\begin{displaymath}
  s'_i = \left\{
    \begin{array}{c@{\quad}l}
      +1 & \mbox{if}\; r \leq \exp[\beta\sum_{\langle ij\rangle} s_j]/Z \\
      -1 & \mbox{otherwise}
    \end{array}
   \right.
\end{displaymath}
with \(Z\defn \exp[\beta\sum_{\langle ij\rangle} s_j] + \exp[-\beta \sum_{
\langle ij\rangle} s_j]\).  The partial ordering is that \(s\succeq t\) if
\(s_i\leq t_j\:\forall i\), so for example \(---+-+\succeq+-++-+\) but
\(+-++-+\) and \(-++--+\) are not comparable. \(\smax\) is the state with all
spins \(-1\), and \(\smin\) that with all spins \(+1\). We just verify that
the heatbath update preserves this ordering: if the nearest neighbours of the
spin \(s_i\) being updated are the same then this is trivial as the new value,
\(s'_i=t'_i\), is the same for both states. If the some of the neighbours
differ then \(s\succeq t \implies \sum_{\langle ij\rangle}
s_j\leq\sum_{\langle ij\rangle} t_j\), and thus \(t'_i=-1 \implies s'_i=-1\)
so \(s'\succeq t'\).\footnote{Of course \(s'_i=-1 \not\implies t'_i=-1\).}

\subsection{Autocorrelations}

Successive states in a Markov chain are correlated in general. There are two
different ways to measure this autocorrection between states: the first is the
\emph{exponential autocorrelation} (\secref{sec:expautocorr}), which is an
intrinsic property of the Markov process itself; and the second is the
\emph{integrated autocorrelation} (\secref{sec:intautocorr}) for some
observable \(\Omega\), which is more useful insofar as it is directly related
to the statistical error of its measured estimator~\(\bar\Omega\).

\subsubsection{Exponential Autocorrelation}
\label{sec:expautocorr}

In \secref{sec:MCMC} we proved that an ergodic Markov process converges to a
unique fixed point. In terms of the transition matrix \(P_{ij} \defn P(i\from
j)\) this corresponds to \(P\) having a unique eigenvector with eigenvalue one
and all its other eigenvalues lying strictly within the unit circle in the
complex plane. In particular, the magnitude of the largest subleading
eigenvalue must be smaller than 1,\footnote{Indeed, from the proof of
\secref{sec:MCMC}, \(|\lambmax|\leq 1-\alpha\) with \(\alpha=\sum_i \min_j
P_{ij}\).} \(|\lambmax|<1\).

The eigenvectors \(u\) of the Markov matrix satisfy \(\int dy\,P(x\from y)u(y)
= \lambda u(x)\), hence
\begin{displaymath}
  \lambda\int dx\,u(x) = \int dy\int dx\,P(x\from y)u(y) = \int dy\,u(y)
\end{displaymath}
and so either \(\lambda=1\) or \(\int dx\,u(x)=0\).  A similar argument
establishes that any vector \(v_j^{(\lambda)}\) belonging to an eigenvalue
\(\lambda\neq1\) satisfies the same condition.  If we expand an arbitrary
initial probability distribution \(Q\) in a complete basis corresponding to a
block diagonal form of the Markov matrix we thus have \(Q = \bar
Q+\sum_{\lambda\neq1}\sum_j c_j^{(\lambda)} v_j^{(\lambda)}\) where the
coefficient of the fixed point probability distribution \(\bar Q\) is one as
\(\int dx\,Q(x)=1\) for probability distributions.  We therefore find that
\begin{displaymath}
  P^NQ-\bar Q = \sum_{|\lambda|<1} \lambda^N
    \sum_j c_j^{(n,\lambda)} v_j^{(\lambda)}
\end{displaymath}
so the leading deviation from the equilibrium state \(\bar Q\) is of magnitude
\(|\lambmax^N| = e^{N\ln|\lambmax|} = e^{-N/\nexp}\), which falls off
exponentially with the number of Markov steps with a characteristic scale or
exponential autocorrelation time of
\begin{displaymath}
 \nexp \defn -\frac1{\ln |\lambda_{\max}|} > 0.
\end{displaymath}

\subsubsection{Integrated Autocorrelation}
\label{sec:intautocorr}

Consider the autocorrelation of some operator \(\Omega\) measured on a
sequence of successive configurations from a Markov chain. Without loss of
generality we may assume \(\langle\Omega\rangle=0\). The variance of the
estimator (sample average) \(\bar\Omega\) is
\begin{eqnarray*}
 \bigl\langle\bar\Omega^2\bigr\rangle
  &=& \frac1{N^2}\sum_{t=1}^N\sum_{t'=1}^N
    \bigl\langle\Omega(\phi_t) \Omega(\phi_{t'})\bigr\rangle \\
  &=& \frac1{N^2} \Bigl\{\sum_{t=1}^{N} 
    \bigl\langle\Omega(\phi_t)^2\bigr\rangle 
    + 2 \sum_{t=1}^{N-1}\sum_{t'=t+1}^N 
      \bigl\langle\Omega(\phi_t) \Omega(\phi_{t'})\bigr\rangle\Bigr\};
\end{eqnarray*}
and if we introduced the \emph{autocorrelation function} (which is independent
of~\(t\))
\begin{displaymath}
  C_\Omega(\ell)
    \defn \frac{\bigl\langle\Omega(\phi_{t+\ell})\Omega(\phi_{t})\bigr\rangle}
    {\bigl\langle\Omega(\phi)^2\bigr\rangle}
\end{displaymath}
this becomes
\begin{displaymath}
  \bigl\langle\bar\Omega^2\bigr\rangle
  = \frac1N\Bigl\{\bigl\langle\Omega^2\bigr\rangle
    + \frac2N\sum_{\ell=1}^{N-1} (N-\ell) 
      C_\Omega(\ell) \bigl\langle\Omega^2\bigr\rangle\Bigr\}.
\end{displaymath}
The autocorrelation function must fall faster than the exponential
autocorrelation
\begin{displaymath}
 \bigl|C_{\Omega}(\ell)\bigr| \leq \lambmax^\ell = e^{-\ell/\nexp},
\end{displaymath}
so, for a sufficiently large number of samples \(N\gg\nexp\),
\begin{eqnarray*}
  \bigl\langle\bar\Omega^2\bigr\rangle
   &=& \Bigl\{1 + 2\sum_{\ell=1}^\infty C_\Omega(\ell)\Bigr\}
    \frac{\bigl\langle\Omega^2\bigr\rangle}N
    \Bigl[1 + O\Bigl(\frac{\nexp}N\Bigr)\Bigr] \\
   &=& (1 + 2A_\Omega) \frac{\bigl\langle\Omega^2\bigr\rangle}N
    \Bigl[1 + O\Bigl(\frac{\nexp}N\Bigr)\Bigr],
\end{eqnarray*}
where we have defined the \emph{integrated autocorrelation} function
\begin{displaymath}
 A_\Omega \defn \sum_{\ell=1}^\infty C_\Omega(\ell).
\end{displaymath}
This result tells us that on average \(1 + 2A_\Omega\) correlated measurements
are needed to reduce the variance of \(\bar\Omega\) by the same amount as a
single truly independent measurement.

\subsection{Hybrid Monte Carlo}
\label{sec:HMC}

In order to carry out Monte Carlo computations that include fermion dynamics
we would like to find an algorithm which has following features:
\begin{itemize}
 \item it updates the fields globally, because single link updates are
       not cheap if the action is not local;
 \item it takes large steps through configuration space, because small-step
       methods carry out a random walk which leads to critical slowing down
       with a \emph{dynamical critical exponent}\footnote{\(z\) relates an
       autocorrelation to a correlation length of the system, \(A_\Omega
       \propto\xi^z\). The correlation length \(\xi\) is a characteristic
       length-scale of the physical system that diverges as the system
       approaches a continuous phase transition such as the continuum limit in
       the case of lattice quantum field theory.}~\(z=2\).
 \item it does not introduce any systematic errors.
\end{itemize}
A useful class of algorithms with these properties is the (Generalized) Hybrid
Monte Carlo (HMC) method.\cite{Duane:1987de} In the HMC algorithm, we
introduce a `fictitious' momentum \(p\) for each dynamical degree of freedom
\(q\), and we construct a Markov chain with fixed point \(e^{-H(q,p)}\) where
\(H\) is the fictitious Hamiltonian \(\half p^2 + S(q)\): here the action
\(S\) of the underlying quantum field theory plays the r\^ole of the potential
in the fictitious classical mechanical system. This generates the desired
distribution \(e^{-S(q)}\) if we ignore the momenta \(p\) (statisticians call
the distribution of \(q\) ignoring \(p\) a \emph{marginal distribution}).

The HMC Markov chain alternates two Markov steps: the first step is
\emph{Molecular Dynamics Monte Carlo} (MDMC) (\secref{sec:MDMC}), and the
second is (Partial) Momentum Refreshment (\secref{sec:momref}). Both have the
desired fixed point, and their composite is clearly\footnote{But it is not
clear that this can be proved rigorously for any but the simplest systems.}
ergodic.

\subsection{MDMC}
\label{sec:MDMC}

If we could integrate Hamilton's equations exactly we would follow a
trajectory \((q,p)\to (q',p')\) of constant fictitious energy, \(H(q,p) =
H(q',p')\), for fictitious time \(\trjlen\): this corresponds to traversing a
set of equiprobable fictitious phase space configurations. Liouville's theorem
tells us that this trajectory preserves the measure\footnote{This is required
for detailed balance to be satisfied for a continuous system.} \(dq\wedge dp =
dq'\wedge dp'\), and reversing the momenta at the end of the trajectory
ensures that it is reversible, \((q',-p')\to (q,-p)\), so such an update
satisfies detailed balance and therefore would provide the desired Markov
step.

Of course in general we cannot integrate Hamilton's equations exactly, but if
we can find an approximate integration scheme that is exactly reversible and
area-preserving (q.v., \secref{sec:sympint}) then it may be used to suggest
configurations to a Metropolis test with acceptance probability \(\min(1,
e^{-\dH})\), where \(\dH\defn H(q',p')-H(q,p)\) is the amount by which our
integrator fails to conserve fictitious energy. This too gives a Markov step
that satisfies detailed balance.

We therefore build the MDMC Markov step out of three parts:
\begin{enumerate}
 \item \emph{Molecular Dynamics} (MD), an approximate integrator \(U(\trjlen):
       (q,p)\mapsto(q',p')\) that is exactly area-preserving
       \begin{displaymath}
 	 \det U_* = \det\left[\pdd{(q',p')}{(q,p)}\right] = 1,
       \end{displaymath}
       and reversible
       \begin{displaymath}
	  F \circ U(\trjlen) \circ F \circ U(\trjlen) = 1;
       \end{displaymath}
 \item a momentum reveral \(F: p\mapsto -p\);
 \item a Metropolis test.
\end{enumerate}
The composition of these, implementing the Metropolis test using a uniform
random number \(0\leq r\leq1\), gives
\begin{displaymath}
  \begin{pmatrix} q \\ p \end{pmatrix}\to
  \begin{pmatrix} q' \\ p' \end{pmatrix}
    = \left[F\circ U(\trjlen)\, \theta(e^{-\dH}-r) + \identity\,
      \theta(r-e^{-\dH})\right] 
        \begin{pmatrix} q \\ p \end{pmatrix}.
\end{displaymath}

\subsection{Partial Momentum Refreshment}
\label{sec:momref}

The MDMC steps enables us to find acceptable candidate points in phase space
far from the starting place, but it is far from ergodic because it almost
stays on a constant fictitious energy surface. We remedy this by alternating
it with a momentum refreshement which updates the momenta \(p\) from a
Gaussian heatbath without touching the \(q\): while this is also manifestly
not ergodic it can easily make large changes in the fictitious energy.

Partial momentum refreshment is a minor generalization of this: it mixes the
old Gaussian distributed momenta \(p\) with Gaussian noise~\(\xi\),
\begin{displaymath}
  \begin{pmatrix} p' \\ \xi' \end{pmatrix}
  = \begin{pmatrix} \cos\theta & \sin\theta \\ 
    -\sin\theta & \cos\theta \end{pmatrix}
 \circ F \begin{pmatrix} p \\ \xi \end{pmatrix}.
\end{displaymath}
The Gaussian distribution of \(p\) is invariant under \(F\). The extra
momentum reversal \(F\) ensures that for small \(\theta\) the momenta are
reversed after a rejection rather than after an acceptance. For
\(\theta=\pi/2\), which is what is chosen for the standard HMC algorithm, all
momentum reversals are irrelevant.

The reason for introducing partial momentum refreshment in GHMC (also known as
second-order Langevin or Kramer's algorithm) was that for small \(\theta\) and
\(\trjlen\) it introduces a small amount of noise frequently into the
classical dynamics, rather than introducing a lot of noise occasionally as in
HMC, where \(\theta=\pi/2\) and \(\trjlen\approx\xi\). Unfortunately any
benefits this may have are negated by the fact that the momentum has to
reversed after each Metropolis rejection, leading to `zitterbewegung' back
and forth along the trajectory.

\subsection{Baker--Campbell--Hausdorff (BCH) formula}
\label{sec:BCH}

Fortunately there is a large class of reversible and area-preserving
integrators: symmetric symplectic integrators (\secref{sec:sympint}). A useful
tool to analyze these is the BCH formula: if \(A\) and \(B\) belong to an
associative algebra then \(e^Ae^B = e^{A+B+\delta}\), where \(\delta\) is
constructed from commutators of \(A\) and~\(B\), i.e., is in the \emph{Free Lie
Algebra} generated by the set \(\{A,B\}\). More precisely, \(\ln (e^Ae^B) =
\sum_{n \geq 1} c_n\), where \(c_1 = A + B\), and
\begin{eqnarray*}
  c_{n+1} = \frac1{n+1} \biggl\{ 
  &-&\half \ad_{c_n}(A-B) \\
  &+& \sum_{m=0}^{\lfloor n/2\rfloor} \frac{B_{2m}}{(2m)!}
      \sum_{\genfrac{}{}{0pt}{1}{k_1,\ldots,k_{2m}\geq1}{k_1+\cdots+k_{2m}=n}}
        \!\!\!\ad_{c_{k_1}} \!\cdots\, \ad_{c_{k_{2m}}}(A+B) \biggr\},
\end{eqnarray*}
where \(\ad_XY\defn [X,Y]\), and the \(B_n\) are Bernoulli
numbers. Explicitly, the first few terms are
\begin{eqnarray*}
  \lefteqn{\ln(e^Ae^B) = \bigl\{A + B\bigr\} + \half[A,B]
      + \rational1{12} \Bigl\{\bigl[A,[A,B]\bigr] 
        - \bigl[B,[A,B]\bigr]\Bigr\}} && \\ 
  && \qquad - \rational1{24}\bigl[B,\bigl[A,[A,B]\bigr]\bigr] \\
  && \qquad \begin{array}{lrrlrrll}
    + \rational1{720}\Bigl\{ 
    & - & 4 & \bigl[B,\bigl[A,\bigl[A,[A,B]\bigr]\bigr]\bigr]
    & - & 6 & \bigl[[A,B],\bigl[A,[A,B]\bigr]\bigr] & \\
   \vphantom{\Bigl\{} 
    & + & 4 & \bigl[B,\bigl[B,\bigl[A,[A,B]\bigr]\bigr]\bigr]
    & - & 2 & \bigl[[A,B],\bigl[B,[A,B]\bigr]\bigr] & \\
   \vphantom{\Bigl\{}
    & - && \bigl[A,\bigl[A,\bigl[A,[A,B]\bigr]\bigr]\bigr]
    & + && \bigl[B,\bigl[B,\bigl[B,[A,B]\bigr]\bigr]\bigr]
    & \Bigr\} + \cdots;
  \end{array}
\end{eqnarray*}
from this we easily obtain the formula for a symmetric product
\begin{eqnarray}
  \lefteqn{\ln(e^{\half A}e^Be^{\half A}) = \bigl\{A + B\bigr\}
      - \rational1{24} \Bigl\{ 2\bigl[B,[A,B]\bigr] 
        + \bigl[A,[A,B]\bigr] \Bigr\}} \nonumber \\
   && \begin{array}{lrrlrrll}
      + \rational1{5760} \Bigl\{
      && 32 & \bigl[B,\bigl[B,\bigl[A,[A,B]\bigr]\bigr]\bigr]
      & - & 16 & \bigl[[A,B],\bigl[B,[A,B]\bigr]\bigr] & \\
      \vphantom{\Bigl\{}
      & + & 28 & \bigl[B,\bigl[A,\bigl[A,[A,B]\bigr]\bigr]\bigr] 
      & + & 12 & \bigl[[A,B],\bigl[A,[A,B]\bigr]\bigr] & \\
      \vphantom{\Bigl\{}
      & + & 8 & \bigl[B,\bigl[B,\bigl[B,[A,B]\bigr]\bigr]\bigr]
      & + & 7 &  \bigl[A,\bigl[A,\bigl[A,[A,B]\bigr] \bigr]\bigr] 
      & \Bigr\} + \cdots.
    \end{array}
  \label{eq:BCHsym}
\end{eqnarray}

\subsection{Symplectic Integrators}
\label{sec:sympint}

We are interested in finding the classical trajectory in phase space of a
system described by the Hamiltonian
\begin{displaymath}
  H(q,p) = T(p)+S(q) = \half p^2 + S(q).
\end{displaymath}
The basic idea of symplectic integrator is to write the time evolution
operator as
\begin{eqnarray*}
  \exp\Bigl(\trjlen\dd{}t\Bigr)
  &\defn& \exp\left(\trjlen\left\{\dd pt\pdd{}p+\dd qt\pdd{}q\right\}\right) \\
  &=& \exp\left(\trjlen\left\{-\pdd Hq\pdd{}p+\pdd Hp\pdd{}q \right\}\right) \\
  &=& \exp\left(\trjlen\left\{-S'(q)\pdd{}p+T'(p)\pdd{}q \right\}\right)
  \defn \exp(\trjlen h), \\
\end{eqnarray*}
which relates a time derivative on the left to a linear differential operator
on phase space, \(h\), on the the right. In differential geometry such an
operator is a \emph{vector field}, and in the particular case where it is
derived from a Hamiltonian function,
\begin{displaymath}
  h = \iota(H) \defn \pdd Hp\pdd{}q - \pdd Hq\pdd{}p,
\end{displaymath}
it is called a \emph{Hamiltonian vector field}. Let us define
\begin{displaymath}
  Q \defn \iota(T) = T'(p)\pdd{}q,\qquad P \defn \iota(S) = -S'(q)\pdd{}p,
\end{displaymath}
so that \(h = P+Q\). Since the kinetic energy \(T\) is a function only of
\(p\) and the potential energy \(S\) is a function only of \(q\), it follows
that the action of \(e^{\trjlen P}\) and \(e^{\trjlen Q}\) may be evaluated
trivially: by Taylor's theorem
\begin{eqnarray*}
  e^{\trjlen Q} &:& f(q,p) \mapsto f\bigl(q+\trjlen T'(p),p\bigr), \\
  e^{\trjlen P} &:& f(q,p) \mapsto f\bigl(q,p-\trjlen S'(q)\bigr).
\end{eqnarray*}
It also means that these maps are area-preserving,
\begin{eqnarray*}
  \pdd{\bigl(e^{\trjlen Q}(q,p)\bigr)}{(q,p)}
   &=& \det
  \begin{pmatrix} 1 & \trjlen T''(p) \\ 0 & 1 \end{pmatrix} = 1, \\[1ex]
  \pdd{\bigl(e^{\trjlen P}(q,p)\bigr)}{(q,p)}
   &=& \det 
  \begin{pmatrix} 1 & 0 \\ -\trjlen S''(q) & 1 \end{pmatrix} = 1.
\end{eqnarray*}
From the BCH formula (\ref{eq:BCHsym}) we find that the \(PQP\) symmetric
symplectic integrator is given by
\begin{eqnarray*}
  \lefteqn{U_0(\dt)^{\trjlen/\dt}
    \defn \left(e^{\half\dt P}e^{\dt Q}e^{\half\dt P}\right)^{\trjlen/\dt}} 
    \quad \\
  &=& \Bigl(\exp\Bigl\{(P+Q)\dt
    - \rational1{24}\bigl(\bigl[P,[P,Q]\bigr] + 2\bigl[Q,[P,Q]\bigr]\bigr)\dt^3
    + O(\dt^5)\Bigr\}\Bigr)^{\trjlen/\dt} \\
  &=& \exp\Bigl\{\trjlen \left( (P+Q)
    - \rational1{24}\bigl( \bigl[P,[P,Q]\bigr]+2\bigl[Q,[P,Q]\bigr]\bigr)\dt^2
    + O(\dt^4)\Bigr)\right\} \\
  &=& e^{\trjlen h'} = e^{\trjlen(P+Q)} + O(\dt^2).
\end{eqnarray*}
Such symmetric symplectic integrators are manifestly area-preserving and
reversible in addition to conserving energy to \(O(\dt^2)\). What is even more
remarkable is that since the vector field \(h'\) is built out of commutators
of Hamiltonian vector fields it is itself a Hamiltonian vector field:
\begin{displaymath}
  h' = \iota(H') \defn \pdd{H'}p\pdd{}q - \pdd{H'}q\pdd{}p.
\end{displaymath}
This means that for each symplectic integrator there exists a Hamiltonian
\(H'\) that is exactly conserved.

In fact we can calculate \(H'\) quite easily. If \(h_1 = \iota(H_1)\) and
\(h_2 = \iota(H_2)\) are two Hamiltonian vector fields derived from
Hamiltonians \(H_1\) and \(H_2\) respectively then their commutator \(h_3
\defn [h_1,h_2]\) is a Hamiltonian vector field derived from the Hamiltonian
that is the \emph{Poisson bracket} of the Hamiltonians,
\begin{displaymath}
  h_3 = [h_1,h_2] = \iota(H_3) \quad \mbox{where} \quad
    H_3 = \{H_1,H_2\} \defn \pdd{H_1}p\pdd{H_2}q - \pdd{H_1}q\pdd{H_2}p.
\end{displaymath}
Poisson brackets satisfy the Jacobi relation \(\bigl\{A,\{B,C\}\bigr\} +
\bigl\{B,\{C,A\}\bigr\} + \bigl\{C,\{A,B\}\bigr\} = 0\), so they endow the
space of Hamiltonian functions with a Lie algebra structure. Indeed, the
exactly conserved Hamiltonian for our PQP integrator may be written using the
BCH formula (\ref{eq:BCHsym}) with the substitution of Poisson brackets for the
corresponding commutators, \([Q,P] = [\iota(T),\iota(S)]\mapsto\{T,S\}\). With
this technology it is easy to see that
\begin{eqnarray*}
  H' = H &+& \rational1{24} (2p^2S'-S'^2) \dt^2 \\
  &+& \rational1{720} \bigl(-p^4S^{(4)}+6p^2(S'S'''+2S''^2)-3S'^2S''\bigr)
    \dt^4 + O(\dt^6).
\end{eqnarray*}
This expansion converges for small values of \(\dt\), and presumably up to the
value of \(\dt\) for which the integrator becomes unstable
(q.v.,~\secref{sec:instability}).

Note that \(H'\) cannot be written as the sum of a \(p\)-dependent kinetic
term and a \(q\)-dependent potential term, so we cannot make use of this to
find an integrator that exactly conserves \(H\). Moreover, as \(H'\) is
conserved, \(\dH = H(q',p') - H(q,p) = [H(q',p') - H'(q',p')] - [H(q,p) -
H'(q,p)]\) is of \(O(\dt^2)\) for trajectories of arbitrary length even if
\(\trjlen=O(\dt^{-k})\) with \(k>1\).

\subsubsection{Integrator Instability}
\label{sec:instability}

What happens if we take the integration step size \(\dt\) to be large? Clearly
the Metropolis acceptance rate will fall as \(\dt\) increases, but this
behaviour undergoes a sudden change at some value of \(\dt\) where the
integrator goes unstable. We can see this for our PQP integrator even for the
simple case where \(S = \half q^2\);\cite{Edwards:1996vs} in this case the
update \(U_0(\dt)\) is a linear mapping
\begin{displaymath}
  \begin{pmatrix} q \\ p \end{pmatrix}
  \mapsto \begin{pmatrix} q' \\ p' \end{pmatrix}
  = \begin{pmatrix}
      1 - \half\dt^2 & dt \\ -\dt + \quarter\dt^3 & 1 - \half\dt^2
    \end{pmatrix}
    \begin{pmatrix} q \\ p \end{pmatrix}.
\end{displaymath}
The determinant of this matrix is unity because of area-preservation, and its
characteristic polynomial has discriminant \(\dt^2(\dt^2-4)\). When \(\dt=2\)
the system changes from having two complex conjugate eigenvalues of magnitude
one, \(e^{\pm i\phi}\:(\phi\in\R)\), to having two real eigenvalues, \(e^{\pm
\nu} \:(\nu\in\R)\); for \(\dt>2\) this means that the errors increase
exponentially with characteristic Liapunov exponent \(\nu\).

\subsection{Multiple Timescales}
\label{sec:multitimescale}

We are not restricted to using simple symmetric symplectic integrators such as
those described so far.\cite{sexton92a} Suppose that the Hamiltonian is split
into pieces
\begin{displaymath}
  H(q,p) = T(p) + S_1(q) + S_2(q),
\end{displaymath}
then we can define
\begin{displaymath}
  Q \defn \iota(T) = T'(p) \pdd{}q, \qquad
  P_i \defn \iota(S_i) = -S'_i(q) \pdd{}p,
\end{displaymath}
so that \(h=P_1 + P_2 +Q\). We may introduce a symmetric symplectic integrator
of the form
\begin{eqnarray*}
  \lefteqn{\USW(\dt)^{\trjlen/\dt} =} \\
    && \left\{\left[
      \exp\left(\frac{\dt}{2n}P_2\right)
      \exp\left(\frac{\dt}{2n}Q\right)
    \right]^n e^{\dt P_1} \left[
      \exp\left(\frac{\dt}{2n}Q\right)
      \exp\left(\frac{\dt}{2n}P_2\right)
    \right]^n\right\}^{\trjlen/\dt}\!\!\!.
\end{eqnarray*}
We have a lot of freedom: all we need do is assemble the pieces symmetrically
and ensure the the leading term in the BCH expansion is~\(H\). The remaining
freedom can be used to reduce or eliminate higher-order errors in \(\dH\), or
to make the step size small for a particular force term so as to avoid
instabilities. For instance, if \(2n\|P_1\|_\infty\approx\|P_2\|_\infty\), then
the instability in the integrator is tickled equally by each sub-step. This
helps if the most expensive force computation does not correspond to the
largest force.

\subsection{Dynamical Fermions}
\label{sec:dynFer}

The direct simulation of Grassmann fields is not feasible: the problem is not
that of manipulating anticommuting Grassmann variables in a computer, but that
\(e^{-\SF}=e^{-\bar\psi\M\psi}\) is not positive definite and this leads to
poor importance sampling and thus a huge variance in measured quantities.

We therefore integrate out the quadratic fermion fields to obtain the fermion
determinant
\begin{displaymath}
  \int d\bar{\psi}\,d\psi\, e^{-\bar\psi\M\psi} \propto \det M.
\end{displaymath}
The overall sign of the exponent is unimportant.

Any operator \(\Omega(\phi,\bar\psi,\psi)\) can be expressed solely in terms
of the bosonic field by Schwinger's technique of adding source terms
\(\bar\psi\eta + \bar\eta\psi\) to \(\SF\) before integrating over the fermion
fields:
\begin{displaymath}
  \Omega'(\phi) = \left.\Omega\left(\phi, \pdd{}{\eta}, \pdd{}{\bar\eta}\right)
    e^{\bar\eta\M(\phi)^{-1}\eta}\right|_{\eta=\bar\eta=0};
\end{displaymath}
e.g., the fermion propagator is
\begin{displaymath}
  G(x,y) = \left\langle\psi(x)\bar{\psi}(y)\right\rangle = \M^{-1}(x,y).
\end{displaymath}

One obvious way of proceeding would be to include the determinant as part of
the observable to be measured and computing a ratio of functional integrals
\begin{displaymath}
  \langle\Omega\rangle =
    \frac{\bigl\langle\det\M(\phi)\Omega(\phi)\bigr\rangle_{\SB}}
      {\bigl\langle\det\M(\phi)\bigr\rangle_{\SB}}
\end{displaymath}
with \(\SB(\phi)\) being the bosonic (gauge field) part of the action;
but this is not feasible because the determinant is extensive in the lattice
volume, and we get hopelessly poor importance sampling.

We therefore proceed by representing the fermion determinant as a bosonic
Gaussian integral with a non-local kernel
\begin{displaymath}
  \det\M(\phi) \propto \int d\bar\chi d\chi\,
    \exp\left[-\bar\chi\M^{-1}(\phi)\chi\right].
\end{displaymath}
The fermion kernel must be positive definite (all its eigenvalues must have
positive real parts) as otherwise the bosonic integral will not converge. The
new bosonic fields are called \emph{pseudofermions}.

It is usually convenient to introduce two flavours of fermion and to write
\begin{displaymath}
  \bigl(\det\M(\phi)\bigr)^2 = \det\left(\M(\phi)\M^{\dagger}(\phi)\right)
    \propto \int d\bar\chi d\chi\,
    \exp\left[-\bar\chi(\M^\dagger\M)^{-1}\chi\right].
\end{displaymath}
This not only guarantees positivity, but also allows us to generate the
pseudofermions from a global heatbath by applying \(\M^\dagger\) to a random
Gaussian distributed field.

The equations for motion for the boson (gauge) fields are
\begin{eqnarray}
 \dot\phi &=& \pi, \nonumber \\
 \dot\pi &=& -\pdd{\SB}\phi - \chi^\dagger\pdd{(\M^\dagger\M)^{-1}}\phi\chi 
   \nonumber \\
 &=& -\pdd{\SB}\phi + \left[(\M^\dagger\M)^{-1}\chi\right]^\dagger
   \pdd{(\M^\dagger\M)}\phi \left[(\M^\dagger\M)^{-1}\chi\right].
 \label{eq:psForce}
\end{eqnarray}
The evaluation of the pseudofermion action and the corresponding force then
requires us to find the solution of a (large) set of linear equations
\(\M^\dagger\M\psi = \chi\).

It is not necessary to carry out the inversions required for the equations of
motion exactly, there is a trade-off between the cost of computing the force
and the acceptance rate of the Metropolis MDMC step. The inversions required
to compute the pseudofermion action for the Metropolis accept/reject step do
need to be computed exactly, however. We usually take `exactly' to by
synonymous with `to machine precision'.

\subsection{Reversibility}
\label{sec:reverse}

We now want to address the question as to whether HMC trajectories are
reversible and area-preserving in practice.\cite{Edwards:1996vs} The only
source of irreversibility is the rounding errors caused by finite precision
floating point arithmetic, so the fundamental reason for irreversibility is
just that floating point arithmetic is not associative. What we are really
studying is how much the MD evolution amplifies this noise.

For fermionic systems we can also introduce irreversibility by choosing the
starting vector for the iterative linear equation solver time-asymmetrically,
as we do if we use a chronological inverter, which takes some extrapolation of
the previous solutions as the starting vector. Of course we do not have to use
chronological inverter, we can start with a zero vector; alternatively we can
find the solution sufficiently accurately that is does not depend on the
initial guess.

A way that we may study the irreversibility is to follow a trajectory for time
\(\trjlen\), then reverse the momenta and follow it back again; in other words
we compute \(U\circ F\circ U\circ F\), which in exact arithmetic should take
us back exactly to where we started. We then measure the distance \(\Delta\)
in phase space from our starting point, using some suitable norm. What we
observe is that the rounding errors are amplified exponentially with the
trajectory length, \(\Delta \propto \trjlen^\nu\), we call the exponent
\(\nu\) the \emph{Liapunov exponent}. In practice if we work with parameters
such that the Liapunov exponent is small then the resulting irreversibility is
not a big problem, because the `seed' rounding errors are of order
\(10^{-7}\) for 32-bit floating point arithmetic and \(10^{-15}\) for 64-bit
precision: rounding errors fall exponentially with the number of bits we use
to represent floating point numbers.

\begin{figure}[hbt]
  \centerline{\epsfxsize=0.7\textwidth \epspdffile{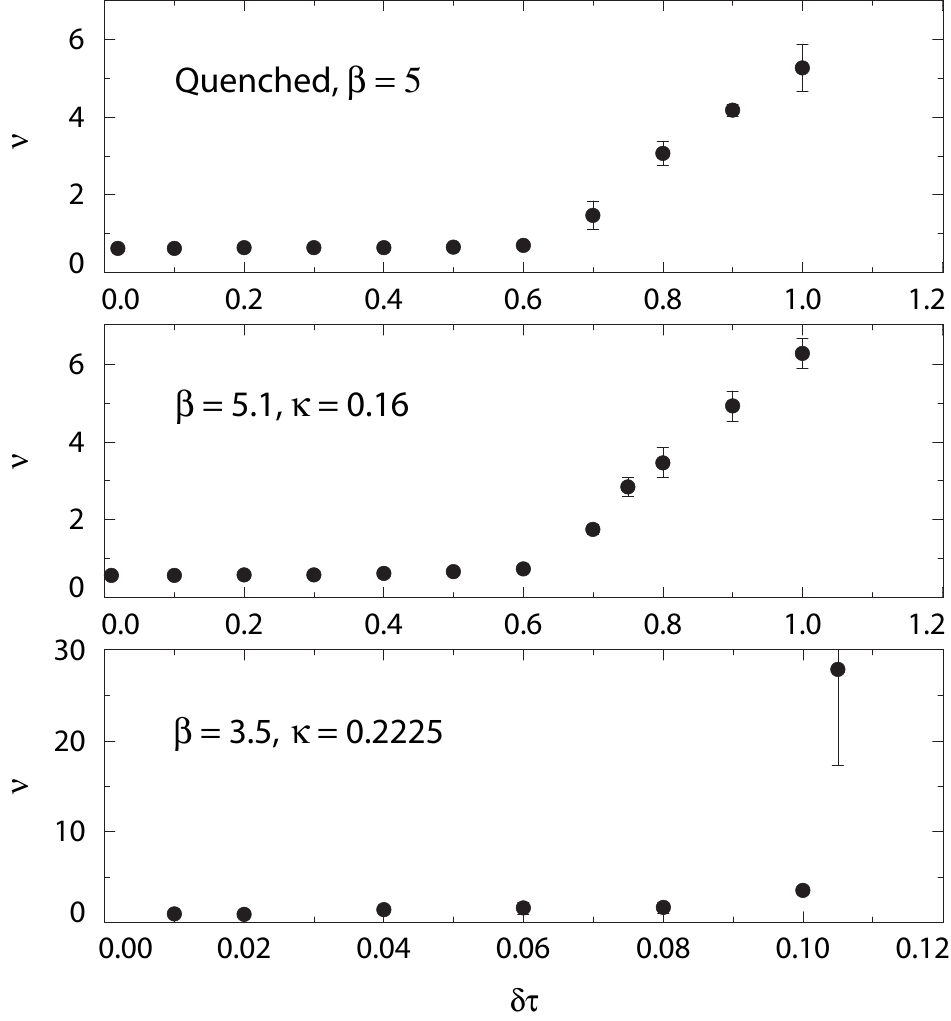}}   
  \caption[SU(3) Instability Plots]{The Liapunov exponent \(\nu\) is shown as a
    function of the integration step size $\dt$. The top graph is for pure
    \(SU(3)\) gauge theory, the middle one is for QCD with heavy dynamical
    Wilson quarks, and the bottom one is for QCD with light dynamical Wilson
    quarks. Note that the scale for the light quark case is quite different
    from the other two. The error bars show the standard deviation of
    measurements made on three independent configurations. All the data is from
    \(4^4\) lattices.}
  \label{fig:reversibility_I}
\end{figure}

In Figure~\ref{fig:reversibility_I} we show data for pure SU(3) gauge theory
and full QCD (both on tiny lattices) where the Liapunov exponent is plotted as
a function of the integration step size \(\dt\).

The fact that for small step size the Liapunov exponent \(\nu\not=0\) appears
to tell us that the underlying continuous-time equations of motion for gauge
field fictitious dynamics are chaotic. In QCD the Liapunov exponent appears to
scale with \(\beta\) as the system approaches the continuum limit, that is
\(\lim_{\beta\to\infty}\nu\xi\approx \mbox{constant}\) where \(\xi\) is the
correlation length as before. This can be interpreted as saying that the
Liapunov exponent characterizes the chaotic nature of the continuum classical
equations of motion, and is not a lattice artefact. If this is so we should
not have to worry about reversibility breaking down as we approach the
continuum limit, as the amplification factor for a trajectory of length
\(\trjlen\approx\xi\) stays fixed. However, beware that this is based on data
from tiny lattices, and is not at all conclusive.

More significantly, at some particular value of the step size, \(\dt\approx
0.6\) for the top two graphs and \(\dt\approx 0.1\) for the bottom one, the
Liapunov exponent start to increase, perhaps linearly. This behaviour is what
we expect when the integrator has become unstable. For the top two graphs this
is not very interesting, as the change in energy along a trajectory of length
\(\trjlen\approx1\) at this step size is very large, \(\dH\gg1\), so the
acceptance rate is essentially zero anyhow. For the bottom graph --- the case
with light fermions --- it is the instability rather than the `bulk'
behaviour of \(\dH\) that limits the step size we can get away with. We shall
investigate on way to circumvent this problem in
\secref{sec:multipseudofermions}.

\section{The RHMC Algorithm}
\label{sec:RHMC}

In this second lecture, we will discuss the Rational Hybrid Monte Carlo (RHMC)
algorithm.\cite{Kennedy:1998cu} For this purpose we shall first take a brief
look at approximation theory.

\subsection{Polynomial Approximation}

We start by consider the problem of approximating a continuous function by a
polynomial. It turns out that the rational approximation theory is simple
generalization of this.

We want to address the question of what is the best polynomial approximation
\(p(x)\) to a continuous function \(f(x)\) over the unit interval
\([0,1]\). To address the question we have to specify an appropriate norm on
the space of functions. The most important class of norms are the \(L_n\)
norms, defined~by
\begin{displaymath}
  \|p-f\|_n = \Bigl(\int_0^1dx\, \bigl|p(x)-f(x)\bigr|^n\Bigr)^{1/n},
  \label{eq:norm}
\end{displaymath}
which satisfies the required axioms\footnote{These are that \(\|f\|\geq
0,\:\|f+g\| \leq \|f\|+\|g\|,\: \|\alpha f\|=|\alpha|\|f\|\:\;\forall f,g\),
where \(\alpha\) is a scalar, and \(\|f\| = 0\) iff \(f=0\) almost everywhere.}
provided \(n \geq 1\). The case \(n=2\), for example, is usual Euclidean
norm. The case \(n=1\) is the \(L_1\) norm which was used in the proof of
convergence of Markov processes in \secref{sec:MCconv}. The \emph{minimax norm}
is defined by taking the limit \(n\to\infty\)
\begin{equation}
  \|p-f\|_\infty = \max_{0\leq x\leq 1} \bigl|p(x)-f(x)\bigr|,
  \label{eq:minimax_norm}
\end{equation}
because when \(n\to\infty\) the integral is dominated by the point at which the
error \(p(x)-f(x)\) is maximal. What we want to do in the following is to find
optimal polynomial approximations with respect to the \(L_\infty\) norm, since
this guarantees a bound on the pointwise error.
We can generalize these definitions by including a positive weight function,
but we will not consider this further in this introduction.

The fundamental theorem on the approximation of continuous functions by
polynomials was given by Weierstrass, who proved that any continuous function
can be arbitrarily well approximated over the unit interval by a
polynomial. This result is very important in functional analysis, where it is
the essential ingredient in many proofs, although it is of less significance
in finding approximations for practical use. The most elegant proof of
Weierstrass' theorem was given by Bernstein, who showed that the
\emph{Bernstein polynomials}
\begin{displaymath}
 p_{n}(x) \defn
 \sum_{k=0}^n f\left(\frac kn\right) \binom nk x^n(1-x)^{n-k}
\end{displaymath}
can arbitrarily well approximate \(f\) over the unit interval by taking \(n\)
to be sufficiently large, that is \(\lim_{n\to\infty}\|p_{n}-f\|_\infty=0\).

\subsection{Chebyshev's theorem}

If we restrict ourselves to considering polynomials of some fixed degree then
the Bernstein polynomial has no reason to be the best approximation
to~\(f\). What we really want is a solution to the `minimax' problem: that is
to find a polynomial \(p\) of fixed degree which gives the minimum \(L_\infty\)
error
\begin{displaymath}
  \Delta = \min_p \|p-f\|_\infty = \min_p \max_{0\leq x\leq1} |p(x)-f(x)|.
\end{displaymath}

A surprisingly elegant solution to the problem of finding the best
approximation of fixed degree was given by Chebyshev, who proved that for any
degree \(d\) there is always a unique polynomial \(p\) that minimizes the
\(L_1\) norm (\ref{eq:minimax_norm}), and it is characterized by the criterion
that the error \(p(x)-f(x)\) attains its maximum absolute value at exactly
\(d+2\) points on the unit interval, and the sign of the error alternates
between successive maxima.

We shall prove Chebyshev's theorem in two steps; we will first establish the
necessity of the criterion stated above, and then prove its sufficiency.

\begin{figure}[htb]
  \centerline{\epsfxsize=0.9\textwidth \epspdffile{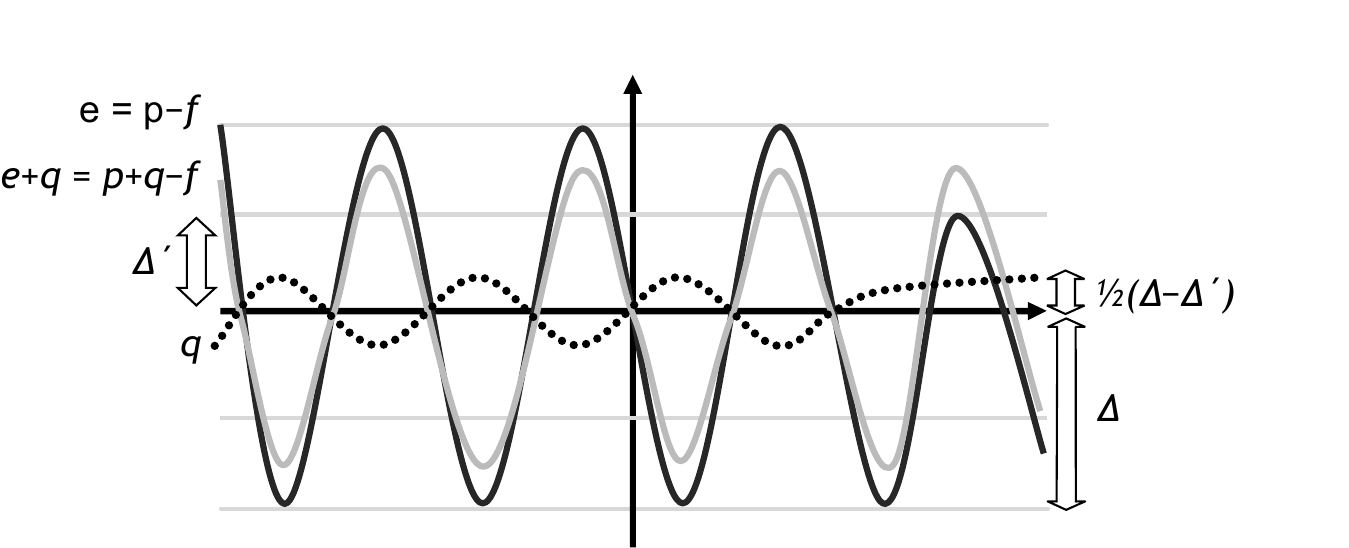}}   
  \caption[Necessity of Chebyshev's criterion]{An illustration of the proof of
    necessity of Chebyshev's condition. \(p\) is a polynomial of degree \(d=7\)
    that approximates a continuous function \(f\). The heavy line shows the
    error \(e=p-f\) that attains its maximum value \(\Delta=\|p-f\|_\infty\) at
    only \(d+1=8\) points (including one end point of the interval). The next
    largest maximum of \(|e(x)|\) has a value of \(\Delta'\), as shown. We
    therefore construct the polynomial \(q\) that passes through a zero of
    \(e\) between each of the points where \(e(x)=\pm\Delta\); there are
    \(d=7\) such points, so \(q\) is also a polynomial of degree \(d=7\). It is
    chosen to have the opposite sign to \(e\) at each of the points where
    \(|e(x)|=\Delta\), and a maximum value of \(\half(\Delta-\Delta')\) over
    the interval. It is shown as the dotted line on the graph. The polynomial
    \(p+q\) is then a better approximation than \(p\), as shown by its error
    \(p+q-f=e+q\) which is the light gray line.}
\end{figure}

Suppose that \(p\) is an optimal polynomial of degree \(d\) for which the error
\(e(x) \defn p(x)-f(x)\) has less than \(d+2\) alternating extrema of equal
magnitude; then at most \(d+1\) of its maxima can exceed some magnitude
\(\Delta'\). This defines a `gap', \(\Delta-\Delta'\). As \(e\) is continuous
it must have at least one zero between two successive maxima; let us suppose
these occur at \(x_i\, (i=1,\ldots,d)\). We can construct a polynomial \(q\) of
degree \(d\) which has the opposite sign to \(e\) at each of the \(d+1\)
maxima,
\(q(x) = k \prod_{i=1}^{d+1} (x-x_i)\), and whose magnitude is smaller than the
`gap' by a suitable choice of the magnitude and sign of the constant \(k\). But
this contradicts the assumption, for the polynomial \(p+q\) is a better
approximation than \(p\) to~\(f\).

\begin{figure}[htb]
  \centerline{\epsfxsize=0.9\textwidth \epspdffile{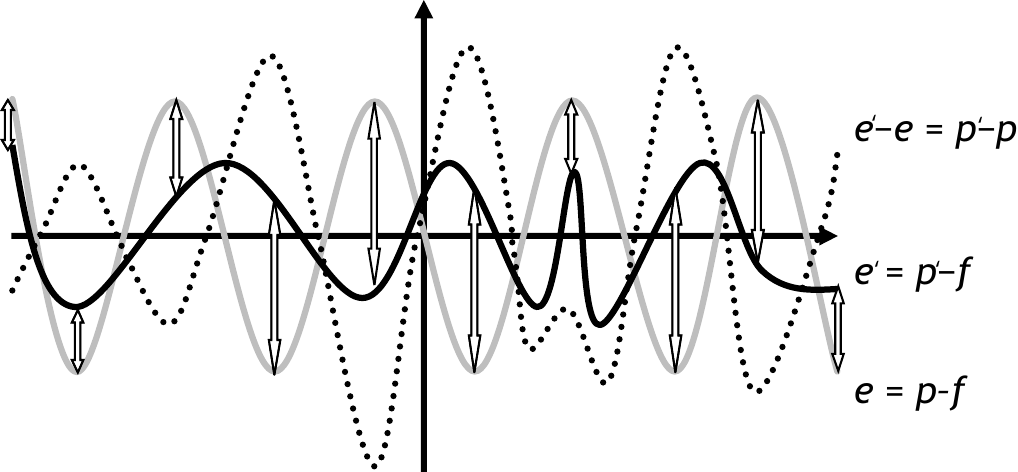}}   
  \caption[Sufficiency of Chebyshev's criterion]{An illustration of the
  sufficiency of Chebyshev's criterion. The polynomial \(p\) of degree \(d=8\)
  satisfies Chebyshev's criterion, as indicated by the graph of its error
  \(e=p-f\) (light gray line) which has \(d+2=10\) alternating extrema of equal
  magnitude on the interval (including the end points). \(p'\) is another
  polynomial also of degree \(d=8\) with a smaller error, \(\|e'\|_\infty
  \leq\|e\|_\infty\), as shown on the graph by the solid line. The difference
  \(p'-p = (p'-f) - (p-f) = e'-e\) shown by the dotted line then must have
  \(d+1=9\) zeros on the interval since \(e(x)-e'(x)\) has an alternating sign
  at each of the extrema of \(e\) (as shown by the arrows), because otherwise
  the maximum error of \(|e'|\) would not be smaller than that of \(e\). Since
  \(p-p'\) is a polynomial of degree \(d=8\) this means it must vanish
  identically by the fundamental theorem of algebra.}
\end{figure}

Sufficiency is shown as following. Suppose that \(p\) satisfies Chebyshev's
criterion, and there is a better polynomial approximation \(p'\) than \(p\),
i.e., \(\|p'-f\|_\infty \leq \|p-f\|_\infty\); then \(|p'(x_i)-f(x_i)| \leq
|p(x_i)-f(x_i)|\) at each of the \(d+2\) extrema. But then \(p'-p\) must have
\(d+1\) zeros on the unit interval, thus \(p'-p=0\) as it is a polynomial of
degree \(d\).

\subsection{Chebyshev Polynomials}
\label{sec:chebyPoly}
 
The r\^ole of Chebyshev polynomials in approximation theory causes some
confusion. The expansion of a function \(f\) in Chebyshev polynomials up to
degree \(d\) does not, in general, give the best approximation of that
degree. What the Chebyshev polynomials do give us it the best approximation of
degree \(d-1\) to the function \(x^d\) over the compact interval \([-1,1]\);
this is given by
\begin{equation}
  p_{d-1}(x) \defn x^d - (\textstyle\half)^{d-1} T_d(x),
  \label{eq:chebPolyApprox}
\end{equation}
where the Chebyshev polynomials are defined by
\begin{displaymath}
  T_d(x) \defn \cos\bigl(d \cos^{-1}(x)\bigr).
\end{displaymath}
The use of the letter \(T\) is from an old transliteration of Chebyshev's
name. Perhaps the most surprising thing about Chebyshev polynomials is that
they are polynomials, despite being defined in terms of transcendental
functions. This is shown by expanding
\begin{eqnarray*}
  \cos(d\xi) &=& \Re e^{id\xi} = \Re (\cos\xi+i\sin\xi)^d
  = \Re \sum_{j=0}^d \binom dj (\cos\xi)^{d-j} (i\sin\xi)^j \\
  &=& \sum_{k=0}^{\lfloor d/2\rfloor} \binom d{2k} (\cos\xi)^{d-2k}
    (-1)^k \bigl[1-(\cos\xi)^2\bigr]^k,
\end{eqnarray*}
from which we immediately see the polynomial nature of \(T_d(x)\): indeed
writing \(x=\cos\xi\) we may proceed to obtain an explicit form
\begin{displaymath}
  T_d(x) = \sum_{k=0}^{\lfloor d/2\rfloor} \sum_{\ell=0}^k
    \binom d{2k} \binom k\ell (-1)^{k+\ell} x^{d-2k+2\ell}.
\end{displaymath}
From this we see that the leading monomial is
\begin{eqnarray*}
  \sum_{k=0}^{\lfloor d/2\rfloor} \binom d{2k} x^d
    &=& \sum_{j=0}^d \binom dj \bigl[1+(-1)^j\bigr] \frac{x^d}2
    = \biggl\{\sum_{j=0}^d \binom dj 
      + \sum_{j=0}^d \binom dj (-1)^j\biggr\} \frac{x^d}2 \\
    &=& \bigl\{(1+1)^d + (1-1)^d\bigr\} \frac{x^d}2 = 2^{d-1}x^d.
\end{eqnarray*}

The error in the approximation of equation~(\ref{eq:chebPolyApprox}) is
\begin{displaymath}
  \bigl\|x^d-p_{d-1}(x)\bigr\|_\infty
  = (\textstyle\half)^{d-1} \bigl\|T_d(x)\bigr\|_\infty
  = 2e^{-d\ln 2},
\end{displaymath}
which obviously satisfies Chebyshev's criterion because \(\cos d\xi\) takes the
values \(\pm1\) exactly \(d+1\) times on the interval \(0\leq\xi\leq\pi\).

In this case the error falls exponentially with the degree \(d\) because the
coefficient of the leading term in the Chebyshev polynomials grows as
\(2^{d-1}\), but in general such exponential convergence does not occur.

\subsection{Chebyshev Optimal Rational Approximation}
\label{sec:chebyRat}

Chebyshev's theorem is easily extended to rational functions (ratios of
polynomials); the criterion that a rational approximation of degree \((n,d)\)
is optimal is that its error must attain its maximum magnitude at \(n+d+2\)
alternating points on the interval.

Rational functions with nearly equal degree numerator and denominator are
usually the best choice. An approximation with \(n<d\) cannot be better than
one with \(n=d\) unless Chebyshev's criterion would lead to the leading \(d-n\)
coefficients of the latter vanishing; this usually only happens when there is
an obvious reason for it, such as trying to approximate a rational function by
one of higher degree, or trying to approximate an odd function in which case
the optimal approximation will generally be of degree \((d\pm1,d)\).

An empirical observation is that convergence is often more often exponential
for optimal rational approximations than for polynomial ones, and that for a
given degree rational functions usually give a much better approximation,
which probably is not surprising.

There are only a few special cases where it is possible to compute optimal
polynomial or rational approximations in closed form: we have seen one in
\secref{sec:chebyPoly} and we will come across another in
\secref{sec:zolotarev}. In general we have to find the optimal approximations
numerically. A simple (but somewhat slow) numerical algorithm for finding the
optimal Chebyshev rational approximation was given by Remez. The Remez
algorithm is basically just an way of implemeting Chebyshev's proof; it
alternates (i)~a phase of searching for an alternating set of \(n+d+2\) points
at which the global maximum of the error occurs, and (ii)~a phase of adjusting
the \(n+d+1\) coefficients of the rational function to make the magnitude of
the error take the same value on the set of points found in the previous
phase. A little though will explain why it is difficult to find the optimal
approximation to a function like \(\sin\frac1x\).

A realistic example of a rational approximation to the function
\(1/\sqrt x\) is
\begin{displaymath}
 \frac1{\sqrt x} \approx\,
  \scriptstyle 0.3904603901 \textstyle
  \frac{(x+2.3475661045)(x+0.1048344600)(x+0.0073063814)}
       {(x+0.4105999719)(x+0.0286165446)(x+0.0012779193)}.
\end{displaymath}
This is accurate to within almost \(0.1\)\% over the range
\([0.003,1]\). Using a partial fraction expansion of such rational functions
allows us to use a multishift linear equation solver, thus reducing the cost
significantly. The partial fraction expansion of the rational function above
is
\begin{displaymath}
 \frac1{\sqrt x} \approx\,
  \scriptstyle 0.3904603901 + \textstyle
  \frac{0.0511093775}{x+0.0012779193} +
  \frac{0.1408286237}{x+0.0286165446} +
  \frac{0.5964845033}{x+0.4105999719}.
\end{displaymath}
This is numerically stable because all the terms are positive and the roots of
the denominators are all real. The fact that this desirable property holds for
optimal rational approximations to \(x^\alpha\) for \(-1\leq\alpha\leq1\) is
one of the principle reasons for the efficacy of the RHMC algorithm, but why it
does is not at all clear.

It is interesting to compare the optimal rational and polynomial approximations
for \(1/\sqrt x\), as this is one of the few cases where we can analyze their
convergence analytically. The optimal rational approximation of degree
\(d\pm1,d\) over the interval \(\varepsilon\leq|x|\leq1\) with a weight
function \(w(x) = \sqrt x\), i.e., where we minimise the maximum relative
error, was found analytically by Zolotarev (\secref{sec:zolotarev}) and has an
\(L_\infty\) error that falls as \(\Delta\asymp\exp(d/\ln\varepsilon)\). The
optimal \(L_2\) approximation with respect to the weight function \(1/\sqrt{1
-x^2}\) is given by the Fourier series
\begin{displaymath}
 \sum_{j=0}^d \frac{(-1)^j4}{(2j+1)\pi}T_{2j+1}(x),
\end{displaymath}
which has \(L_2\) error that falls as \(O(1/d)\). The optimal \(L_\infty\)
approximation cannot be too much better, or it would lead to a better \(L_2\)
approximation, so it certainly only converges as \(1/d\). Note that we are not
being entirely fair in this comparison, as the weights, as well as the
approximation intervals, are different in the two cases.

It is also worth noting in comparing rational and polynomial approximations
that using optimal polynomial approximation to \(1/x\) to compute the inverse
of a matrix (\secref{sec:ratMatApprox}) is essentially the same as using a
Jacobi iteration scheme, and this is well-known to be inferior to Gauss-Seidel
iteration or Krylov space methods.

\subsection{Non-Linearity of CG Solver}

We have now completed our survey of approximation theory,\cite{Kennedy:2004tj}
and we turn to considering its application in the RHMC algorithm. To motivate
the need for this let us consider the following suggestion of a method to
accelerate matrix inversion.

Suppose we want to solve a system of linear equations \(A^2x=b\) where \(A\)
is a positive Hermitian matrix using the conjugate gradient (CG) method. It is
well known that it is better to solve \(Ay=b\) and \(Ax=y\) successively,
because the condition number\footnote{The \emph{condition number} of a matrix
is defined to be the ratio of its largest and smallest eigenvalues. To a first
approximation the cost of solving a system of linear equations is proportional
to the condition number.} is reduced from \(\kappa(A^2) = \kappa(A)^2\) to
\(2\kappa(A)\).

Suppose now that we want to solve \(Ax=b\). Why don't we solve \(\sqrt Ay=b\)
and \(\sqrt Ax=y\) successively? The square root of \(A\) is uniquely defined
as \(A>0\), and we can even choose it to be positive too. Indeed, for
application to quantum field theory most of our time is spent finding
\((\M^\dagger \M)^{-1} \chi\), and the fermion kernel is manifestly positive,
\(\M^\dagger \M>0\). All this generalizes trivially to \nth~roots, so it seems
that we have the opportunity of reducing the cost even further.

The question that needs to be addressed is how do we apply the square root of
a matrix?

\subsection{Rational Matrix Approximation}
\label{sec:ratMatApprox}

Before we can investigate how to evaluate functions of matrices we need to
define what we mean by them.  For our purposes we only need to define a
function of a Hermitian matrix \(H\), and this can be diagonalized by unitary
transformation \(H = UDU^{-1}\).  We define the function \(f(H)\) just by
changing to the basis where \(H\) is diagonal, applying \(f\) to the diagonal
elements (eigenvalues), and then transforming back to the original basis,
\begin{displaymath}
  f(H) = f(UDU^{-1}) \defn Uf(D)U^{-1}.
\end{displaymath}
The great advantage of rational functions of matrices is that they do not
require this diagonalisation to be carried out explicitly, since by linearity
\begin{displaymath}
 \alpha H^m + \beta H^n = U(\alpha D^m + \beta D^n)U^{-1}
 \quad\mbox{and}\quad H^{-1} = UD^{-1}U^{-1}.
\end{displaymath}
It suffices to compute the rational function using the operations of matrix
addition, matrix multiplication, and matrix inversion in place of their scalar
analogues. Furthermore, if we only want the result of applying the matrix
function to a vector, \(f(H)v\), then we can use the appropriate matrix-vector
operations rather than the more costly matrix-matrix ones; in particular we
can use a linear equation solver rather than finding a complete matrix
inverse.

\subsection{`No Free Lunch' Theorem}
\label{sec:NFL}

We now have an efficient way of computing matrix functions by finding a good
rational approximation to the function and applying it using matrix
operations. Let us return to our problem of solving \(Ax=b\) by solving \(n\)
systems of the form \(A^{1/n}x=y\) successively: for each of these we use a
rational approximation to compute\footnote{We save a lot of work here by
noticing that it is just as easy to approximate \(A^{-1/n}\) as it is to
approximate \(A^{1/n}\), so there is no need for nested CG solvers.}
\(x=A^{-1/n}y\). To do this, we must apply the rational approximation of the
\nth~root \( r(A)\approx A^{1/n}\). This is done efficiently by expanding the
rational function in partial fractions and then applying all the terms at once
using a multishift CG solver. The condition number for each term in the
partial fraction expansion is approximately that of the original matrix
\(\kappa(A)\), so the cost of applying \(A^{1/n}\) is proportional to
\(\kappa(A)\). We have therefore found out why our suggested inversion method
fails: even though the condition number \(\kappa(A^{1/n})=\kappa(A)^{1/n}\)
and \(\kappa\bigl(r(A)\bigr)=\kappa(A)^{1/n}\) the cost of applying \(r(A)\)
is \(\kappa(A)\), so unfortunately we don't win anything.

\subsection{Multiple Pseudofermions}
\label{sec:multipseudofermions}

Let us return to the topic we were considering in the previous lecture
(\secref{sec:dynFer}), where we introduced pseudofermions as a means of
representing the fermion determinant. Recall that we rewrote the determinant
as a bosonic functional integral over a pseudofermion field \(\phi\) with
kernel \(\M^{-1}\)
\begin{displaymath}
  \det \M \propto \int d\phi^\dagger d\phi\,
    \exp\left[-\phi^\dagger\M^{-1}\phi\right].
\end{displaymath}
What we are doing is to evaluate functional integrals that include the fermion
determinant \(\det\M\) by using a stochastic estimate of the determinant,
namely we are approximating the integral over the pseudofermion fields by a
single configuration --- in other words evaluating the integrand at only one
point --- yet this produces a Markov process with exactly the correct fixed
point. We again seem to be getting something for nothing, so perhaps we should
be suspicious that we are paying a hidden price somewhere.

The hidden price is that we are introducing extra noise into the system by
using a single pseudofermion field to sample this functional integral. This
noise manifests itself as fluctuations in the force exerted by the
pseudofermions on the gauge fields, and in turn
\begin{itemize}
 \item this increases the maximum fermion force;
 \item which triggers the integrator instability;
 \item which requires decreasing the integration step size.
\end{itemize}

A better estimate is to write the fermion determinant as
\begin{displaymath}
  \det\M = \bigl[\det\M^{1/n}\bigr]^n,
\end{displaymath}
and introduce a separate pseudofermion field for each factor
\begin{displaymath}
  \det\M = \bigl[\det\M^{1/n}\bigr] 
  \propto \prod_{j=1}^n\int d\phi_j^\dagger d\phi_j\,
    \exp\left[-\phi_j^\dagger\M^{-1/n}\phi_j\right].
\end{displaymath}
 
\subsubsection{The Hasenbusch Method}
\label{sec:hasenbuschery}

Before we go further into the RHMC implementation of multiple pseudofermions we
ought to note that the idea of using multiple pseudofermions was originally due
to Hasenbusch.\cite{Hasenbusch:2002ai} He considered a theory with the Wilson
fermion action \(\M(\kappa)\) where \(\kappa\) is the \emph{hopping parameter}
which controls the fermion mass. He introduced a heavy fermion kernel
\(\M'=\M(\kappa')\), and made use of the trivial identity following from the
associativity of matrix multiplication \(\M=\M'(\M'^{-1} \M)\) to write the
fermion determinant as \(\det\M=\det\M' \det(\M'^{-1} \M)\). He then introduced
separate pseudofermions for each determinant, and tuned \(\kappa'\) to minimise
the cost of the overall computation. This can be all be easily generalized to
more than two pseudofermions and to the clover-improved Wilson action.

\subsection{Violation of NFL theorem}

Let us now return to using our \nth~root trick to implement multiple
pseudofermions. As before we observe that the condition number of the \nth~root
\(\kappa\bigl(r(\M)\bigr)=\kappa(\M)^{1/n}\), and we may use the simple model
that the largest contribution to the force acting on the gauge fields due to
one of our pseudofermion fields is inversely proportional to the smallest
eigenvalue of the fermion kernel, at least when the fermion mass is
sufficiently small. This follows by considering equation~(\ref{eq:psForce}),
and recalling that we used \(\M^\dagger\M\) where we now have \(\M^{1/n}\). In
fact we might even expect the force to grow faster than this, but let us stick
with our simple model. As the largest eigenvalue of \(\M\) is more-or-less
fixed this force is proportional to \(\kappa(\M)^{1/n}\); we have \(n\)
pseudofermions each contributing to the force, and if we are conservative and
assume that these contributions add coherently we get a total force
proportional to \(n\kappa(\M)^{1/n}\). This is to be compared to the original
single pseudofermionic force of \(\kappa(\M)\), so we expect that the maximum
force is reduced by a factor of \(n\kappa(\M)^{\frac1n-1}\). This is a good
approximation if the condition number is dominated by a few isolated tiny
eigenvalues, which is just the what happens in the cases of interest. If the
force is reduced by this factor the according to our previous considerations
(\secref{sec:reverse}) we can increase the step size by the reciprocal of this
factor, and all other things being equal, the overall cost is reduced by a
factor of \(n\kappa(\M)^{\frac1n-1}\).

If we take this model seriously then we can easily calculate the optimal
number of pseudofermions by minimising the cost, and we find that the optimal
value is \(\nopt\approx\ln\bigl(\kappa(M)\bigr)\), and the corresponding
optimal cost reduction is \(e\ln\kappa/\kappa\).

So, by introducing a small number of pseudofermion fields (of order
\(\ln\kappa(\M)\)) we expect to get a cost reduction of order \(1/\kappa(\M)\),
and this works in practice --- at least there is a significant cost reduction,
the exact scaling laws of our simple model are undoubtledly only followed very
approximately at best --- thereby violating the infamous `no free lunch'
theorem.

The advantage that this method has over the Hasenbusch method is that it is
trivially applicable to any fermion kernel and that no parameter tuning is
required: the condition number is automatically equipartitioned between the
pseudofermion fields. On the other hand, the Hasenbusch method has the
advantage that one of the pseudofermions may be cheap to apply because it is
heavy.

\subsection{Rational Hybrid Monte Carlo}

Let us now go through the details of the Rational Hybrid Monte Carlo (RHMC)
algorithm for the fermion kernel\footnote{This corresponds to \(1/\nth\) of a
fermion `flavour'; we take advantage of the fact that it is no more work to
take the \(2\nth\) root of \(\M^\dagger\M\) than to take the {\nth} root of
\(\M\), so we can ensure that the fermion kernel is manifestly positive for no
extra cost.} \((\M^\dagger \M)^{1/2n}\). In \secref{sec:HMC} we explained how
the HMC algorithm alternates two Markov steps --- momentum refreshment and
MDMC --- both of which have the desired fixed point and together are
ergodic. When we include pseudofermion fields we need to add a third Markov
step to ensure ergodicity, which is sampling the pseudofermion fields from a
heatbath. This is easy to do because the (pseudo)fermions only appear
quadratically in the action:\footnote{We could not get away with this
simplification if we wanted to improve the fermion action by including
four-fermion operators or higher.} we sample the pseudofermions from a
Gaussian heatbath by applying the square root of the kernel to
Gaussian-distributed random noise,
\begin{displaymath}
  \chi_j =(\M^\dagger\M)^{1/4n}\xi_j \quad\mbox{with}\quad
  P(\xi_j) \propto e^{-\xi_j^\dagger\xi_j},
\end{displaymath}
as then
\begin{eqnarray*}
  P(\chi_j) &\propto& \int_{-\infty}^\infty d\xi_j\, e^{-\xi_j^\dagger\xi_j}\,
    \delta\Bigl(\chi_j-(\M^\dagger\M)^{\frac1{4n}}\xi_j\Bigr) \\
  &\propto& \exp\Bigr[-\chi_j^\dagger(\M^\dagger\M)^{-\frac1{2n}} \chi_j\Bigl].
\end{eqnarray*}
We then refresh the momenta\footnote{There is no \emph{a priori} reason why we
have to refresh the pseudofermions and the momenta with equal frequency, but
there is no obvious benefit from not doing so.}, and integrate Hamilton's
equations for the gauge fields using the force (c.f., equation
(\ref{eq:psForce})) 
\begin{eqnarray*}
  \dot\pi &=& -\pdd{\SB}\phi
    - \sum_{j=1}^n \chi_j^\dagger\pdd{(\M^\dagger\M)^{-1/2n}}\phi\chi_j \\
  &=& -\pdd{\SB}\phi
    + \sum_{\chi=1}^n \left[(\M^\dagger\M)^{-1/2n}\chi_j\right]^\dagger
     \pdd{(\M^\dagger\M)^{1/2n}}\phi \left[(\M^\dagger\M)^{-1/2n}\chi_j\right].
\end{eqnarray*}
Finally we complete the MDMC step by applying a Metropolis test.

Of course we apply all the fractional powers of the matrices that occur using
optimal rational approximations expressed in partial fraction form. Note that
\begin{itemize}
 \item we use an accurate rational approximation \(r(x) \approx \sqrt[4n]{x}\)
       for the pseudofermion heatbath;
 \item we use a less accurate approximation for the MD evolution,
       \(\tilde{r}(x) \approx \sqrt[2n]{x}\);
 \item we use an accurate approximation for Metropolis acceptance step.
\end{itemize}
This is because any errors in the pseudofermion heatbath or Metropolis test
would effect the fixed point distribution, whereas errors in the MD only
effect the acceptance rate. If the rational approximations used in the
pseudofermion heatbath and the Metropolis test are accurate to machine
precision then the algorithm is still as exact as HMC: that is, the only
systematic errors are the rounding errors from using finite precision floating
point arithmetic, and from the fact that the Markov process may not have
converged well enough to its fixed point. It is not hard to generate optimal
rational approximations that are good to machine precision: this should not be
surprising as this is how most scalar functions (exponential, logarithm,
fractional powers, etc.) are evaluated anyhow.

Let us summarize the key properties of the RHMC algorithm.
\begin{itemize}
  \item We apply all rational approximations using their partial fraction
	expansions:
	\begin{itemize}
	  \item the denominators occuring in this are all just shifts of the
                original fermion kernel;
	  \item all poles of optimal rational approximations are real and
        	positive for cases of interest (Miracle~\#1);
	  \item only simple poles appear (by construction, a double pole can
                only occur if we try to approximate something that was itself
                a square, which would not be a sensible thing to do).
	\end{itemize}
  \item We a use multishift solver to invert all the partial fractions using a
	single Krylov space: the cost is dominated by the construction of the
	Krylov space; updating the \(n\) solution vectors is comparatively
	cheap, at least for \(O(20)\) shifts.
  \item The result is numerically stable, even in 32-bit precision. This is
        because all the partial fractions have positive coefficients
        (Miracle~\#2).
  \item The MD force term is of the same form as for conventional HMC except
        for a constant shift for each partial fraction; therefore the method
        is immediately applicable to any fermion kernel for which we can
        implement conventional HMC.
\end{itemize}

\subsection{Comparison with R Algorithm}

Let us now look at some empirical (numerical) studies\cite{Clark:2005sq} that
compare the performance of the RHMC and R~algorithms for non-zero temperature
QCD with staggered quarks (\secref{sec:finiteTemperature}) and for chirally
symmetric domain wall fermions (\secref{sec:DWF}), where the R algorithm has
been the method of choice in the past.

\subsubsection{Finite Temperature QCD}
\label{sec:finiteTemperature}

First we compare the of the algorithms' performance near the chiral transition
point. The aim of this study was to see how the RHMC algorithm behaves in this
regime, and to compare its cost with that of the R algorithm. 

\begin{table}[pht]
  \tbl{Values of the Binder cumulant for \(\langle\bar\psi\psi\rangle\),
    \(B_4\), and the RHMC acceptance rate~\(\Pacc\).}{
  \begin{tabular}{cccc}
    \hline
    Algorithm & \(\dt\) & \(\Pacc\) & \(B_4\) \\
    \hline
    R & 0.0019 & & 1.56(5) \\
    R & 0.0038 & & 1.73(4) \\
    RHMC & 0.055 & 84\% & 1.57(2) \\
    \hline
  \end{tabular}
  \label{tab:FTcomparison}}
\end{table}

For these tests we used \(2+1\) flavours of na\"ive staggered fermions with
\(\ml=0.0076\) and \(\ms=0.25\), and the Wilson gauge action on an \(8^3\times
4\) lattice. The MD trajectory length was \(\trjlen=1\). The results are given
in Table~\ref{tab:FTcomparison}, where the Binder cumulant of the chiral
condensate, \(B_4\), is shown as a function of step size for the two
algorithms. The RHMC acceptance rate, \(\Pacc\), is also given. The correct
value of the Binder cumulant is obtained with RHMC using a step size
\(\approx29\)~times larger than is necessary for the R~algorithm.

\begin{figure}[htb]
  \centerline{\epsfxsize=0.8\textwidth \epspdffile{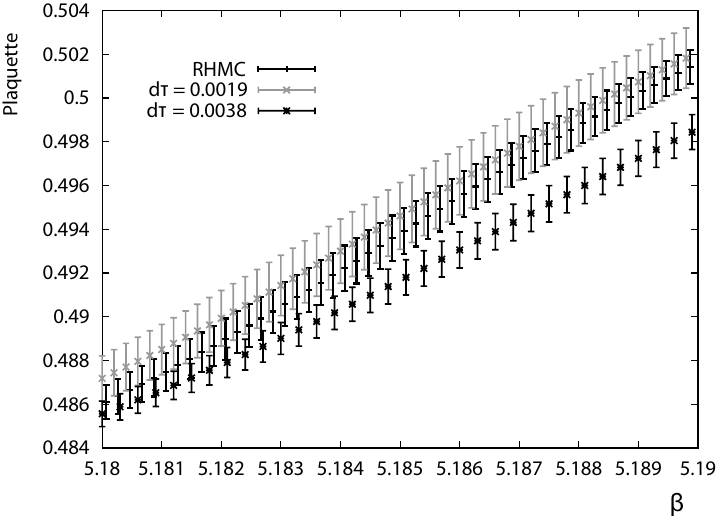}}   
  \caption[Plaquette versus beta]{The mean plaquette is shown as function of
    \(\beta\) for the RHMC and R~algorithms at the parameters given in the text
    (near the chiral transition point). The RHMC points have been displaced
    right so that they can be seen more clearly. For the R algorithm only the
    data at the smaller step size is within one standard deviation of the RHMC
    results.}
  \label{fig:comparison_with_R_I}
\end{figure}

\begin{figure}[htb]
  \centerline{\epsfxsize=0.8\textwidth \epspdffile{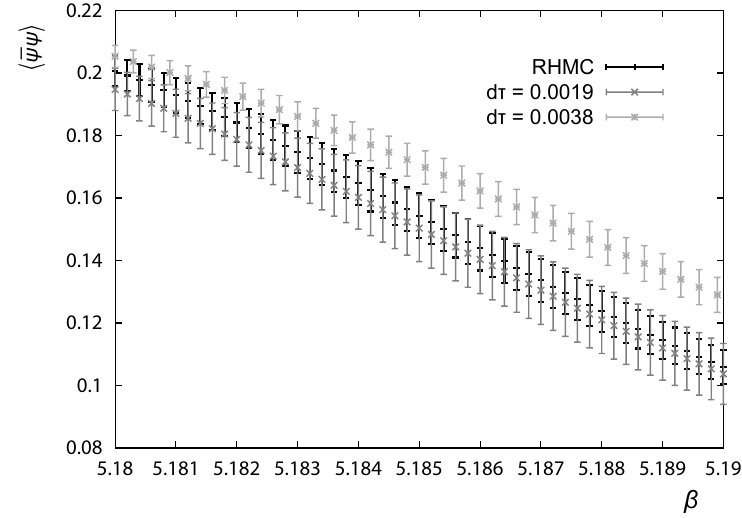}}   
  \caption[Chiral condensate versus beta]{The chiral condensate
    \(\langle\bar\psi\psi\rangle\) is shown as a function of \(\beta\) for the
    RHMC and R~algorithms at the parameters given in the text (near the chiral
    transition point). For the R algorithm only the data at the smaller step
    size is within one standard deviation of the RHMC results.}
  \label{fig:comparison_with_R_II}
\end{figure}

Figures~\ref{fig:comparison_with_R_I} and \ref{fig:comparison_with_R_II} are
plots of the plaquette and the chiral condensate \(\langle\bar\psi\psi
\rangle\) as a function of \(\dt\): only for the smaller step size is the R
algorithm result consistent with that from RHMC.

It is clear that RHMC is superior to the R algorithm in this case, principally
because it is vital to keep the systematic step size errors under control when
studying the properties of phase transitions. Even though the errors in the
equilibrium distribution of the R algorithm Markov chain are of \(O(\dt^2)\),
even the introduction of a small admixture of unwanted operators in the action
can significantly change the behaviour of a system near criticality; indeed,
they can even change the order of the phase transition if we are not extremely
careful when we take the zero step size limit.

\subsubsection{2+1 Domain Wall Fermions}
\label{sec:DWF}

Lattice calculations with on-shell chiral dynamical fermions are still in
their infancy, but it seems clear that they will become increasingly important
as sufficient computing power becomes available. There are several methods to
carry out such calculations, as we shall discuss in the next lecture
(\secref{sec:GW}), but at present the only such computations carried out on
large lattices with fairly light quarks have used the domain wall formulation.
The results shown here were obtained as part of the UKQCD--Columbia--RBRC
collaboration, and the parameters used are summarized in
Table~\ref{tab:DWFparams}.

\begin{table}[htp]
  \tbl{The different masses at which the domain wall results were gathered,
    together with the step sizes \(\dt\), acceptance rates \(\Pacc\), and mean
    plaquette values \(\langle P\rangle\). The runs were all performed on
    \(16^3\times32\) lattices with \(n_5=8\) with the DBW2 gauge action at
    \(\beta=0.72\)}{
  \begin{tabular}{cccccc}
    \hline
     Algorithm  & \(\ml\) & \(\ms\) & \(\dt\) & \(\Pacc\) & \(\langle
       P\rangle\) \\
     \hline
     R    & 0.04    & 0.04    & 0.01    & - & 0.60812(2) \\
     R    & 0.02    & 0.04    & 0.01    & - & 0.60829(1) \\
     R    & 0.02    & 0.04    & 0.005   & - & 0.60817(3) \\
     RHMC & 0.04    & 0.04    & 0.02    & 65.5\% & 0.60779(1) \\
     RHMC & 0.02    & 0.04    & 0.0185  & 69.3\% & 0.60809(1) \\
     \hline
  \end{tabular}
  \label{tab:DWFparams}}
\end{table}

\begin{figure}[htb]
  \centerline{\epsfxsize=0.9\textwidth\epspdffile{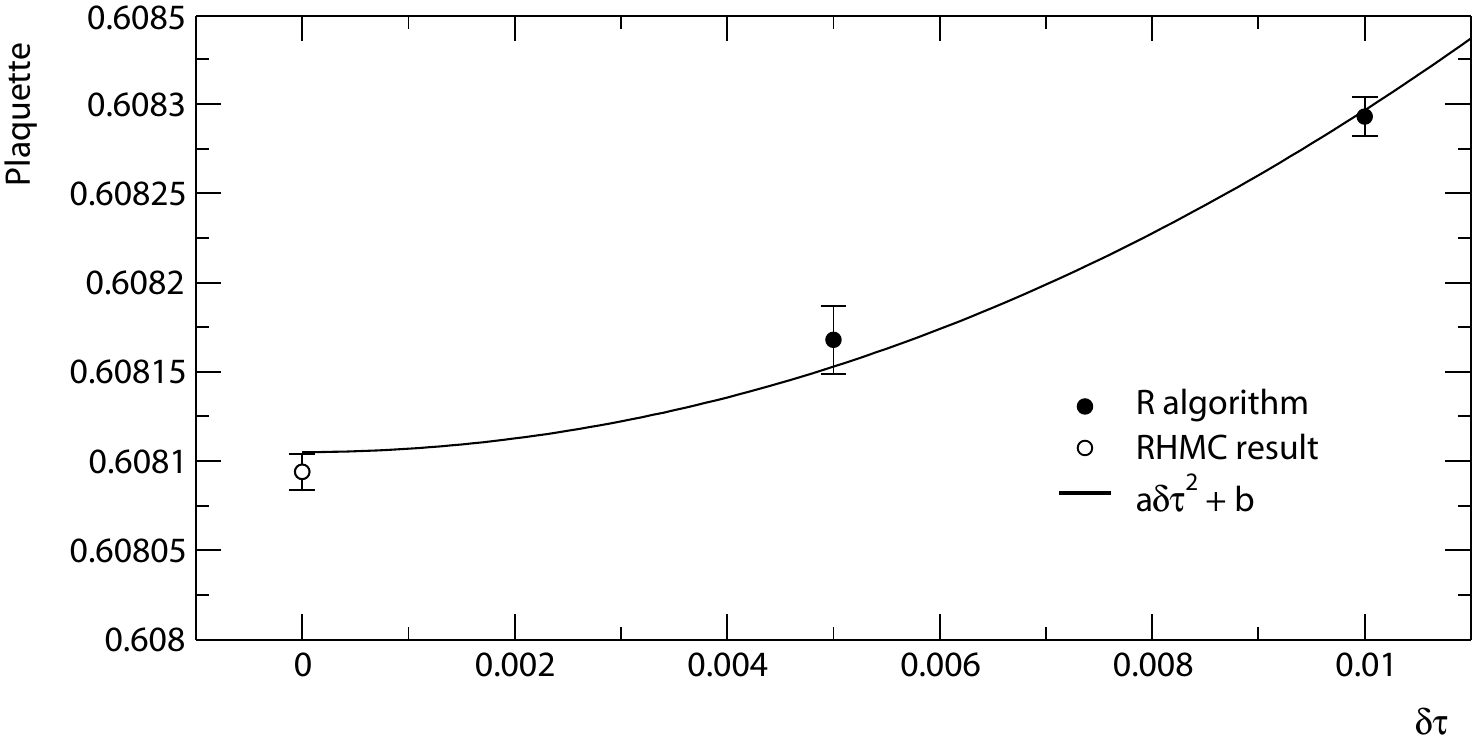}}   
  \caption[Step size dependence]{The step size dependence of the plaquette for
    domain wall fermions with the parameter values given in
    Table~\ref{tab:DWFparams}. Note that the RHMC data point shown at \(\dt=0\)
    (because it has no step size errors) is really at \(\dt=0.04\), which is
    far to the right of the graph. The quadratic extrapolation seems to work
    reasonably for the R~algorithm data points, but the need to carry out this
    extrapolation significantly increases the estimated error.}
  \label{fig:DWFstepsize}
\end{figure}

The step size dependence of the plaquette is shown in
Figure~\ref{fig:DWFstepsize}. From this we can see that in order to obtain a
result consistent within error bars the R~algorithm needs an integration step
size more than 10 times smaller than that used for~RHMC, although the zero step
size extrapolation using a quadratic fit to the R~algorithm data gives a
reasonable value.

\begin{figure}[htb]
  \centerline{\epsfxsize=0.8\textwidth \epspdffile{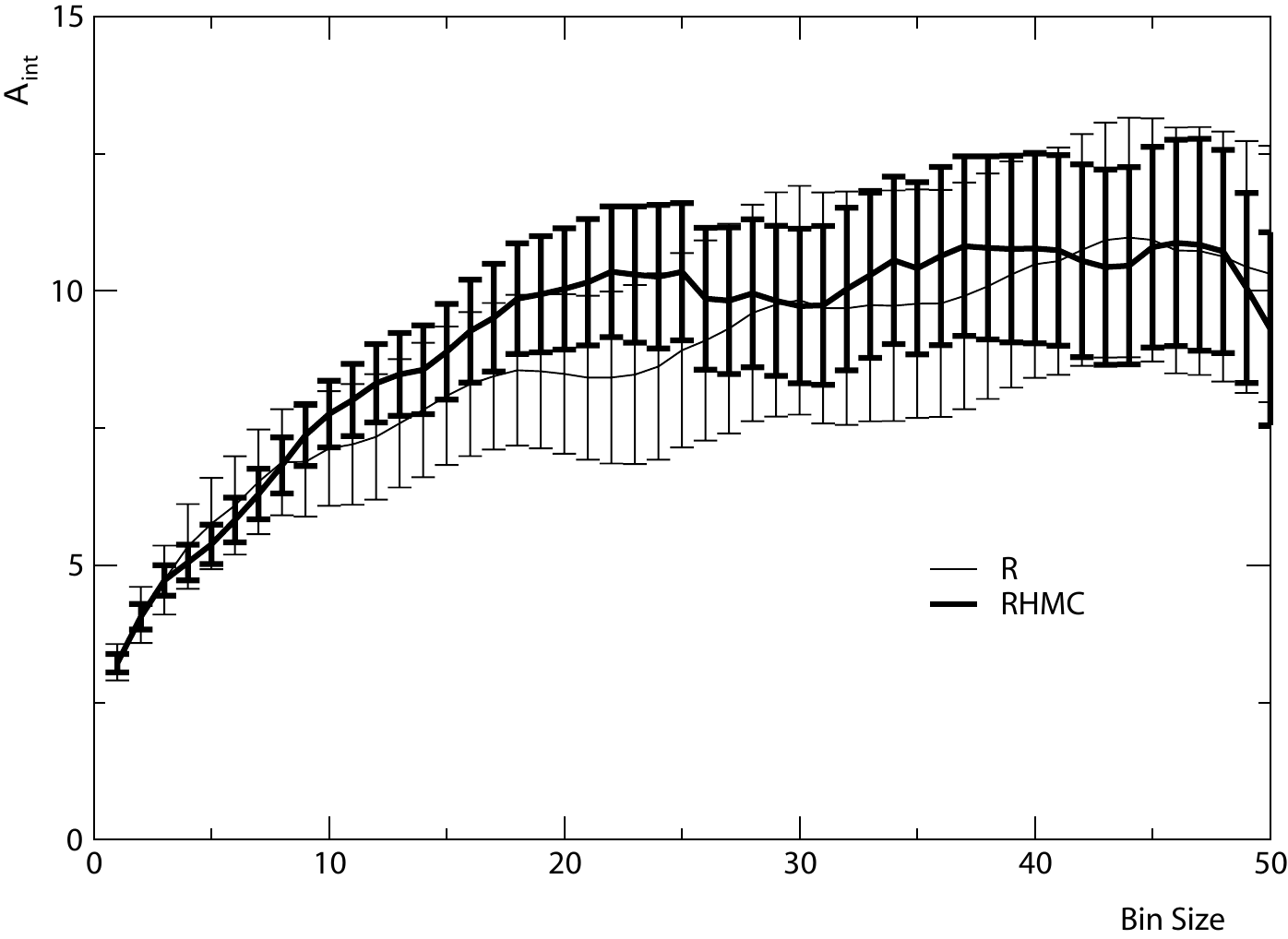}}   
  \caption{The integrated autocorrelation of time slice 13 of the \(\pi\)
    propagator for domain wall fermions.}
  \label{fig:DWFautocorrelation}
\end{figure}

We expect that the autocorrelations for RHMC and R~algorithms should be
similar if carried out with the same trajectory lengths, except of course that
the RHMC autocorrelations should be longer by a factor of the inverse
acceptance rate. The integrated autocorrelation time for the
\(\pi\)~propagator from the domain wall test with \(m_{ud}=0.04\) is shown in
Figure~\ref{fig:DWFautocorrelation}, and the data seems to bear this out.

The conclusion from this study too is that the RHMC algorithm leads to a
significant cost reduction in this case too.

\subsection{Multiple Pseudofermions with Multiple Timescales}

\begin{figure}[htbp]
  \centerline{\epsfxsize=0.9\textwidth \epspdffile{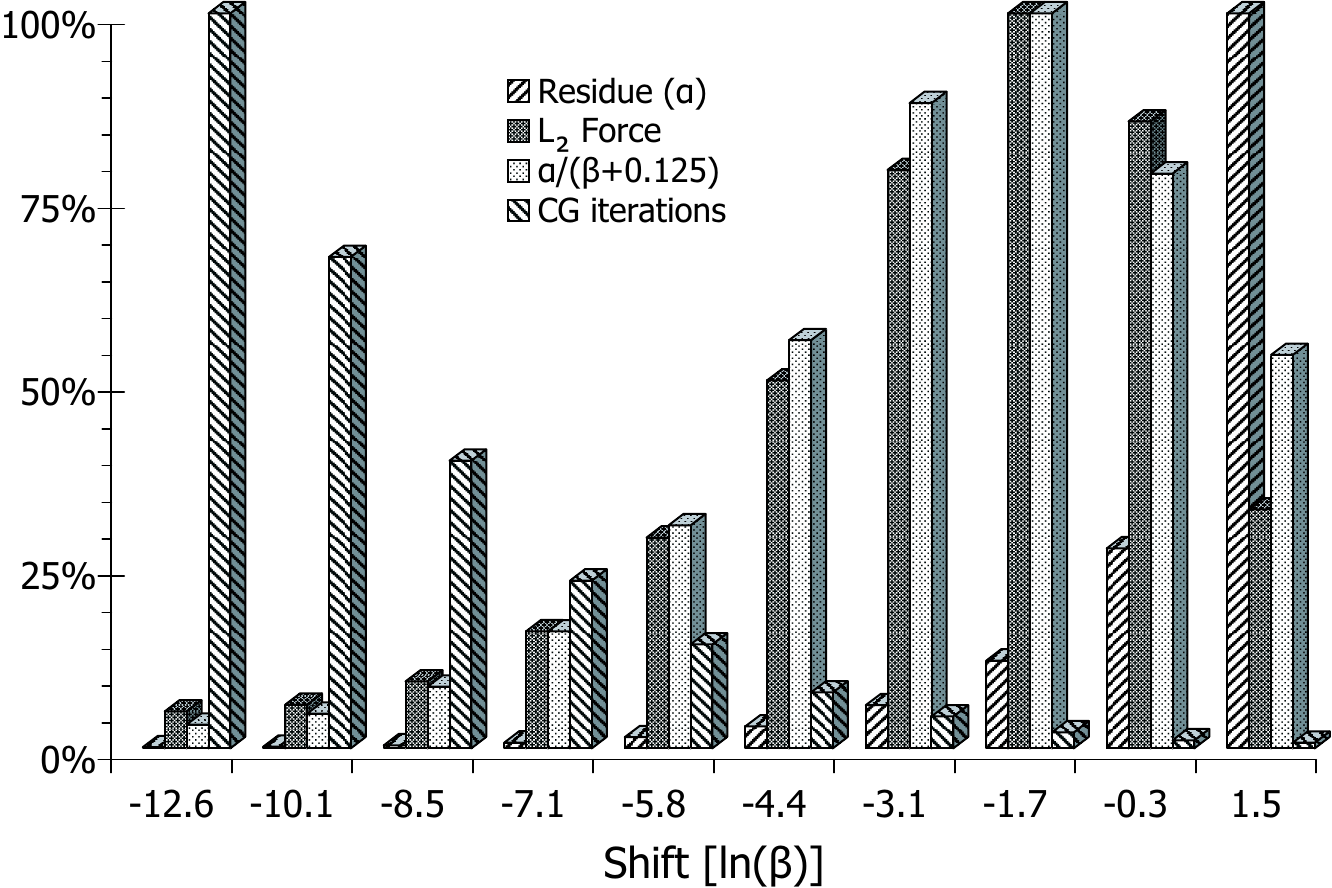}}
  \caption[Force and CG per pole]{In this graph we plot the \(L^2\) force as a
    function of the partial fraction \(\alpha/(x+\beta)\). The value of the
    logarithm of the shift, \(\ln\beta\), is plotted on the horizontal axis,
    and it is clear that the values of \(\ln\beta\) are fairly unformly
    distributed. The coefficients of each pole, the residues \(\alpha\), are
    also plotted, and it can be seen that they increase rapidly with increasing
    values of the shift \(\beta\). We also show the measured \(L_2\) norm of
    the force contribution from each pole, and these also increase with
    \(\beta\) for small \(\beta\), but reach a maximum at
    \(\ln\beta=-1.7\). The \(L_\infty\) norm of the force contributions behaves
    similarly. We also show the values of \(\alpha/(\beta+0.125)\), which can
    be taken as a simple model of the behaviour of the force
    contributions. Finally we plot the number of CG iterations required to
    reach a fixed residual for each pole: this is proportional to the cost of
    computing the force, and decreases rapidly with \(\beta\). It is clear that
    the largest force contributions come from the poles that require only a
    cheap CG computation, and the expensive poles only contribute a small
    amount to the total force acting on the gauge field. We can taken advantage
    of this either by putting the small shift poles onto a coarser timescale,
    or by evaluating them much less accurately.}
\label{fig:multitimescales}
\end{figure}

We now turn to a way that the RHMC algorithm allows us to implement \emph{UV
filtering}, that is to separate the short-distance ultraviolet contributions
to the pseudofermion force from the long-distance infrared ones. The
semi-empirical observation is that the largest contributions to the gauge
field force from the pseudofermions do not come from the partial fractions
with small shifts, but from those with large ones. To see why this may be so,
look at the numerators in the partial fraction expansion that we exhibited
earlier
\begin{displaymath}
  \frac1{\sqrt x} \approx \:\scriptstyle
    0.3904603901 + \textstyle
    \frac{0.0511093775}{x+0.0012779193} +
    \frac{0.1408286237}{x+0.0286165446} +
    \frac{0.5964845033}{x+0.4105999719}.
\end{displaymath}
Even for this low-order approximation the coefficient of the largest shift is
more than an order of magnitude larger than that of the smallest shift. A
large shift corresponds (in some sense) to a larger fermion mass, which is why
the effects of the large shift partial fractions may be thought of as short
distance or ultraviolet modes. Figure~\ref{fig:multitimescales} shows the
situation for a more realistic situation.

There are a couple of ways we can make use of this observation: we can use a
coarser timescale for the more expensive smaller shifts, as explained in
\secref{sec:multitimescale}, or we can be less sophisticated (but sometimes
more cost effective) and significantly increase the CG residual used in
computing the force from the small shift partial fractions. We stress that
this does not lead to any systematic errors provided we use a time-symmetric
initial guess in constructing the Krylov space.

It is worth noting that we are putting the force contribution from different
partial fractions for each pseudofermion field on different timescales: the
different pseudofermions themselves are treated entirely equally.

\subsection{\(L_{2}\) versus \(L_{\infty}\) Force Norms}

\begin{figure}[htb]
  \centerline{\epsfxsize=0.8\textwidth \epspdffile{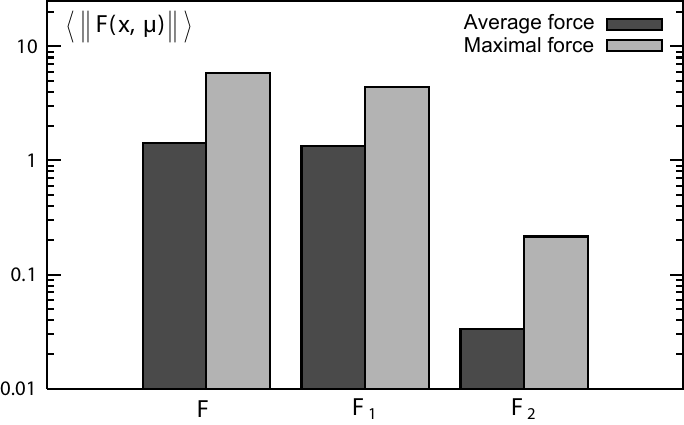}}   
  \caption{\(L_2\) versus \(L_\infty\) force norms.}
  \label{fig:L_2_vs_L_infty_force_norms}
\end{figure}

Figure.~\ref{fig:L_2_vs_L_infty_force_norms} is Wilson fermion forces taken
from the paper by Urbach~\textit{et al.}\cite{Urbach:2005ji} Here the authors
have used Hasenbusch's method (q.v., \secref{sec:hasenbuschery}), and as can be
seen there is a clear difference between the \(L_2\) norms of the force
contributions and the \(L_\infty\) norms. From our discussion of instabilities
in \secref{sec:instability} we may expect that the \(L_\infty\) is the more
appropriate one to detect when a particular force contribution causes the
integrator to become unstable. However, while the values of the two norms are
very different the relative magnitudes of the force contributions are still
fairly similar, so the difference may be moot.

\subsection{Berlin Wall for Wilson fermions}

\begin{figure}[htb]
  \centerline{\epsfxsize=0.8\textwidth \epspdffile{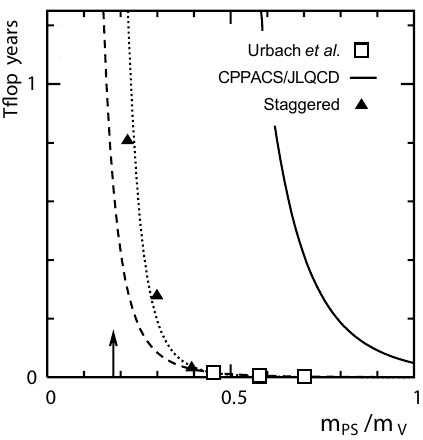}}   
  \caption[Berlin wall]{The Berlin wall for Wilson fermions moves to much
           lighter quarks when multiple pseudofermions are introduced.}
  \label{fig:Berlin_Wall}
\end{figure}

Figure~\ref{fig:Berlin_Wall} shows the `Berlin wall' for Wilson fermions, and
is also taken from the paper by Urbach~\textit{et al.},\cite{Urbach:2005ji}.
The estimated cost of generating \(1,000\) independent configurations on a
\(24^3\times40\) lattice is plotted in units of Tflop years (\(3\times10^{19}\)
floating point operations) as a function of \(m_\pi/m_\rho\). The solid line is
the result for a conventional two-flavour HMC simulation with Wilson fermions
performed by CP-PACS and JLQCD; the squares are results of work of Urbach
\textit{et al.}, where they introduced two or three pseudofermion fields using
Hasenbusch's method and used different integrator step sizes for the various
force contributions such that the force times step size \(\dt\) is roughly
constant. The dashed and dotted lines are the extrapolations of these latter
results using ansatze of \((m_\pi/ m_\rho)^{-4}\) and \((m_\pi/ m_\rho)^{-6}\)
respectively. The triangles are the corresponding results for Staggered
fermions, and the arrow indicates the physical pion to rho meson mass ratio.

They show that the performance of Hasenbusch trick and multiple timescale is
comparable to L\"uscher's SAP algorithm, which from our present point of view
probably speeds up dynamical Wilson fermion computations because it too splits
the pseudofermion force into three components.

\subsection{Conclusions (RHMC)}

In summary, the advantages of RHMC are that it
\begin{itemize}
 \item is exact, there are no step size errors: hence no step size
       extrapolations are required;
 \item is significantly cheaper than the R algorithm;
 \item allows easy implementation of Hasenbusch-like tricks using multiple
       pseudofermion fields;
 \item is amenable to futher improvements, such as using multiple timescales
       for different terms in the partial fraction expansion.
\end{itemize}
So far RHMC has no obvious disadvantages.

\section{5D Algorithms for Chiral Lattice Fermions}
\label{sec:GW}

This final lecture section is devoted to five-dimensional formulations of
chiral fermions. We begin by introducing on-shell lattice chiral fermions,
following a logical rather than historical approach.

\subsection{On-shell chiral symmetry}

We shall write the various forms of the Dirac operator as \(\Dslash=D_\mu\cdot
\gamma_\mu\) where the Dirac \(\gamma\) matrices in Euclidean space are
necessarily Hermitian. We shall also assume all Dirac operators are
\(\gamma_5\) Hermitian, \(\Dslash^\dagger=\gamma_5 \Dslash\gamma_5\). Note that
while this is an assumption it does not seem to be very strong one, as almost
all lattice Dirac operators satisfy it (except for the twisted-mass
formulation).

The key idea is that it is possible to have chiral symmetry on the lattice
without doublers if we only insist that the symmetry holds on-shell. L\"uscher
observed that on-shell chiral symmetry transformations should be of the
form\cite{Luscher:1998pq}
\begin{displaymath}
  \psi \to e^{i\alpha\gamma_5 [1-2\zeta a\Dslash]} \psi \quad\mbox{and}\quad
  \bar\psi \to \bar\psi e^{i\alpha[1-2(1-\zeta) a\Dslash]\gamma_5},
\end{displaymath}
where we have introduced a free parameter \(\zeta\), for which you may choose
your favourite value. As always, the choice of chiral transformation properties
of a field is independent of the choice of Lorentz transformation properties,
and we can assign \(\psi\) and \(\bar\psi\) independent choices of
transformation as they are independent fields, despite the notation: the dagger
just indicates that \(\bar\psi=\psi^\dagger\) has the same Lorentz
transformation as the Hermitian conjugate of \(\psi\). In order that this be a
symmetry the Dirac operator occuring in it must be invariant under the
transformation
\begin{displaymath}
 \Dslash \to
  e^{i\alpha [1-2(1-\zeta)a\Dslash]\gamma_5} \Dslash
    e^{i\alpha\gamma_5 [1-2\zeta a\Dslash]} 
  = \Dslash.
\end{displaymath}
As this must hold identically for all values of \(\alpha\) the coefficient of
\(\alpha\) in its Taylor expansion must vanish, namely
\begin{displaymath}
  [1-2(1-\zeta)a\Dslash]\gamma_5 \Dslash
    + \Dslash\gamma_5 [1-2\zeta a\Dslash]=0,
\end{displaymath}
which is nothing but the Ginsparg--Wilson relation
\begin{equation}
  \gamma_5 \Dslash + \Dslash \gamma_5 = 2a \Dslash \gamma_5 \Dslash.
  \label{eq:GW_relation}
\end{equation}

\subsection{Neuberger's Operator}

We can find the unique\footnote{Up to the choice of kernel, and assuming
\(\gamma_5\) hermiticity.} solution of the Ginsparg--Wilson relation
equation~(\ref{eq:GW_relation}) in the following way. We choose to parametrize
the lattice Dirac operator by writing
\begin{displaymath}
 a\Dslash = \half(1+\gamma_5\hat\gamma_5);
\end{displaymath}
this is no more that the definition of \(\hat\gamma_5 \defn \gamma_5 (2a\Dslash
- 1)\). The \(\gamma_5\) hermiticity of the Dirac operator immediately implies
that \(\hat\gamma_5\) is a Hermitian operator \(\hat\gamma_5^\dagger =
\hat\gamma_5\). If we also require the Dirac operator to satisfy the
Ginsparg--Wilson relation, we immediately find that \(\hat\gamma_5^2=1\), so
\(\hat\gamma_5\), like \(\gamma_5\), is both Hermitian and unitary. \(\Dslash\)
must also have the correct continuum limit \(\Dslash \to Z_N\dslash + O(a)\),
so
\begin{displaymath}
  \hat\gamma_5\ = \gamma_5(2aZ_N\dslash-1) + O(a^2)
    = \gamma_5\left(\frac{\Dslash_W}M - 1\right)+O(a^2),
\end{displaymath}
where we have defined \(M\defn Z_W/2aZ_N\) and \(\Dslash_W \to Z_W\dslash +
O(a)\). Here \(\Dslash_W\) is some lattice Dirac operator, such as the Wilson
operator, and \(Z_N\) and \(Z_W\) are wavefunction renormalizations. The large
negative mass \(M\) is called the domain wall height in the language of domain
wall fermions, and is an unphysical cut-off-scale quantity. Its value is to be
chosen to distinguish cleanly between the physical fermions and the doublers,
the two getting sent to opposite ends of the Neuberger operator spectrum. Both
of these conditions are satisfied if we define
\begin{displaymath}
  \hat\gamma_5
   = \gamma_5\frac{\Dslash_W-M}{\sqrt{(\Dslash_W-M)^\dagger(\Dslash_W-M)}}
   = \sgn\bigl[\gamma_5(\Dslash_W-M)\bigr].
\end{displaymath}
This is the form given by Neuberger.\cite{Neuberger:1997fp,Neuberger:1998wv}

\begin{figure}[htb]
  \centerline{\epsfxsize=0.45\textwidth \epspdffile{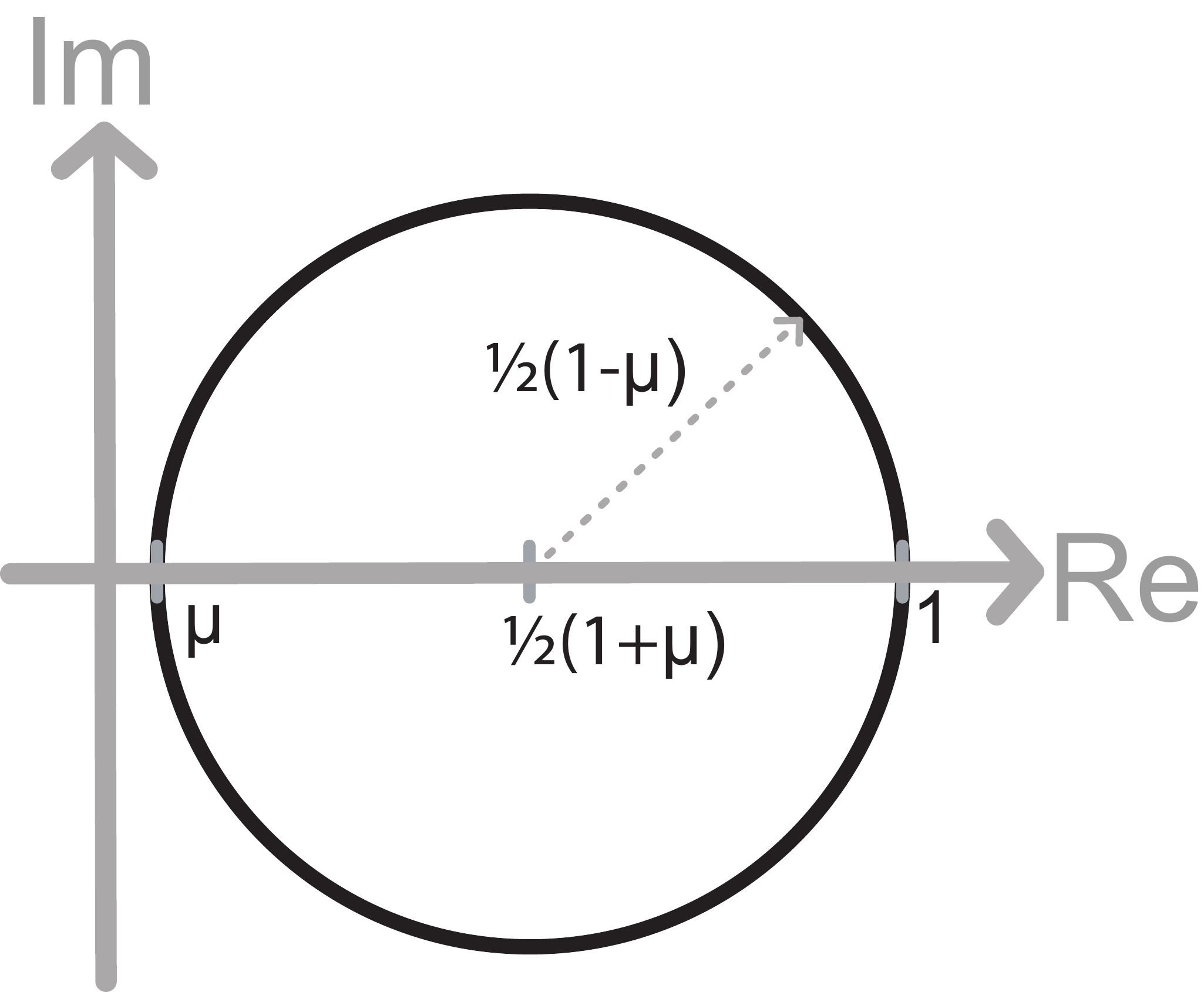}}   
  \caption[Neuberger spectrum]{The operator \(\gamma_5\hat\gamma_5\) is
    unitary and thus its spectrum lies on the unit circle. This means that the
    spectrum of the massive Neuberger operator defined by
    equation~(\ref{eq:Neuberger}) lies on a circle of radius \(\half(1-\mu)\)
    whose centre is at \(\half(1+\mu)\), i.e., a circle whose minimum real
    part is \(\mu=O(a)\) and whose maximum real part is one. The physical
    states all lie near \(\mu\) and the doublers are near one.}
  \label{fig:spectrum_of_Neuberger_operator}
\end{figure}

It is often convenient to break chiral symmetry explicitly by adding a physical
fermion mass. The massive Neuberger operator is
\begin{equation}
  \Dslash_N(\mu,H) = \half\bigl[1+\mu+(1-\mu)\gamma_5\sgn(H)\bigr],
  \label{eq:Neuberger}
\end{equation}
where \(\mu\) is the physical mass and \(H\) is the Hermitian Dirac operator
kernel including the negative mass, \(H = H(-M) = \gamma_5(\Dslash-M)\). As
shown in Figure~\ref{fig:spectrum_of_Neuberger_operator}, all the eigenvalue of
the Neuberger operator lie on the unit circle.

The immediate question is whether \(\Dslash_N\) is a local operator. Although
it is manifestly not ultralocal, it can be shown to be a local operator if the
kernel \(\Dslash_W\) has a gap in its spectrum.\cite{Hernandez:1998et} The
Wilson kernel \(\Dslash_W\) certainly has a gap if the gauge fields are smooth
enough to satisfy the admissibility condition that \(\langle P\rangle<1/30\):
this is a sufficient but not a necessary condition. Moreover, it seems
reasonable that good approximations to the Neuberger operator will be local if
\(\Dslash_N\) is and vice versa.

\subsection{5D Chiral Fermions}

In order to measure propagators, or to compute the pseudofermionic force in
HMC, we have to find the inverse of the Neuberger operator \(\Dslash_N\). If
we approximate the \(\sgn\) function by a rational function of some sort then
just applying it using our usual partial fraction technique requires the
construction of a Krylov space. When we were considering RHMC before we noted
(\secref{sec:NFL}) that it is just as cheap to compute \(A^{-1/n}\) as it is
to compute \(A^{1/n}\), since the inverse of a rational function is another
rational function. Sadly, this does not help us in the present situation, as
the Neuberger operator is constructed from two non-commuting operators,
\(\Dslash\) and~\(\gamma_5\).

It transpires that one can trade-off this non-commuting property with a kind of
dimensional reduction from five dimensions to four (this is of course a totally
different fifth dimension from the fictitious time we introduced in
\secref{sec:MDMC}). We are thus led to study five-dimensional formulations of
chiral fermions where in exchange for having an extra dimension we get away
with inverting the Neuberger operator within a single Krylov space.

We shall introduce a class of five-dimensional algorithms that are
described by four independent choices:
\begin{itemize}
  \item \emph{Kernel} --- what Dirac kernel is used as the argument of the
       \(\sgn\) function in Neuberger's operator (\secref{sec:kernel}).
 \item \emph{Constraint} --- whether we use the five-dimensional system solely
       as a means of computing propagators, or use it to construct a
       five-dimensional functional integral formulation
       (\secref{sec:constraint}).
 \item \emph{Approximation} --- what rational approximation of the \(\sgn\)
       function is used (\secref{sec:approximation}).
 \item \emph{Representation} --- how we map the rational approximation into a
       five-dimensional system of linear equations
       (\secref{sec:representation}).
\end{itemize}

\subsection{Choice of Kernel}
\label{sec:kernel}

The simplest kernel is the Wilson kernel with a large negative mass, \(H_W
\defn \gamma_5(\Dslash_W-M)\). Another popular kernel which arises naturally in
the domain wall formalism\cite{Kaplan:1992bt,Shamir:1993zy} is the Shamir
kernel \(H_T\defn \gamma_5 \Dslash_T\) where \[a_5\Dslash_T=\frac{a_5(\Dslash_W
-M)}{2+a_5(\Dslash_W-M)}.\] A third kernel which interpolates between the
preceding kernels is the \emph{M\"obius kernel} \(H_M \defn\gamma_5\Dslash_M\),
where \[a_5\Dslash_M \defn \frac{a_5(b_{+}+b_{-}) (\Dslash_W-M)}{2 +
a_5(b_{+}-b_{-}) (\Dslash_W-M)}.\] Special cases corresponding to some choice
of the two parameters \(b_{+}\) and \(b_{-}\) are shown in
Table~\ref{tab:moebius}.

\begin{table}[pht]
  \tbl{Special cases of the M\"obius kernel.}{
  \begin{tabular}{ccc}
    \hline
    Kernel & \(b_{+}\) & \(b_{-}\) \\
    \hline
    Bori\c ci/Wilson & 1 & 1 \\
    Shamir/DWF & 1 & 0 \\
    \hline
  \end{tabular}
  \label{tab:moebius}}
\end{table}

\subsection{Schur Complement}
\label{sec:schur}

The key to understanding the connection between four and five-dimensional
formulations is the \emph{Schur complement} of a matrix. Let us consider the
five-dimensional block matrix
\begin{displaymath}
  M_5 = \begin{pmatrix} A & B \\ C & D \end{pmatrix},
\end{displaymath}
where the blocks \(A\), \(B\), \(C\) and \(D\) are themselves matrices.
 \(M_5\) may be block diagonalized by an LDU factorisation (also known as
 Gaussian elimination)
\begin{eqnarray*}
  \begin{pmatrix} A & B \\ C & D \end{pmatrix}
  &=& \begin{pmatrix} 1 & 0 \\ CA^{-1} & 1 \end{pmatrix}
    \begin{pmatrix} A & B \\ 0 & D-CA^{-1}B \end{pmatrix} \\
  &=& \begin{pmatrix} 1 & 0 \\ CA^{-1} & 1 \end{pmatrix}
    \begin{pmatrix} A & 0 \\ 0 & D-CA^{-1}B \end{pmatrix}
    \begin{pmatrix} 1 & A^{-1}B \\ 0 & 1 \end{pmatrix}.
\end{eqnarray*}
This decomposition can be done by inspection, and is unique. The bottom right
block is the Schur complement. In particular, the determinant of \(M_5\) is
readily calculated by
\begin{displaymath}
  \det \begin{pmatrix} A & B \\ C & D \end{pmatrix}
  = \det\bigl(AD-ACA^{-1}B\bigr).
\end{displaymath}

\subsection{Constraint}
\label{sec:constraint}

How do we make use of a representation of the Neuberger operator as the Schur
complement of some five-dimensional matrix? The first way is to use a five
dimensional linear system to compute the required four-dimensional propagator.
Consider the five-dimensional system of linear equations
\begin{displaymath}
 D_5 \tilde\Phi \defn D_5
  \begin{pmatrix}
    \tilde\phi_1 \\
    \tilde\phi_2 \\
    \vdots\\
    \tilde\phi_{n-2} \\ 
    \tilde\phi_{n-1} \\ 
    \psi 
  \end{pmatrix} = 
  \begin{pmatrix} 0 \\ 0 \\ \vdots\\ 0 \\ 0 \\ \chi \end{pmatrix}
  \defn\tilde X,
\end{displaymath}
where \(D_5\) is an \(n\times n\) five-dimensional matrix, and \(\tilde\Phi\)
and \(\tilde X\) are five-dimensional vectors constructed from \(n\)
four-dimensional vectors as indicated. Using the block \(LDU\) factorisation
(\secref{sec:schur}) our linear system reduces to
\begin{displaymath}
  L^{-1}D_5 \tilde\Phi = L^{-1}LDU \tilde\Phi = D\Phi
  = L^{-1}\tilde X = \tilde X,
\end{displaymath}
where
\begin{displaymath}
  \Phi \defn U\tilde\Phi =
  \begin{pmatrix} \phi_1 \\ \phi_2 \\ \vdots\\ \phi_{n-2} \\ \phi_{n-1} \\
    \psi \end{pmatrix}
\end{displaymath}
is the five-dimensional column matrix analogous to \(\tilde\Phi\) but with the
unphysical constraint fields \(\tilde\phi\) replaced with the equally
unphysical \(\phi\), and with the same bottom component \(\psi\). Note that the
inverse of a triangular matrix is of the same shape as the original matrix, so
\(L^{-1}\) is a lower triangular matrix 
As \(D\) is block diagonal we have thus reduced the five-dimensional system to
a family of uncoupled four-dimensional ones, the bottommost of which involves
the Schur complement, which is just the Neuberger operator, \(D_{n,n}\psi =
\Dslash_N\psi=\chi\). Therefore by solving the five-dimensional system
\(D_5\tilde\Phi=\tilde X\) the bottom component \(\psi\) of \(\tilde \Phi\)
is the solution of the four-dimensional system we are interested in. Moreover,
we shall see that we can construct \(D_5\) such that it is linear in the kernel
\(\Dslash_W\) and has as its Schur complement any rational approximation to the
Neuberger operator we desire.

Alternatively, we may introduce a five-dimensional pseudofermion field \(\Phi
= \begin{pmatrix} \phi_1 & \phi_2 & \cdots & \phi_{n-1} & \psi
\end{pmatrix}^T\) and write the five-dimensional pseudofermion functional
integral as
\begin{displaymath}
  \int d\Phi^\dagger d\Phi\, e^{-\Phi^\dagger D_5^{-1}\Phi}
  \propto \det D_5 = \det LDU = \det D = \prod_{j=1}^{n}\det D_{j,j}.
\end{displaymath}
The fermion determinant thus obtained is the product of the determinants of all
the diagonal blocks \(D_{j,j}\), wherease we only want the determinant of the
Schur complement, \(D_{n,n}\approx \Dslash_N\). We therefore also introduce
\(n-1\) Pauli--Villars fields\footnote{Perhaps they would be better called
pseudo-pseudofermion fields, as they are boson fields with a local kernel
introduced to cancel the unwanted extra pseudofermion determinants.} to cancel
them,
\begin{displaymath}
  \prod_{j=1}^{n-1}\int d\xi_j^\dagger d\xi_j\, e^{-\xi_j^\dagger D_{j,j}\xi_j}
  \propto \prod_{j=1}^{n-1}\det D_{j,j}^{-1}.
\end{displaymath}

\subsection{Approximation}
\label{sec:approximation}

\subsubsection{Hyperbolic Tangent}
\label{sec:tanh}

The simplest approximation to the \(\sgn\) function is \(\tanh\bigl(n
\tanh^{-1}(x) \bigr)\). In some ways this is quite similar to the situation
for Chebyshev polynomials (\secref{sec:chebyPoly}): there we had an apparently
transcendental function that was in fact a polynomial, here we have an
apparently transcendental function that is in fact a rational function. We
will write the relevant formul{\ae} for even values of \(n\); there is no
problem with using odd values of \(n\), but the formul{\ae} are just a little
messier if we try to be too general. We first write note that
\begin{displaymath}
  x \defn \tanh z = \frac{\sinh z}{\cosh z} = \frac{e^z-e^{-z}}{e^z+e^{-z}}
  = \frac{1-e^{-2z}}{1+e^{-2z}},
\end{displaymath}
solving this equation gives
\begin{displaymath}
  e^{-2z} = \frac{1-x}{1+x},
\end{displaymath}
so we may write
\begin{equation}
  \varepsilon_{n-1,n}(x) = \tanh(n\tanh^{-1} x)
  = \frac{1-e^{-2nz}}{1+e^{-2nz}}
  = \frac{\displaystyle1-\left(\frac{1-x}{1+x}\right)^n}
    {\displaystyle1+\left(\frac{1-x}{1+x}\right)^n}.
  \label{eq:tanh}
\end{equation}
It should be noted that the leading term in the numerator is \(x^{n-1}\) and
not \(x^n\), as that term's coefficient vanishes for \(n\) even; this should
not be surprising, as \(\varepsilon\) is an odd function of~\(x\). The
denominator vanishes when
\begin{displaymath}
    \left(\frac{1-x}{1+x}\right)^n = -1 = e^{(2k+1)i\pi}
    \quad\mbox{with \(k\in\Z_n\)} \quad
    \implies \quad \frac{1-x}{1+x} = e^{(2k+1)i\pi/n},
\end{displaymath}
or
\begin{displaymath}
  x^2 + \left(\tan\frac{(k+\half)\pi}n\right)^2 = 0.
\end{displaymath}
Using a similar argument to find the roots of the numerator we obtain the
factored form
\begin{displaymath}
  \varepsilon_{n-1,n}(x)
    = xn\frac{\prod_{k=1}^{\frac n2-1}
        \left[x^2 + \left(\tan\frac{k\pi}n\right)^2\right]}
      {\prod_{k=0}^{\frac n2-1}
	\left[x^2 + \left(\tan\frac{(k+\half)\pi}n\right)^2\right]}.
\end{displaymath}
If we define \(\xi\defn e^{-2z}\) then
\begin{displaymath}
  \varepsilon_{n-1,n}(x) + 1 = \frac{1-\xi^n}{1+\xi^n} + 1 = \frac2{1+\xi^n}
    = \frac2{\displaystyle\prod_{k=0}^{n-1}(\xi-\xi_k)}
    = \sum_{k=0}^{n-1} \frac{\mu_k}{\xi-\xi_k},
\end{displaymath}
where \(\xi_k\defn \zeta\omega^k\) satisfies \(\xi_k^n=-1\) with \(\zeta\defn
e^{\pi i/n}\) and \(\omega\defn\zeta^2\) being a primitive \nth~root of unity,
and
\begin{displaymath}
  \mu_k \defn
    \frac2{\displaystyle\prod_{\genfrac{}{}{0pt}{1}{j=0}{j\neq k}}^{n-1}
      (\xi_k-\xi_j)} 
    = \frac2{\displaystyle\xi_k^{n-1} \prod_{\ell=1}^{n-1} (1-\omega^\ell)}
    = -\frac{2\xi_k}n.
\end{displaymath}
The last equality follows from the fact that \(\prod_{\ell=1}^{n-1} (x -
\omega_\ell)\) is a cyclotomic polynomial,
\begin{displaymath}
  \frac{x^n-1}{x-1} = \sum_{k=0}^{n-1} x^k
    = \prod_{\ell=1}^{n-1}(x-\omega^\ell)
  \implies \prod_{\ell=1}^{n-1}(1-\omega^\ell) = n.
\end{displaymath}
We thus have
\begin{eqnarray*}
  && \varepsilon_{n-1,n}(x) + 1
  = -\frac2n \sum_{k=0}^{n-1} \frac{\xi_k}{\xi-\xi_k}
  = -\frac2n \sum_{k=0}^{\frac n2-1} \left\{\frac{\xi_k}{\xi-\xi_k} +
    \frac{\xi_{n-1-k}}{\xi-\xi_{n-1-k}}\right\} \\
  &&= -\frac2n \sum_{k=0}^{\frac n2-1} \left\{\frac{\xi_k}{\xi-\xi_k} +
    \frac{1/\xi_k}{\xi-1/\xi_k}\right\}
  = -\frac2n \sum_{k=0}^{\frac n2-1} \frac{(\xi_k+1/\xi_k)\xi-2}
    {\xi^2-(\xi_k+1/\xi_k)\xi+1} \\
  &&= \frac2n \sum_{k=0}^{\frac n2-1} \left[1
    + \frac{x\,\left(\sec\frac{(k+\half)\pi}n\right)^2}
        {x^2 + \left(\tan\frac{(k+\half)\pi}n\right)^2} \right]
  = 1 + \frac{2x}n 
    \sum_{k=0}^{n-1} \frac{1 + \left(\tan\frac{(k+\half)\pi}n\right)^2}
	{x^2 + \left(\tan\frac{(k+\half)\pi}n\right)^2},
\end{eqnarray*}
which is the partial fraction expansion of \(\varepsilon_{n-1,n}(x)\).

\subsubsection{Zolotarev's Formula}
\label{sec:zolotarev}

The hyperbolic tangent is simple, but it is by no means the optimal
approximation of a given order. The analytic form of the optimal approximation
that satisfies Chebyshev's criterion (\secref{sec:chebyRat}) was found by
Zolotarev, who was a student of Chebyshev. Again, Zolotarev's formula is most
simply written in terms of transcendental functions, this time \emph{elliptic
functions}
\begin{displaymath}
 \sn(z/M;\lambda)
  = \frac{\sn(z;k)}M\prod_{m=1}^{\lfloor n/2\rfloor}
    \frac{1-\cfrac{\sn(z;k)^2}{\sn\left(2iK'm/n;k\right)^2}}
	 {1-\cfrac{\sn(z;k)^2}{\sn\left(2iK'(m-\frac{1}{2})/n;k\right)^2}}.
\end{displaymath}
The way to understand this formula is that it expresses the elliptic function
\(\sn(z/M;\lambda)\) as a function of \(\sn(z;k)\), in just the same way that
we expressed \(\tanh nx\) as a function of \(\tanh x\), or \(\cos nx\) as a
function of \(\cos x\) previously. Elliptic functions are doubly periodic
analytic functions, and the Jacobi elliptic function \(\sn(z;k)\) is
defined~by\footnote{This is the analogue of defining \(\sin\) as the inverse
of \(\sin^{-1}(x) \defn \int_0^x dt/\sqrt{1-t^2}\). The inverse of \(\sn\) is
called the elliptic integral of the first kind, \(E(z;k)\).}
\begin{displaymath}
  z \defn \int_0^{\sn(z;k)} \frac{dt}{\sqrt{(1-t^2)(1-k^2t^2)}}
\end{displaymath}
so as to have a real period of \(4K(k)\) and imaginary period of
\(2iK(k')\), where \(k^2+k'^2=1\) and \(K\) is the complete elliptic integral
\begin{displaymath}
  K(k) \defn \int_0^1 \frac{dt}{\sqrt{(1-t^2)(1-k^2t^2)}}.
\end{displaymath}
\(M\) and \(\lambda\) are chosen such that the imaginary period of
\(\sn(z,k)\) is \(n\) times that of \(\sn(z/M;\lambda)\), \(2iMK(\lambda') =
2iK(k')/n\), while the real periods are the same, \(4MK(\lambda)=4K(k)\).

\begin{figure}[htb]
  \centerline{\epsfxsize=0.9\textwidth \epspdffile{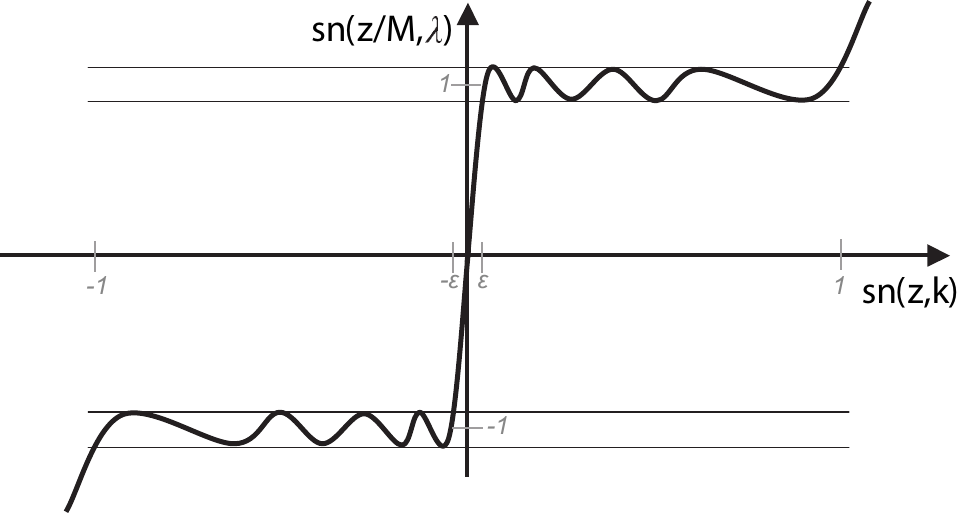}}   
  \caption[Zolotarev approximation]{The degree Zolotarev approximation to
           \(\sgn x\) over the interval \(\varepsilon\leq |x|\leq1\). The real
           period of the elliptic function corresponds to the step in the
           \(\sgn\) function, and the imaginary period corresponds to the
           equal height oscillations about \(\pm1\).}
  \label{fig:approximation_zolotarev}
\end{figure}

Figure~\ref{fig:approximation_zolotarev} shows \(\sn(z/M;\lambda)\) as a
function of \(\sn(z;k)\). The r\^ole of the double periodicity is now
apparent: the real period corresponds to the step in the \(\sgn\) function at
the origin, and the imaginary period corresponds to the equal height
oscillations about \(\pm1\) required by Chebyshev's criterion.

\begin{figure}[htb]
  \centerline{\epsfxsize=0.75\textwidth \epspdffile{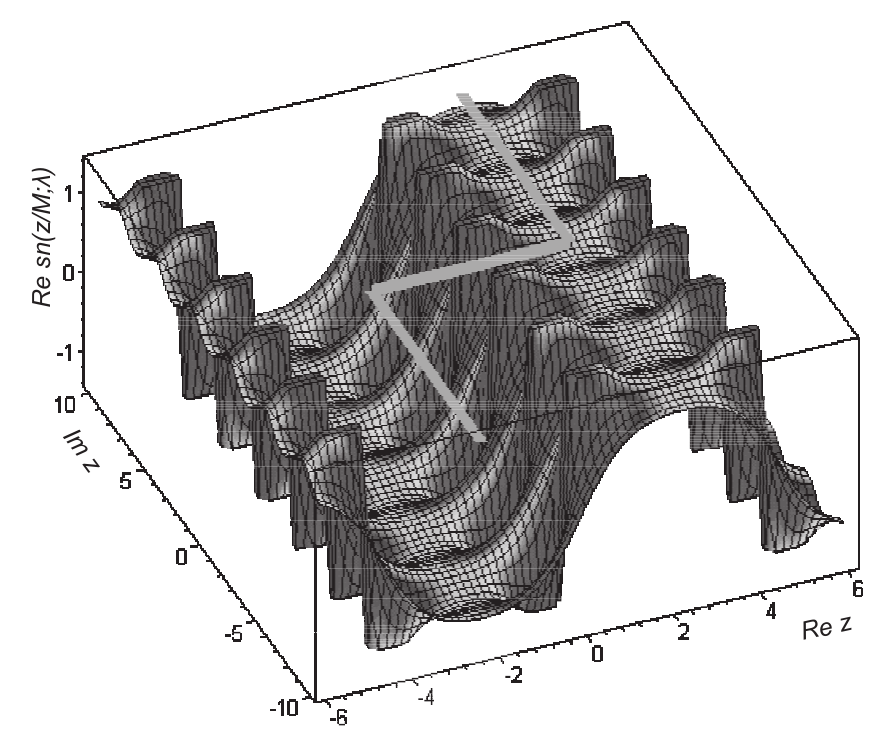}}   
  \caption{Surface plot showing the real part of sn(\(z/M\);\(\lambda\)) over
           the fundamental region of sn(\(z\);\(k\)). The imaginary part
           vanishes along the contour indicated.}
\label{fig:surface_plot}
\end{figure}

Figure~\ref{fig:surface_plot} shows a surface plot of the real part of
\(\sn(z/M;\lambda)\) for \(n=6\); the shading of the surface indicates the
imaginary part which is zero along the indicated contour. The entire region of
the complex plane shown is the fundamental region for \(\sn(z;k)\), and
corresponds to six fundamental periods stacked in the imaginary direction for
\(\sn(z/M;\lambda)\).

\subsubsection{Errors}

\begin{figure}[htb]
  \centerline{\epsfxsize=0.9\textwidth \epspdffile{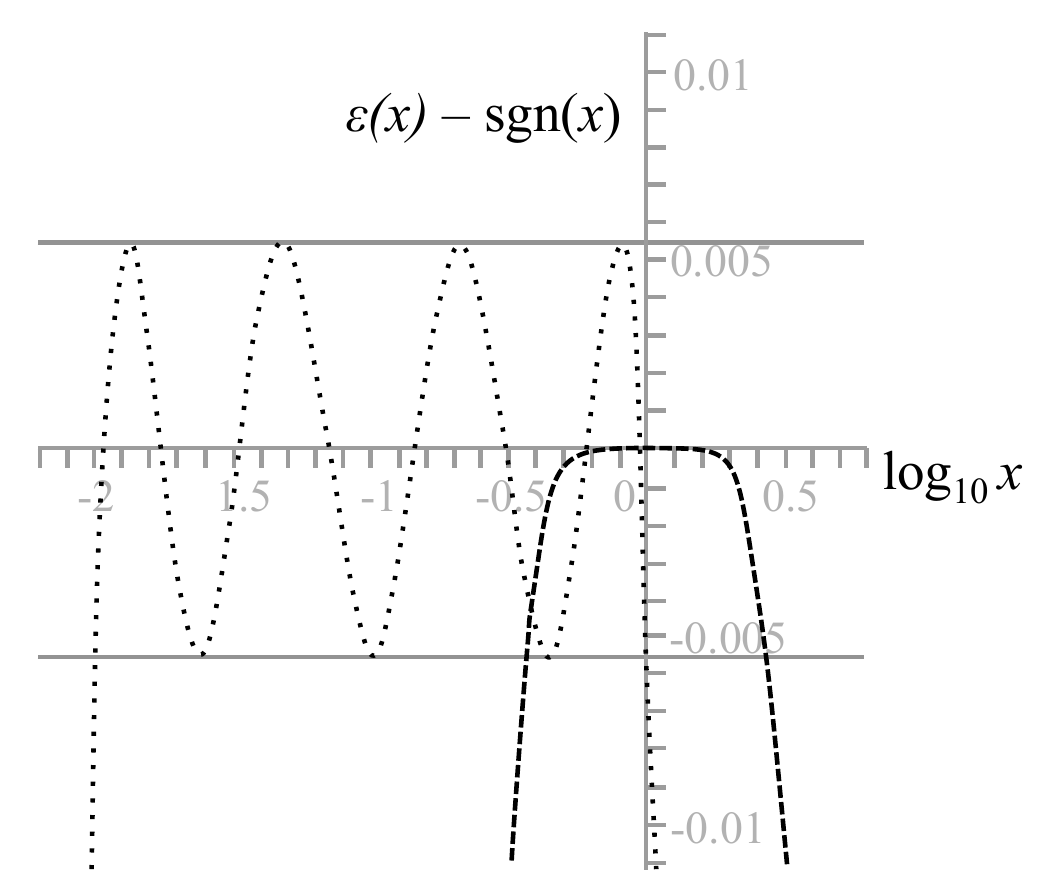}}   
  \caption{The errors in approximations to the sign function,
           \(\varepsilon(x)-\sgn(x)\), as a function of \(x\). The dashed
           curve is \(\tanh\bigl(8 \tanh^{-1}x\bigr) - \sgn x\) and the
           dotted curve is the corresponding error for the optimal (Zolotarev)
           approximation of the same degree over the interval
           \(10^{-2}\leq|x|\leq1\). The \(\tanh\) approximation does not
           depend on the choice of interval.}
  \label{fig:approximation_errors}
\end{figure}

We next compare these approximations to the sign function over the interval
\(10^{-2}\leq|x|\leq1\). For both Zolotarev and \(\tanh\) approximations we
take rational functions of degree \((7,8)\) and compare them in
Figure~\ref{fig:approximation_errors} on a log--linear plot. We can see that
\(\tanh\) is a better approximation near \(|x|=1\), but becomes a much worse
approximation when \(x\) is small. Indeed, the errors in the \(\tanh\)
approximation are symmetric about \(|x|=1\): this was what motivated the
introduction of the M\"obius kernel, as a suitable choice of the parameters
shift the spectrum of the kernel to better overlap with the region where the
error for the \(\tanh\) approximation is small.

\subsection{Representation}
\label{sec:representation}

The formalism developed in \secref{sec:constraint} requires that we have a
(five-dimensional) matrix which has the desired approximation
(\secref{sec:approximation}) as its Schur complement. Constructing a variety
of such matrices is the goal of this section.

\subsubsection{Continued Fraction}

We begin with the continued fraction representation.\cite{Neuberger:1999re,%
Borici:2001ua} Consider the block (five-dimensional) matrix with
four-dimensional blocks \(A_0,\ldots, A_n\) on the principal diagonal and unit
matrices on the adjacent block diagonals, for example with \(n=4\) this is
\begin{displaymath}
  \begin{pmatrix}
   A_0 & 1 & 0 & 0 \\
   1 & A_1 & 1 & 0 \\
   0 & 1 & A_2 & 1 \\
   0 & 0 & 1 & A_3
  \end{pmatrix};
\end{displaymath}
it is straightforward to write its block LDU decomposition,
\begin{displaymath}
  \begin{pmatrix}
   1 & 0 & 0 & 0 \\
   S_0^{-1} & 1 & 0 & 0 \\
   0 & S_1^{-1} & 1 & 0 \\
   0 & 0 & S_2^{-1} & 1
  \end{pmatrix}
  \begin{pmatrix}
   S_0 & 0 & 0 & 0 \\
   0 & S_1 & 0 & 0 \\
   0 & 0 & S_2 & 0 \\
   0 & 0 & 0 & S_3
  \end{pmatrix}
  \begin{pmatrix}
   1 & S_0^{-1} & 0 & 0 \\
   0 & 1 & S_1^{-1} & 0 \\
   0 & 0 & 1 & S_2^{-1} \\
   0 & 0 & 0 & 1
  \end{pmatrix},
\end{displaymath}
where \(S_0=A_0\) and \(S_n+1/S_{n-1}=A_n\). The Schur complement of the
matrix is thus the \emph{continued fraction}
\begin{displaymath}
  S_3 =
  A_3 - \cfrac1{S_2} =
  A_3 - \cfrac1{A_2 - \cfrac1{S_1}} =
  A_3 - \cfrac1{A_2 - \cfrac1{A_1 - \cfrac1{A_0}}}.
\end{displaymath}
We may use this representation to linearize our rational approximations to the
\(\sgn\) function expressed as a continued fraction,
\begin{displaymath}
  \varepsilon_{n-1,n}(H) = \beta_0 H
    + \cfrac1{\beta_1H + \cfrac1{\beta_2H + \ddots + \cfrac1{\beta_nH}}},
\end{displaymath}
by setting \(A_{n-j}=(-1)^j\beta_jH\) for \(j=0,\ldots,n\). We can improve the
condition number of our five-dimensional matrix by making use of the freedom to
introduce arbitrary parameters into the continued fraction expansion by
multiplying the numerator and denominator at level \(j\) by a
constant~\(c_{j+1}\),
\begin{displaymath}
  \varepsilon_{n-1,n}(H) = \beta_0 H
    + \cfrac{c_1}{c_1\beta_1H
      + \cfrac{c_1c_2}{c_2\beta_2H
	+ \ddots
	+ \cfrac{c_{n-1}c_n}{c_n\beta_nH}}},
\end{displaymath}
which is the Schur complement of the block tridiagonal five-dimensional matrix
\begin{displaymath}
  \setlength{\arraycolsep}{0.5em}
  \begin{pmatrix}
    c_n\beta_nH & \sqrt{c_{n-1}c_n} & 0 & & & \\
    \sqrt{c_{n-1}c_n} & -c_{n-1}\beta_{n-1}H & \ddots & 0 & & \\
    0 & \ddots & \ddots & \ddots & 0 & \\
    & 0 & \ddots & c_2\beta_2H & \sqrt{c_1c_2} & 0 \\
    & & 0 & \sqrt{c_1c_2} & -c_1\beta_1H & \sqrt{c_1} \\
    & & & 0 & \sqrt{c_1} & \beta_0H
  \end{pmatrix}.
\end{displaymath}

\subsubsection{Partial Fraction}

Next we consider the partial fraction representation.\cite{Neuberger:1998my,%
Edwards:1998yw} To this end we consider a matrix of the form
\begin{displaymath}
  \setlength{\arraycolsep}{0.5em}
  \setbox\strutbox=\hbox{\vrule height3.5ex depth2ex width0pt}
  \left(\begin{array}{cc|cc|c}
    \displaystyle\frac x{p_1} & 1 & 0 & 0 & 1 \\
    1 & \displaystyle\frac{p_1x}{q_1^2} & 0 & 0 & 0 \\[1ex] \hline
    0 & 0 & \displaystyle\frac x{p_2} & 1 & 1 \\
    0 & 0 & 1 & \displaystyle\frac{p_2x}{q_2^2} & 0 \\[1ex] \hline
    -1 & 0 & -1 & 0 & R
  \end{array}\right),
\end{displaymath}
which in general has \(n\) \(2\times2\) blocks along the principal diagonal
with the corresponding \(\mp1\) in the bottom row and rightmost column, and
compute its block LDU decomposition:
\begin{displaymath}
  \setlength{\arraycolsep}{0.5em}
  \setbox\strutbox=\hbox{\vrule height3.5ex depth2ex width0pt}
  L = \left(\begin{array}{cc|cc|c}
    1 & 0 & 0 & 0 & 0 \\
    \displaystyle\frac{p_1}{x} & 1 & 0 & 0 & 0 \\[1ex] \hline
    0 & 0 & 1 & 0 & 0 \\
    0 & 0 & \displaystyle\frac{p_2}{x} & 1 & 0 \\[1ex] \hline
    -\displaystyle\frac{p_1}{x} &
    \displaystyle\frac{q_1x}{x^2-q_1^2} - \frac{q_1}{x+q_1} &
    -\displaystyle\frac{p_2}{x} &
    \displaystyle\frac{q_1x}{x^2-q_2^2} -\frac{q_2}{x+q_2} & 1
  \end{array}\right),
\end{displaymath}
\begin{displaymath}
  \setlength{\arraycolsep}{0.5em}
  \setbox\strutbox=\hbox{\vrule height3.5ex depth2ex width0pt}
  D = \left(\begin{array}{cc|cc|c}
    \displaystyle\frac x{p_1} & 0 & 0 & 0 & 0 \\
    0 & \displaystyle\frac{p_1(x^2-q_1^2)}{xq_1^2} & 0 & 0 & 0 \\[1ex] \hline
    0 & 0 & \displaystyle\frac x{p_2} & 0 & 0 \\
    0 & 0 & 0 & \displaystyle\frac{p_2(x^2-q_2^2)}{xq_2^2} & 0 \\[1ex] \hline
    0 & 0 & 0 & 0 &
    \displaystyle R+\frac{p_1x}{x^2-q_1^2}+\frac{p_2x}{x^2-q_2^2}
  \end{array}\right),
\end{displaymath}
and
\begin{displaymath}
  \setlength{\arraycolsep}{0.5em}
  \setbox\strutbox=\hbox{\vrule height3.5ex depth2ex width0pt}
  U = \left(\begin{array}{cc|cc|c}
    1 & \displaystyle\frac{p_1}{x} & 0 & 0 & \displaystyle\frac{p_1}{x} \\
    0 & 1 & 0 & 0 & -\displaystyle\frac{q_1x}{x^2-q_1^2} + \frac{q_1}{x+q_1}
    \\[1ex] \hline
    0 & 0 & 1 & \displaystyle\frac{p_2}{x} & \displaystyle\frac{p_2}{x} \\
    0 & 0 & 0 & 1 & -\displaystyle\frac{q_2x}{x^2-q_2^2}+\frac{q_2}{x+q_2}
    \\[1ex] \hline
    0 & 0 & 0 & 0 & 1
  \end{array}\right);
\end{displaymath}
so its Schur complement is
\begin{displaymath}
  R + \frac{p_1x}{x^2-q_1^2} + \frac{p_2x}{x^2-q_2^2}.
\end{displaymath}
This allows us to represent the partial fraction expansion of our rational
function as the Schur complement of a five-dimensional linear system
\begin{displaymath}
  \varepsilon_{n-1,n}(H) = \sum_{j=1}^{n} \frac{p_jH}{H^2-q_j^2}.
\end{displaymath}

\subsubsection{Euclidean Cayley Transform}
\label{sec:ECT}

Finally we consider a five-dimensional matrix of the following
form\cite{Borici:1999zw,Edwards:2000qv,Chiu:2002ir,Brower:2004xi}
\begin{displaymath}
  \setlength{\arraycolsep}{0.5em}
  \begin{pmatrix}
     1 & 0 & 0 & -T_1^{-1}C_{+} \\
     -T_2^{-1} & 1 & 0 & 0 \\
     0 & -T_3^{-1} & 1 & 0 \\
     0 & 0 & -T_4^{-1} & C_{-}
  \end{pmatrix},
\end{displaymath}
and compute its block LDU decomposition
\begin{displaymath}
  \setlength{\arraycolsep}{0.5em}
  L =\begin{pmatrix}
    1 & 0 & 0 & 0 \\
    -T_2^{-1} & 1 & 0 & 0 \\
    0 & -T_3^{-1} & 1 & 0 \\
    0 & 0 & -T_4^{-1} & 1
  \end{pmatrix},
\end{displaymath}
\begin{displaymath}
  \setlength{\arraycolsep}{0.5em}
  D = \begin{pmatrix}
    1 & 0 & 0 & 0 \\
    0 & 1 & 0 & 0 \\
    0 & 0 & 1 & 0 \\
    0 & 0 & 0 & C_{-} - T_4^{-1}T_3^{-1}T_2^{-1}T_1^{-1}C_{+}
 \end{pmatrix},
\end{displaymath}
and
\begin{displaymath} 
  \setlength{\arraycolsep}{0.5em}
  U = \begin{pmatrix}
    1 & 0 & 0 & -T_1^{-1}C_{+} \\
    0 & 1 & 0 & -T_2^{-1}T_1^{-1}C_{+} \\
    0 & 0 & 1 & -T_3^{-1}T_2^{-1}T_1^{-1}C_{+} \\
    0 & 0 & 0 & 1
  \end{pmatrix},
\end{displaymath}
so its Schur complement, in the general \(n\times n\) case, is
\begin{displaymath}
  D_{n,n} = C_{-} - T_n^{-1} T_{n-1}^{-1} \cdots T_2^{-1} T_1^{-1}C_{+}.
\end{displaymath}
Note that \(L\) does not depend on \(C_\pm\). If we define \(T = T_1 \cdots
T_n\) and \(C_{\pm} = \half(1-\mu) \pm \half(1+\mu) \gamma_5\), then this Schur
complement becomes
\begin{displaymath}
  D_{n,n}(\mu) = -(1+T^{-1})\gamma_5 \times
    \half\left[(1+\mu)+(1-\mu)\gamma_5\frac{1-T}{1+T}\right].
\end{displaymath}
The rational approximation to the Neuberger operator can thus be written~as
\begin{eqnarray*}
  \Dslash_N(H)
  &\approx& \half\bigl[(1+\mu)+(1-\mu)\gamma_5\varepsilon(H)\bigr] \\
  &=& D_{n,n}(1)^{-1}D_{n,n}(\mu)
  = \half\left[(1+\mu)+(1-\mu)\gamma_5\frac{1-T}{1+T}\right],
\end{eqnarray*}
if we define \(T(x)\) to be the \emph{Euclidean Cayley transform}\footnote{A
Cayley transform is the mapping \[U = \frac{1-iH}{1+iH}, \qquad H = i\frac{U-
1}{U+1},\] between a Hermitian matrix \(H\) and a unitary matrix \(U\).} of
\(\varepsilon(x) \approx \sgn(x)\)
\begin{displaymath}
  \varepsilon(x) = \frac{1-T(x)}{1+T(x)};
\end{displaymath}
it is trivial to find \(T(x)\) because this relation is its own inverse,
that~is
\begin{displaymath}
  T(x) = \frac{1-\varepsilon(x)}{1+\varepsilon(x)}.
\end{displaymath}
For an odd function such as \(\varepsilon(x)\) we have
\begin{displaymath}
  \varepsilon(-x) = -\varepsilon(x)
  \quad\Leftrightarrow\quad  T(-x) = \frac1{T(x)};
\end{displaymath}
this means that if \(\omega_j\) is a zero of \(T(x)\) then \(-\omega_j\) must
be a pole, so \(T(x)\) can be written as a product of factors of the form
\[T_j(x) = \frac{\omega_j-x}{\omega_j+x}.\] Moreover, if \(\varepsilon(0) =
0\) then \(T(0) = 1\), so \(T = T_1T_2\cdots T_n\). 

The Euclidean Cayley transform of the hyperbolic tangent approximation
(\secref{sec:tanh}) is immediately apparent from equation~(\ref{eq:tanh}),
\begin{displaymath}
  T(x) = \left(\frac{1-x}{1+x}\right)^n,
\end{displaymath}
so all the roots are degenerate, \(\omega_j=1\:(\forall j)\). For the Zolotarev
approximation (\secref{sec:zolotarev}) the roots of its Euclidean Cayley
transform are the values of \(x\) for which the \(\varepsilon_{n-1,n}(x)=1\),
which are readily found in terms of elliptic integrals.

We want to solve the four-dimensional linear system \[\Dslash_N(\mu)\phi\approx
\half\left[(1+\mu) + (1-\mu)\gamma_5\frac{1-T}{1+T}\right] \phi =
D_{n,n}(1)^{-1}D_{n,n}(\mu) \phi = \chi,\] and we can do this by solving the
five-dimensional sytem \(D_5(\mu)\tilde\Phi = D_5(1)\tilde X\) with the vectors
\(\tilde\Phi\) and \(\tilde X\) as in \secref{sec:constraint}. The block LDU
decomposition of this~gives \(LD(\mu)U(\mu)\tilde\Phi = LD(1)U(1)\tilde X\)
where \(L\) is independent of \(\mu\) (recall that we noted above that it did
not depend upon \(C_\pm\), which is only source of \(\mu\)-dependence).
Multiplying on the left by \(L^{-1}\), and substituting \(U(\mu)\tilde\Phi =
\Phi\) and \(U(1)\tilde X = X\) (with \(X_n=\chi\)), we obtain the block
diagonal equation \(D(\mu)\Phi = D(1)X\), whose bottom block is
\(D_{n,n}(\mu)\phi = D_{n,n}(1) \chi\). Since \(D_{n,n}\) is the Schur
complement this gives \(D_{n,n}(1)^{-1} D_{n,n}(\mu) \phi = \chi\), so as
before (q.v., \secref{sec:constraint}) we find the bottom four-dimensional
block \(\phi\) of the five-dimensional solution vector \(\tilde\Phi\) is the
solution of our four-dimensional problem.

\subsubsection{Relation to Domain Wall Fermions}

The Euclidean Cayley transform representation is a generalization of the usual
domain wall fermion formulation. We consider the generalised domain wall with
the M\"obius kernel, for which one solves the five-dimensional linear system
\(D_5(\mu) \tilde\Phi = D_5(1) \tilde X\) where
\begin{eqnarray}
  D_5(\mu) &=& \begin{pmatrix}
    D_{+}^{(1)} &\quad
    & D_{-}^{(1)}P_{+} &\quad
    & 0 &\quad
    & 0 &\quad
    & -\mu D_{-}^{(1)}P_{-} \\
    D_{-}^{(2)}P_{-} &\quad
    & D_{+}^{(2)} &\quad
    & D_{-}^{(2)}P_{+} &\quad
    & 0 &\quad
    & 0 \\
    0 &\quad
    & D_{-}^{(3)}P_{-} &\quad
    & D_{+}^{(3)} &\quad
    & D_{-}^{(3)}P_{+} &\quad
    & 0 \\
    0 &\quad
    & 0 &\quad
    & D_{-}^{(4)}P_{-} &\quad
    & D_{+}^{(4)} &\quad
    & D_{-}^{(4)}P_{+} \\
    -\mu D_{-}^{(5)}P_{+} &\quad
    & 0 &\quad
    & 0 &\quad
    & D_{-}^{(5)}P_{-} &\quad
    & D_{+}^{(5)}
  \end{pmatrix} \nonumber \\
  &=&
  \begin{pmatrix}
    D_{+}^{(1)} &\quad & 0 &\quad & 0 &\quad & 0 &\quad &-\mu D_{-}^{(1)} \\
    D_{-}^{(2)} &\quad & D_{+}^{(2)} &\quad & 0 &\quad & 0 &\quad & 0 \\
    0 &\quad & D_{-}^{(3)} &\quad & D_{+}^{(3)} &\quad & 0 &\quad & 0 \\
    0 &\quad & 0 &\quad & D_{-}^{(4)} &\quad & D_{+}^{(4)} &\quad & 0 \\
    0 &\quad & 0 &\quad & 0 &\quad & D_{-}^{(5)} &\quad & D_{+}^{(5)}
  \end{pmatrix} P_{-} \nonumber \\
  && \qquad +
  \begin{pmatrix}
    D_{+}^{(1)} &\quad & D_{-}^{(1)} &\quad & 0 &\quad & 0 &\quad & 0 \\
    0 &\quad & D_{+}^{(2)} &\quad & D_{-}^{(2)} &\quad & 0 &\quad & 0 \\
    0 &\quad & 0 &\quad & D_{+}^{(3)} &\quad & D_{-}^{(3)} &\quad & 0 \\
    0 &\quad & 0 &\quad & 0 &\quad & D_{+}^{(4)} &\quad & D_{-}^{(4)} \\
    -\mu D_{-}^{(5)} &\quad & 0 &\quad & 0 &\quad & 0 &\quad & D_{+}^{(5)}
  \end{pmatrix} P_{+}.
 \label{eq:DWF}
\end{eqnarray}
Here \(P_\pm = \half(1\pm\gamma_5)\) are the chiral projectors, and
\begin{displaymath}
  a_5D_{\pm}^{(s)} = \alpha_s \bigl[b_{\pm}^{(s)}a_5(\Dslash_W-M) \pm 1\bigr],
\end{displaymath}
with
\begin{displaymath}
  b_{+}^{(s)} + b_{-}^{(s)} = \frac{b_{+} + b_{-}}{\omega_s}, \qquad
  b_{+}^{(s)} - b_{-}^{(s)} = b_{+} - b_{-},
\end{displaymath}
and \(\alpha_s\) is a new free parameter that we may use to minimize the
condition number of~\(D_5\) as it does not effect the Schur complement.
Shamir's domain wall is the special case with \(b_{+}=1\), \(b_{-}=0\),
\(\omega_s=1\), and \(\alpha_s=1\). For Chiu's Zolotarev variant we set
\(\omega_s\) to be the roots of the Euclidean Cayley transform of the
Zolotarev approximation to the \(\sgn\) function.

We now continue our derivation from equation~(\ref{eq:DWF}) by cyclically
shifting the columns of the right-handed part left, \(D_5\to D_5\cshift\),
where
\begin{displaymath}
  \cshift \defn
  \begin{pmatrix}
    0 & 0 & 0 & 0 & 1 \\
    1 & 0 & 0 & 0 & 0 \\
    0 & 1 & 0 & 0 & 0 \\
    0 & 0 & 1 & 0 & 0 \\
    0 & 0 & 0 & 1 & 0 \\
  \end{pmatrix} P_{+} +
  \begin{pmatrix}
    1 & 0 & 0 & 0 & 0 \\
    0 & 1 & 0 & 0 & 0 \\
    0 & 0 & 1 & 0 & 0 \\
    0 & 0 & 0 & 1 & 0 \\
    0 & 0 & 0 & 0 & 1
  \end{pmatrix} P_{-},
\end{displaymath}
which is a unitary transformation, \(\cshift^\dagger=\cshift^{-1}\). This gives
\begin{eqnarray}
  D_5(\mu)\cshift &=&
  \begin{pmatrix}
    D_{+}^{(1)} &\quad& 0 &\quad& 0 &\quad& 0 &\quad& -\mu D_{-}^{(1)} \\
    D_{-}^{(2)} &\quad& D_{+}^{(2)} &\quad& 0 &\quad& 0 &\quad& 0 \\
    0 &\quad& D_{-}^{(3)} &\quad& D_{+}^{(3)} &\quad& 0 &\quad& 0 \\
    0 &\quad& 0 &\quad& D_{-}^{(4)} &\quad& D_{+}^{(4)} &\quad& 0 \\
    0 &\quad& 0 &\quad& 0 &\quad& D_{-}^{(5)} &\quad& D_{+}^{(5)}
  \end{pmatrix} P_{-} \nonumber \\
  && + \begin{pmatrix}
    D_{-}^{(1)} &\quad& 0 &\quad& 0 &\quad& 0 &\quad& D_{+}^{(1)} \\
    D_{+}^{(2)} &\quad& D_{-}^{(2)} &\quad& 0 &\quad& 0 &\quad& 0 \\
    0 &\quad& D_{+}^{(3)} &\quad& D_{-}^{(3)} &\quad& 0 &\quad& 0 \\
    0 &\quad& 0 &\quad& D_{+}^{(4)} &\quad& D_{-}^{(4)} &\quad& 0 \\
    0 &\quad& 0 &\quad& 0 &\quad& D_{+}^{(5)} &\quad& -\mu D_{-}^{(5)}
  \end{pmatrix} P_{+}.
  \label{eq:DWF1}
\end{eqnarray}
If we define \(Q_{\pm}^{(s)} \defn D_{+}^{(s)}P_{\pm} + D_{-}^{(s)}P_{\mp}\),
then
\begin{displaymath}
  \begin{array}{lr@{\qquad}lr}
    D_{+}^{(s)} P_{+} &= Q_{+}^{(s)} P_{+}, &
    D_{+}^{(s)} P_{-} &= Q_{-}^{(s)} P_{-}, \\
    D_{-}^{(s)} P_{+} &= Q_{-}^{(s)} P_{+}, &
    D_{-}^{(s)} P_{-} &= Q_{+}^{(s)} P_{-};
  \end{array}
\end{displaymath}
so equation~(\ref{eq:DWF1}) becomes
\begin{displaymath}
  D_5(\mu)\cshift =
  \begin{pmatrix}
    Q_{-}^{(1)} &\quad& 0 &\quad& 0 &\quad& 0 &\quad& Q_{+}^{(1)} C_{+} \\   
    Q_{+}^{(2)} &\quad& Q_{-}^{(2)} &\quad& 0 &\quad& 0 &\quad& 0 \\
    0 &\quad& Q_{+}^{(3)} &\quad& Q_{-}^{(3)} &\quad& 0 &\quad& 0 \\
    0 &\quad& 0 &\quad& Q_{+}^{(4)} &\quad& Q_{-}^{(4)} &\quad& 0 \\
    0 &\quad& 0 &\quad& 0 &\quad& Q_{+}^{(5)} &\quad& Q_{-}^{(5)} C_{-}
  \end{pmatrix},
\end{displaymath}
where \(C_{\pm} \defn \half(1-\mu) \pm \half(1+\mu)\gamma_5 = P_{\pm} - \mu
P_{\mp}\) as before. Scaling out the matrix
\begin{displaymath}
  \Q \defn \begin{pmatrix}
    Q_{-}^{(1)} & 0 & 0 & 0 & 0 \\
    0 & Q_{-}^{(2)} & 0 & 0 & 0 \\
    0 & 0 & Q_{-}^{(3)} & 0 & 0 \\
    0 & 0 & 0 & Q_{-}^{(4)} & 0 \\
    0 & 0 & 0 & 0 & Q_{-}^{(5)}
  \end{pmatrix},
\end{displaymath}
the domain wall operator reduces to the form introduced before
\begin{displaymath}
  \Q^{-1} D_5(\mu) \cshift = \begin{pmatrix}
    1 & 0 & 0 & 0 & -T_1^{-1}C_{+} \\
    -T_2^{-1} & 1 & 0 & 0 & 0 \\
    0 & -T_3^{-1} & 1 & 0 & 0 \\
    0 & 0 & -T_4^{-1} & 1 & 0  \\
    0 & 0 & 0 & -T_5^{-1} & C_{-}
  \end{pmatrix},
\end{displaymath}
where \(T_s^{-1} = -{Q_{-}^{(s)}}^{-1}Q_{+}^{(s)}\) and hence
\begin{eqnarray*}
  T_s &\defn& -{Q_{+}^{(s)}}^{-1} Q_{-}^{(s)} \\
  &=& - \bigl[D^{(s)}_{+}P_{+}+D^{(s)}_{-}P_{-}\bigr]^{-1}
    \bigl[D^{(s)}_{+}P_{-}+D^{(s)}_{-}P_{+}\bigr] \\
  &=& - \left[
    \alpha_s\bigl(b^{(s)}_{+}a_5(\Dslash_W-M) + 1\bigr) P_{+}
    + \alpha_s\big(b^{(s)}_{-}a_5(\Dslash_W-M) - 1\bigr) P_{-}
    \right]^{-1} \\
  && \quad \times \left[
    \alpha_s\bigl(b^{(s)}_{+}a_5(\Dslash_W-M) + 1\bigr) P_{-}
    + \alpha_s\bigl(b^{(s)}_{-}a_5(\Dslash_W-M) - 1 \bigr) P_{+}
    \right] \\
  &=& - \left[
    (b^{(s)}_{+} + b^{(s)}_{-})a_5 (\Dslash_W-M) +
    \bigl((b^{(s)}_{+} - b^{(s)}_{-})a_5 (\Dslash_W-M) + 2\bigr) \gamma_5
    \right]^{-1} \\
  && \quad \times \left[
    (b^{(s)}_{+} + b^{(s)}_{-})a_5 (\Dslash_W-M) -
    \bigl((b^{(s)}_{+} - b^{(s)}_{-})a_5 (\Dslash_W-M) + 2\bigr) \gamma_5
    \right] \\
  &=& - \left[
    \frac1{\omega_s} (b_{+} + b_{-})a_5 (\Dslash_W-M) +
    \bigl((b_{+} - b_{-})a_5 (\Dslash_W-M) + 2\bigr) \gamma_5 
    \right]^{-1} \\
  && \quad \times \left[
    \frac1{\omega_s} (b_{+} + b_{-})a_5 (\Dslash_W-M) -
    \bigl((b_{+} - b_{-})a_5 (\Dslash_W-M) + 2\bigr) \gamma_5 
    \right] \\
  &=& - \gamma_5 \left[
    \frac{(b_{+} + b_{-})a_5 (\Dslash_W-M)} {2 + (b_{+} 
      - b_{-})a_5 (\Dslash_W-M)}
      \gamma_5 + \omega_s \right]^{-1} \gamma_5\\ 
  && \quad \times \gamma_5 \left[
    \frac{(b_{+} + b_{-})a_5 (\Dslash_W-M)} {2 + (b_{+}
      - b_{-})a_5 (\Dslash_W-M)}
      \gamma_5 - \omega_s \right] \gamma_5 \\
  &=& \frac{\omega_{s} - a_5H_M}{\omega_{s} + a_5H_M},
\end{eqnarray*}
\(H_M\) being the M\"obius kernel introduced in \secref{sec:kernel}.  We choose
\(a_5\) such that the spectrum of \(a_5H_M\) is within \([-1,1]\).

Note that \(\DDW^{-1} \defn \Dslash_N^{-1}-a\) satisfies
\begin{displaymath}
  \{\DDW^{-1},\gamma_5\} = \{\Dslash_N^{-1},\gamma_5\} - 2a\gamma_5
  = \Dslash_N^{-1} \{\Dslash_N,\gamma_5\} \Dslash_N^{-1} - 2a\gamma_5 = 0,
\end{displaymath}
so for valence measurements the `external' propagator \(\DDW^{-1}\) is very
convenient, as it satisfies the off-shell anticommutation relation. However,
for dynamical calculations this would violate the Nielsen--Ninomaya theorem,
and indeed as shown by Pelisetto renormalisation induces unwanted ghost
doublers that prevent the off-shell Ward identity from being preserved, so we
cannot use \(\DDW\) for dynamical (`internal') propagators: we must use
\(D_{N}\) in the quantum action instead.

\subsection{Chiral Symmetry Breaking}

In order to measure chiral symmetry breaking, we define Ginsparg--Wilson
\emph{defect}
\begin{displaymath}
  \gamma_5\Dslash + \Dslash\gamma_5 - 2a\Dslash\gamma_5\Dslash
  \defn \gamma_5\Delta_L.
\end{displaymath}
For the approximate Neuberger operator \(a\Dslash = \half[1+\gamma_5
\varepsilon(H)]\) the defect is \(a\Delta_L = \half[1-\varepsilon^2(H)]\).  If
we use a Zolotarev approximation for \(\varepsilon\) that covers the spectrum
of the kernel \(H\) with a minimax error of \(\Delta\) this shows
that\footnote{Ignoring terms of \(O(\Delta^2)\).}  \(\|\Delta_L\| \leq
\Delta\), which enables us to bound the matrix elements of \(\Delta_L\) that
must be present in all (on-shell) chiral symmetry violating corrections. There
is no such simple bound if the \(\tanh\) approximation is used.

The \emph{residual mass} may be defined by
\begin{displaymath}
 \mres =
 \frac{\langle\tr G^\dagger\Delta_L G\rangle}{\langle\tr G^\dagger G\rangle},
\end{displaymath}
where \(G\) is the quark propagator. For DWF this has been shown by Brower
\textit{et al.} to be exactly the usual domain wall residual mass. The
residual mass \(\mres\) is just one moment of \(\Delta_L\).

\subsection{Numerical Studies}

Preliminary numerical studies have been carried out comparing the various
five-dimensional on-shell chiral fermion formulations introduced in this
lecture.\cite{Edwards:2005an} These were carried out on an ensemble of 15
gauge configurations from the RBRC two-flavour domain wall fermion
dataset. The lattice size was \(V=16^3\times 32\), and the parameters used to
generate the ensemble were \(n_s=12\) (the size of fifth dimension), the DBW2
action with \(\beta=0.8\), a domain wall height of \(M=1.8\), and a fermion
mass of \(\mu=0.02\).

\begin{figure}[htb]
  \centerline{\epsfxsize=0.9\textwidth \epspdffile{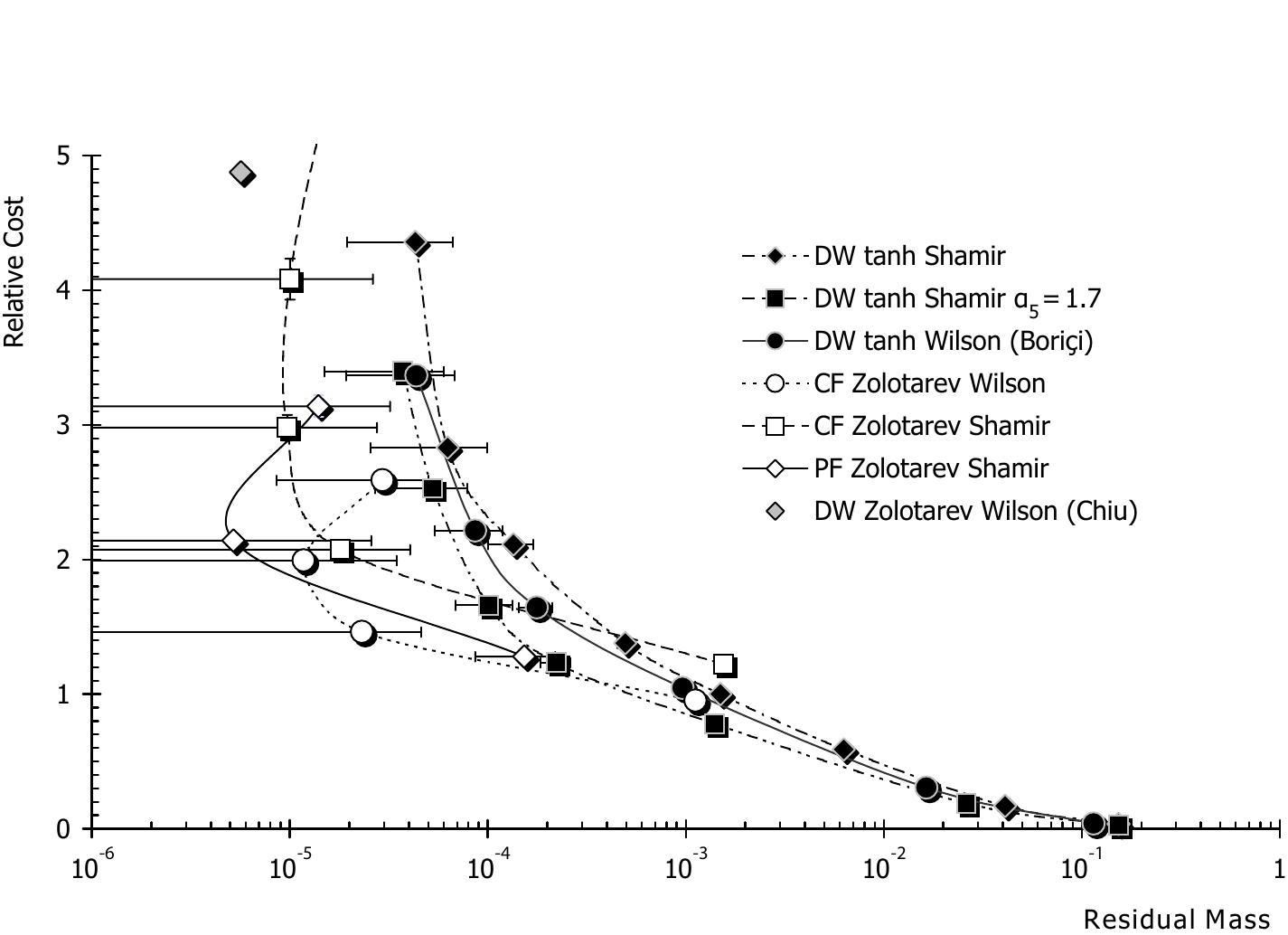}}   
  \caption{Relative cost in units of Wilson--Dirac matrix applications as a
           function of the residual mass \(\mres\) for some five-dimensional
           actions. Note the Zolotarev domain wall data point was only measured
           on a single configuration, and therefore is plotted without any
           error bars. The Zolotarev approximations were over an interval of
           \(0.01\leq|x|\leq1\), which does not completely cover the spectrum
           of the kernel: if the approximation was chosen to cover the
           spectrum, or the eigenvectors corresponding to eigenvalues
           \(|\lambda|<0.01\) were projected out and handled exactly, then we
           would expect \(\mres\) to be much smaller, probably for little extra
           cost.}
  \label{fig:cost_vs_m_res}
\end{figure}

The costs of inverting the various five-dimensional operators were compared as
a function of the residual mass \(\mres\), and in
Figure~\ref{fig:cost_vs_m_res} we plot the results for the most promising
actions. When comparing different valence kernels the \(\pi\) mass was
matched. In order to compare like with like all the Dirac operators were
even-odd preconditioned, and no projection of the subspace corresponding to
small eigenvalues of the kernel was attempted.

\subsection{Future Work}

There are still a lot of things to investigate about algorithms for dynamical
on-shell chiral fermions:
\begin{itemize}
 \item \emph{Five-dimensional or four-dimensional dynamics.} Should we
       introduce five-dimensional pseudofermions as in
       \secref{sec:constraint}, or only have four-dimensional pseudofermion
       fields with a five-dimensional linear equation solver? Five-dimensional
       pseudofermions would seem to introduce extra noise into the system for
       no obvious benefit, so the latter seems to be a more promising
       choice. The Pauli--Villars fields that need to be introduced to cancel
       the effects of the unwanted five-dimensional pseudofermions also
       introduce extra noise, unless they are generated from the same noise
       fields.
 \item \emph{Multiple pseudofermion acceleration.} Do multiple pseudofermion
       fields allow for a significant increase in step size, as discussed in
       the second lecture? This has been studied for DWF, where the answer is
       affirmative, but it would be interesting to investigate it for
       four-dimensional pseudofermion dynamics.
 \item \emph{Five-dimensional multishift solver.} If we want to apply the RHMC
       multiple pseudofermion method discussed in the second lecture to
       four-dimensional pseudofermion dynamics then we would like to be able to
       use a multishift solver to evaluate all the partial fractions in the
       same Krylov space. Unfortunately we have no idea how do this, as a
       constant shift of the four-dimensional system, which is the Schur
       complement of our five-dimensional systems, is not a constant shift of
       the five-dimensional matrix. This makes either a Hasenbusch-like scheme
       (\secref{sec:hasenbuschery}) or a nested four-dimensional solver look
       more promising at the moment.
 \item \emph{Nested Krylov solvers.} Instead of using a five-dimensional Schur
       complement approach as discuss in this lecture we could use a nested
       `inner--outer' solver. For the inner solve we use a multishift technique
       to apply the partial fractions of a rational representation of the
       \(\sgn\) function, for the outer we may use it to implement multiple
       pseudfermion acceleration (as above), or multiple valence masses if we
       are carrying out valence measurements. The disadvatage is that we keep
       building four-dimensional inner Krylov spaces and then discarding them,
       which seems wasteful. At present, for a single solve (as opposed to a
       multishift outer solve), the five-dimensional solvers may be 5--10 times
       faster. There have been some studies of how the residuals for the inner
       solver should be chosen to ensure the convergence of the outer solver.
 \item \emph{Tunneling between different topological sectors.} Whatever
       algorithm we use for dynamical GW fermions it will suffer from severe
       critical slowing down as \(\mu\to0\). The potential barriers between
       different topological charge sectors grow as \(\mu\) gets small, and the
       sectors decouple for \(\mu=0\). It may be thought that for \(\mu=0\) we
       would never want to tunnel from the zero topological charge sector to
       one of a different topology, but we must be careful as it is not
       manifestly obvious that the topological charge zero sector is connected.
 \item \emph{Reflection/refraction} algorithms have been introduced to
       alleviate the problems with topology change. As the problems have so far
       only been seen on tiny lattices, where the integration step size \(\dt\)
       may be large compared to the width of the barriers, it remains to be
       seen how necessary or effective these approaches are.
\end{itemize}

\section*{Acknowledgments}

I would like to thank Dr. Naoto Tsutsui who prepared the excellent first draft
of the write-up of these lectures, and to the organizers of the ILFTN workshop
in Nara for arranging such a stimulating and interesting meeting.

\end{document}